\documentclass[preprint,aps]{revtex4}
\usepackage{units}
\usepackage{graphicx}
\usepackage{amsmath}
\usepackage{hyperref}
\usepackage{natbib}
\usepackage{xcolor}
\usepackage{subfig}

\begin{document}

\title{Modeling the iron oxides and oxyhydroxides for the prediction of 
environmentally sensitive phase transformations}
\author{Haibo Guo}
\author{Amanda S. Barnard}
\affiliation{Virtual Nanoscience Laboratory, Division of Materials Science and
Engineering, CSIRO, Australia}
\date{\today}

\begin{abstract}
Iron oxides and oxyhydroxides are challenging to model computationally as
competing phases may differ in formation energies by only several
kJ\,mol$^{-1}$, they undergo magnetization transitions with temperature, their
structures may contain partially occupied sites or long-range ordering of
vacancies, and some loose structures require proper description of weak
interactions such as hydrogen bonding and dispersive forces. If structures and
transformations are to be reliably predicted under different chemical
conditions, each of these challenges must be overcome simultaneously, while
preserving a high level of numerical accuracy and physical sophistication. Here
we present comparative studies of structure, magnetization, and elasticity
properties of iron oxides and oxyhydroxides using density functional
theory calculations with plane-wave and locally-confined-atomic-orbital
basis sets, which are implemented in VASP and SIESTA packages, respectively. We
have selected hematite ($\alpha$-Fe$_{2}$O$_{3}$), maghemite
($\gamma$-Fe$_{2}$O$_{3}$), goethite ($\alpha$-FeOOH), lepidocrocite
($\gamma$-FeOOH), and magnetite (Fe$_{3}$O$_{4}$) as model systems from a total
of 13 known iron oxides and oxyhydroxides; and use same convergence criteria and
almost equivalent settings in order to make consistent comparisons. 
Our results show both basis sets can reproduce the energetic stability and 
magnetic ordering, and are in agreement with experimental observations.
There are advantages to choosing one basis set over the other, depending on the
intended focus. In our case, we find the method using PW basis set most
appropriate, and combine our results to construct the first phase diagram of 
iron oxides and oxyhydroxides in the space of competing chemical potentials, 
generated entirely from first principles.

keywords: iron oxides and oxyhydroxides; phase diagram; density functional
theory; modeling and simulation.
\end{abstract}

\maketitle

\section{Introduction}

Iron oxides and oxyhydroxides are abundant in nature; they are widespread
in soils, waters and rocks, and are also found in living organisms,
air dusts, meteorites, and Martian soils.~\cite[p. 1-7]{Cornell2003}
Iron oxides and oxyhydroxides have been the focus of numerous studies
in the fields of geology, materials, soil, biology, and environmental
sciences, and have broad applications in pigments, magnetic recording
devices, medical imaging contrast agents, heavy metal sequestration
absorbents.~\cite[p. 2, 509-523]{Cornell2003} To date, 13 natural
and synthetic iron oxides and oxyhydroxides (in addition to 2 hydroxides,
see Ref.~\cite[p. 2]{Cornell2003}) have been identified. These polymorphs
have complicated structures (poor crystallization, ordering of vacancies,
partial site occupancy), undergo a range of phase transformations,
have characteristic magnetization states, and participate in a number
of different types of interactions with contaminants and adsorbates.
Given their ubiquity, it is surprising to find that the structures
of some iron oxides and oxyhydroxides remain poorly understood, even
after years of studies and numerous debates. In addition to this,
size effects introduce a further complication, especially when we
approach nanometer regimes, as shown in a recent review of the structure
complexity,~\cite{Navrotsky2008} in which the authors showed particle
size, hydrous and hydrated environments, and synthesis processes all
affect the observed structure. Collectively, these complicated issues
have fueled constant interests in iron oxides and oxyhydroxides over the past
decades.

Like many materials, the development of characterization technologies
and new samples often sparked renewed debates and led to new questions.
One example is the debate on the origins of magnetite found in meteorites
and magnetotactic bacteria. The magnetite nanocrystals from the Martian
meteorite ALH84001 share many features with that from magnetosomes
in terrestrial magnetotactic bacteria.~\cite{McKay1996} The similarities
include unusual morphology, chemical purity, and crystallographic
perfection. The similarities led to the proposal that the magnetite
nanocrystals from the Martian meteorite were produced by biogenic
processes, therefore provided strong evidence of lives in early 
Mars.~\cite{Thomas-Keprta2000,Thomas-Keprta2001}
This proposal was later questioned~\cite{Gibson2001,Buseck2001}
and even dismissed~\cite{Golden2004} because inorganic processes
can also produce similar morphologies. However, the debate triggered
new studies seeking reliable methods to identify origins or magnetite
nanocrystals, and crystal size distributions~\cite{Arato2005} and
oxygen isotope fractionation~\cite{Faivre2006} have now been proposed
to discriminate inorganic from organic origins. 

In recent years, computational modeling has opened up another potential
way to solve the pending questions about iron oxides and oxyhydroxides.
It generally requires electronic-level modeling methods to capture
the magnetization states of iron oxides and oxyhydroxides, and density
functional theory (DFT)~\cite{KS1965} is able to solve electronic
structures with desired accuracy at affordable computational cost. While DFT
implementations have been routinely used to solve a wide range of 
problems in materials
science, iron oxides and oxyhydroxides are particularly challenging
for a number of reasons. Firstly, the energy differences between different
solid phases or magnetization states may be as low as several kJ\,mol$^{-1}$,
which is close to the resolutions of most DFT calculations, and necessitates
energetic convergence criteria on the order of a few meV. Secondly,
the underestimation of band gaps by DFT makes it difficult to depict
the correct electronic structures of iron oxides and oxyhydroxides,
most of which are semiconductors. A remedy to this problem is to include
on-site Coulomb interaction to describe the strongly-correlated $3d$
electrons.~\cite{Anisimov1991a} Thirdly, the structures of iron
oxides and oxyhydroxides may have partially occupied sites, or long-range
ordering of vacancies (as in maghemite), which need large super cells
and, accordingly, heavy computation loads. Fourthly, the charge ordering
and associated symmetry change in magnetite below the Verwey transition
temperature~\cite{Cullen1971,Yoshida1977,Iizumi1982,Zuo1990,Zhang1991,
Madsen2005,Mazo-Zuluaga2005}
are computationally intractable. Working models proposed for charge
ordering generally go beyond most DFT implementations. Fifthly, some
iron oxyhydroxides have loose structure, eg. lepidocrocite ($\gamma$-FeOOH),
where the binding between layers relies on week hydrogen bonds and
dispersive forces which, however, are poorly described in DFT. And
finally, various magnetization states in iron oxides and oxyhydroxides
usually lead to slow convergence in calculations.

The challenges of iron oxides and oxyhydroxides make computational
modeling and simulation non-trivial tasks, and work in this area tends
to be sparse and sporadic. Despite the difficulties, DFT calculations
have been applied to some iron oxides and oxyhydroxides in the 
past.~\cite{Zhang1991,Punkkinen1999,Rosso2001,Rollmann2004,Chamritski2005,
Madsen2005,Shiroishi2005,Kubicki2008,Pentcheva2008,Martin2009,Pinney2009,
Russell2009,Wilson2009,Grau-Crespo2010}
These calculations incorporated different approximations, basis sets,
and computational settings. It is therefore difficult to compare their
accuracy and assess the methodology and algorithms, even though such
a comparison is highly desirable for selecting computation tools in
studying this difficult system. It also means that a systematic comparison
between studies is not necessarily reliable, and cross-comparisons
of different materials (such as those provided in phase diagrams)
is not possible. However, when we seek to overcome this problem, we
are confronted with the question of which is the most appropriate
technique to employ.

In this study, we will present a comparative study between
two implementations of DFT in calculating thermodynamic,
magnetic and elastic properties of iron oxides, and assess the efficiency,
accuracy and convergence, based on five iron oxides and oxyhydroxides,
including hematite ($\alpha$-Fe$_{2}$O$_{3}$), 
maghemite ($\gamma$-Fe$_{2}$O$_{3}$),
goethite ($\alpha$-FeOOH), lepidocrocite ($\gamma$-FeOOH), and magnetite
(Fe$_{3}$O$_{4}$). Based on this large and consistent set of results,
we are in a position to present the first environmentally sensitive
phase diagram of iron oxides and oxyhydroxides, generated entirely
from first principles, for predicting the thermodynamically stable
structure as a function of the supersaturation of oxygen and/or hydrogen.

\section{Computational methods}

A major difference among implementations of DFT is the choice of basis
sets to expand the state space.  Electronic wave functions can be constructed
by linear combination of delocalized plane waves (PW's), or locally-confined
atomic orbitals (LCAO's). The two basis sets have their advantages and
disadvantages.~\cite{Artacho2008,Poulet2003} PW's have definite
mathematical forms, are easy to implement, and have systematic convergence
over cutoff energies, but their delocalized nature prevents linear
scaling with the system size. LCAO's are flexible in terms of shape,
size, and range, require much less number of orbitals compared to
PW's, are localized thus suitable for spatial partition and linear
scaling algorithms, but lack a systematic convergence and require
extra effort to tune the LCAO parameters. 
We choose the DFT implementations in VASP (Vienna ab initial simulation
package)~\cite{Kresse1996:VASP1,Kresse1996:VASP2} for the PW basis set,
and SIESTA (Spanish initiative for electronic simulations with thousands
of atoms)~\cite{Artacho1999,Soler2002} for the LCAO basis set.

It is a known failure of local density approximation (LDA) or local spin
density approximation (LSDA) to accurately predict the ground state
of bulk iron, while generalized gradient approximation (GGA) can reproduce
the ferromagnetic BCC (body-centered cubic) ground state.~\cite{Leung1991}
Therefore, in our study, we choose GGA (in the form of Perdew, Burke
and Ernzerhof~\cite{Perdew1996}) for describing electron-electron
interactions. For consistency, we use the same exchange-correlation
functionals for both PW- and LCAO-based implementations, thereby enabling 
a complementary and detailed comparison between the two basis sets 
to assist others in this field. 

\subsection{Pseudopotentials}
\label{ss:pp}

In this study, we use pseudopotentials to describe core electrons
and nuclei. For the PW basis set, we use the projector augmented wave
(PAW) potentials from the pseudopotential libraries shipped with VASP.
The reference states of valence electrons for generating the pseudopotentials
of Fe is $3d^{7}4s^{1}$. The core radii are 2.30 Bohr, 1.1 Bohr,
and 1.52 Bohr for Fe, H, and O, respectively. Nonlinear core corrections
are included for Fe with radius of 2.0 Bohr. 

For Fe, the $3d$ electron orbital overlaps with $3s$ and $3p$ core
orbitals in real space, and has a small core radius of approximately
0.7 Bohr. $4s$ and $4p$ orbitals extrude further away from the nucleus,
with radii of approximately 2 Bohr. These different core radii make
it difficult to assign a common cutoff to all the orbitals, due to
the short core radius of $3d$ orbital ($\sim$0.7 Bohr), which requires
very large cutoff of plane-waves (about 11000 eV) to converge the
energy in 3 meV/atom.~\cite{Latham2006} One practice to eliminate
the difference in core radii is to include $3s$ and $3p$ as semi-core
states, in place of the $4s$ and $4p$ states respectively. In this
way, the reference state is not neutral (one $4s$ electron or two
$4s$ electrons are excluded, assuming the ground state is $3d^{7}4s^{1}$
or $3d^{6}4s^{2}$), which is acceptable under the pseudopotential
scheme. It is therefore possible to generate high-quality pseudopotentials
with small core radii of around 0.6--0.9\,Bohr. The hard pseudopotentials
can accurately reproduce all-electron calculations to excited states,
but are computationally demanding. However, it has been previously
shown the gain in quality of the calculation is not apparent when
semi-cores are included in Ti and Cu.~\cite{Trinite2008} In order
to reduce the computation cost, settings of $\sim$2\,Bohr radii
have been found to be good compromise between efficiency and cost.
The soft pseudopotentials often produce acceptable results in calculating
lattice parameters, magnetization and electronic structures. 

For the LCAO basis set, we generate norm-conserving pseudopotentials
according to the revised scheme of Troullier and Martins.~\cite{Troullier1991}
A potential generated with the reference valence state of $3d^{6}4s^{2}$,
core radii of 2.0\,Bohr, and partial core radius of 0.7\,Bohr was
used in previous studies.~\cite{Fagan2003,Izquierdo2000} Since we
wish to compare with our PW calculations, we have chosen the same
reference states for Fe ($3d^{7}4s^{1})$. According to our convergence
tests, the core radii are 2.0 Bohr for Fe, 1.1 Bohr for O, and 0.8
Bohr for H, smaller than those core radii of the PAW potentials for
the PW basis set. Nonlinear core corrections are included for both Fe
and O. We found the nonlinear core radius of 0.7 Bohr provides the
best match between pseudocore electron density and all-electron core-electron
density. The pseudopotential of Fe generated with the same configurations
has been used in studies to structure and magnetic properties of 
iron.~\cite{Izquierdo2000,Garcia-Suarez2008}
The nonlinear core radius of O is 0.7 Bohr, which is same as that
in ref.~\cite{Ortega-Castro2009}. We test the transferability of
the pseudopotentials by comparing atomic energies of excited states
from pseudopotentials and from all-electron calculations.

It is important to point out that a different nonlinear core radius
is used for Fe, and nonlinear core correction is excluded for O in
PAW potentials. These differences reflects to the degrees of compromise
between efficiency and transferability. Fortunately, the provision
of a standard pseudopotential database (by VASP) allows for considerable
testing in a large variety of situations, and the norm-conserving
pseudopotentials for LCAO have been tested in the above-mentioned
references. Therefore, we are confident that both sets of pseudopotentials
should represent core electrons of Fe, H, and O, and are adequate
for this comparative study.

\subsection{Basis sets}
\label{ss:basis}

PW's have a definite mathematical formula with no adjustable parameters.
LCAO-based basis sets use the so-called pseudo-atomic orbitals (PAO's)
whose shape, size, and range are configurable. The PAO's are mathematical
functions with adjustable parameters, which must be optimized for
specific systems, and the quality of the PAO's are critical to the
simulation results of LCAO basis sets. In the present study we
have optimized our PAO's by comparing simulated and known properties
of simple structures, specifically, the lattice parameter of bulk
body-centered cubic (BCC) Fe, and the bond lengths of $\mathrm{H}_{2}$
and $\mathrm{O}_{2}$ molecules.

The PAO's in the present study (for all the three elements) are of
double-$\zeta$ plus polarization (DZP). The dimensionless parameter
split-norm, which determines the splitting of different $\zeta$ functions,
was set to 0.28 for Fe, 0.24 for O, and 0.65 for H. The large split-norm
of H is in accordance with the large variation in the effective spatial
extent of hydrogen in charged states. A similar value of 0.5 was employed
during a study of the pressure effects on hydrogen bonds as reported
in Ref.~\cite{Winkler2008}. Soft confinement has been applied according
to the scheme proposed in Ref.~\cite{Junquera2001} to avoid discontinuity
of the functions at the cutoff distance. The parameters for generating
the PAO's are summarized in Table~\ref{tab:pao}. In the DZP scheme,
the numbers of PAO's per atom are 17 for Fe, 13 for O, and 3 for H. 
The results of bulk Fe and the gas molecules (H$_{2}$,
O$_{2}$, and H$_{2}$O) used to construct our basis sets, are provided
in the Section~\ref{sec:results}.

\begin{table}
\caption{PAO parameters of Fe, O and H. The $r_{\mathrm{c1}}$ and 
$r_{\mathrm{c2}}$are radii of double-$\zeta$ and polarization orbitals; 
$V$ is soft-confinement potential, and $r_{\mathrm{i}}$ is inner radius of 
the soft confinement; ``Polar.'' represents polarization orbitals.}
\label{tab:pao}
\begin{tabular}{ccccc}
\hline 
 & $r_{\mathrm{c1}}$(Bohr) & $r_{\mathrm{c2}}$ (Bohr) & $V$ (Ry) & $r_{\mathrm{i}}$ (Bohr) \\
\hline
Fe $3d$ & 4.229 & 2.292 & 50 & 3.81\\
Fe $4s$ & 6.800 & 5.363 & 150 & 6.12\\
Fe Polar. & 6.800 & - & 150 & 6.12\\
O $2s$ & 5.000 & 2.580 & 0 & -\\
O $2p$ & 6.500 & 2.497 & 0 & -\\
O polar. & 3.923 & - & 104.3 & 0.00\\
H $1s$ & 4.971 & 1.771 & 2.07 & 0.00\\
H polar. & 4.988 & - & 0.89 & 0.00\\
\hline
\end{tabular}
\end{table}

\subsection{\texorpdfstring{GGA$+U$}{GG+U} parameter}

The strong correlation effects of iron $3d$ electrons lead to splitting
of $d$ bands. Depending on the relative positions of oxygen $2p$
and iron $3d$ orbitals in valence bands, iron oxides and oxyhydroxides
may be semiconducting or metallic.~\cite[p. 115-117]{Cornell2003} 
Both GGA and LDA tend to over-delocalize electrons and underestimate
correlation effects and band gaps. Model Hamiltonian approaches are
often used in such strongly correlated systems.~\cite{Anisimov1991} 
In these models, electrons hopping between atoms experience the effective 
Coulomb interaction $U$, which is defined as the energy cost for
moving an electron between two atoms that both initially had the same
number of electrons, or $U=E_{n+1}+E_{n-1}-2E_{n}$, where $E_{n}$
is the energy of an atom with $n$ $3d$ (for transition metals) or
$4f$ (for rare earth elements) electrons.~\cite{Anisimov1991}
This energy fluctuations result in the formation of band gaps. As
implementation of the model Hamiltonian approaches in DFT, the LDA+$U$
(or GGA$+U$) method~\cite{Anisimov1991a,Dudarev1998} includes
on-site Coulomb interactions among strongly correlated electrons.

We point out that there exist alternative approaches to solve or alleviate the
band-gap problem of DFT, including hybrid HF-DFT
functionals~\cite{Becke1993,Perdew1996a}, and self-interaction
correction~\cite{Perdew1981,Svane1990}. These approaches (including the
aforementioned DFT$+U$) are being extensively tested in a large variety of
chemical environments and becoming widely implemented. Particularly for strongly
correlated systems, hybrid functionals have been shown to properly describe the
magnetic coupling in and band gaps of NiO~\cite{Moreira2002},
UO$_2$~\cite{Kudin2002}, CeO$_2$ and Ce$_2$O$_3$~\cite{Hay2006}, plutonium
oxides~\cite{Prodan2005}, and several strongly correlated
solids~\cite{Rivero2009}. Various hybrid functionals have been developed and
actively tested with other hybrid functionals and pure
functionals.~\cite{Csonka2009,Yang2010} Among the recent developments of hybrid
functionals, the range-separated
hybrids~\cite{Vydrov2006,Krukau2008,Henderson2008,Rivero2009} and
Heyd-Scuseria-Ernzerhof hybrid
functional~\cite{Heyd2003,Heyd2004,Peralta2006,Brothers2008} are very promising
in tackling the correlation effects in solids. Hybrid functionals often give
acceptable thermochemical results owing partly to their semi-empirical nature
and the fitting procedure (such that the amount of exact exchange can be tuned
to fit known physical and chemical properties). With the increasingly available
options, it is, however, desirable to select those density functional
approximations of nonempirical constraint satisfactory with least fitting
parameters.~\cite{Perdew2005} Choices of the approaches may depend on the
availability of implementations or computational cost. In this study, we have
chosen DFT$+U$ to account for the band-gap problems of DFT because it is
implemented in both computation packages (VASP and SIESTA) which are analyzing
herein. There are, of course, many other computation packages that use PW and
LCAO basis sets with different compromise between accuracy and computation cost.
Both the two packages we use for this study have users of broad interests,
ranging from physics, chemistry, materials science, and biology. They thus
should serve as robust computational tools for our study.

For the strongly-correlated systems, the improvements to band-structure
calculations provided by DFT+$U$ are substantial.~\cite{Dudarev1998} To
demonstrate this, we test the GGA+$U$ methods in the calculations of bulk Fe,
iron oxides and oxyhydroxides in both PW and LCAO methods. Both packages have
implemented the GGA+$U$ method based on a simplified rotationally invariant
formulation by Dudarev et al.~\cite{Dudarev1998}. In this implementation, only
the effective Coulomb repulsion $U_{\mathrm{eff}}=U-J$ is significant. In our
study, the on-site Coulomb interactions are included only for strongly
correlated Fe $3d$ electrons, but not for the electrons of O or H, or other
types of electrons of Fe.

The DFT$+U$ method has previously been employed to study
magnetite~\cite{Madsen2005,Zhang1991},
hematite~\cite{Punkkinen1999,Rollmann2004,Rohrbach2004}, 
goethite~\cite{Russell2009},
and maghemite~\cite{Grau-Crespo2010}, for which the parameter $U$ varies
between 2 eV to 5 eV. Cococcioni and Gironcoli suggested $U_{\mathrm{eff}}$ of
bulk iron to be $\sim2.2$ eV using a linear-response
approach;~\cite{Cococcioni2005} Anisimov and Gunarsson gave rather large
$U_{\mathrm{eff}}$ of about 6\,eV.~\cite{Anisimov1991} Rollmann et al
recommended $U_{\mathrm{eff}}$ of 3.0\,eV through their study of the electronic
structure of hematite.~\cite{Rollmann2004} Punkkinen et al suggested a much
smaller value ($\sim1.0$\,eV) for hematite, designed to reproduce experimentally
observed features of the electronic structure, such as the crystal field induced
band splitting.~\cite{Punkkinen1999} Grau-Crespo et al. used
$U_{\mathrm{eff}}=4.0$\,eV in the study of vacancy ordering of
maghemite.~\cite{Grau-Crespo2010} The difference may originate from
implementations of the DFT$+U$ method, pseudo-core configurations, and even
basis sets. In the present study, the parameter of $U_{\mathrm{eff}}$ is chosen
so that the calculated band gaps and lattice parameters both match the
experimental values. We found $U_{\mathrm{eff}}=4.5$\,eV provides best
match the experimental band gaps of hematite (see section~\ref{ss:hmt}) and
goethite (not shown), and acceptable lattice parameters of all the iron oxides.
The same $U_{\mathrm{eff}}$ was used for all the iron oxides for consistency.

It is worth noting that standard DFT (LDA or GGA) reproduces thermodynamic
properties very well, sometimes exceeding the predictions of the DFT$+U$
method in comparison to experiments. However, DFT$+U$ methods provide
much more accurate predictions of electronic structures. Ideally,
first principle methods should accurately predict both thermodynamic
and electronic properties, but this remains a goal of those involved
in the development of new density functionals. In our calculations,
we compare the results of GGA$+U$ with GGA, to assist others in selecting 
the most appropriate approach for their
work. 

\subsection{Computational Settings}

To facilitate a cross-comparison, we have used consistent settings
for all the iron oxides in both basis sets.  
The $k$-points for sampling over the Brillouin zone were generated
using Monkhorst-Pack scheme.~\cite{Monkhorst1976} For a primitive
cell of BCC Fe, a $k$-mesh grid of $23\times23\times23$, which corresponds
to 364 irreducible $k$-points in the first Brillouin zone, can achieve
convergence of total-energy in 2 meV/atom when using the PW basis set.
For the LCAO basis set, a $k$-grid of $17\times17\times17$ can reach
the same convergence of energy, and the number of $k$-points is 2457.
One immediately notices the large difference in the numbers of $k$-points
in the PW and LCAO basis sets. This is due to the different symmetrization
treatments in the two programs. The VASP code utilizes crystal symmetries
to calculate the charge density, forces, and stresses. The symmetry
elements of the crystal structure greatly reduce the number of necessary
$k$-points for adequate sampling. The SIESTA code is designed for
large systems, as its name indicates, and symmetry constraints are
usually excluded. SIESTA only trims a small amount of redundant $k$-points
from the constructed grid. Alternatively, SIESTA uses molecular dynamics
(MD) algorithms for geometry optimization over an auxiliary supercell.
This difference in symmetrization leads to a very different number
of $k$-points used in sampling the band energies, however, the convergence
criteria of $k$-mesh density with respect to total-energies are set
to 1-2 meV/atom in both basis sets. The sizes of $k$-mesh and numbers
of $k$-points used in the calculations are shown in Table~\ref{tab:nkpt}.

\begin{table}
\caption{Sizes of $k$-meshes and numbers of $k$-points in the calculations
using the PW and LCAO basis sets. The numbers in the column of PW are
the numbers of irreducible $k$-points in the first Brillouin zone,
and the numbers in the column of LCAO are the numbers of trimmed $k$-points.\label{tab:nkpt}}
\begin{tabular}{cccc}
\hline 
 & $k$-grid & PW & LCAO\\
\hline
magnetite & $4\times4\times4$ & 10 & 44\\
hematite & $4\times4\times4$ & 13 & 64\\
maghemite & $2\times2\times2$ & 1 & 8\\
goethite & $4\times6\times4$ & 24 & 60\\
lepidocrocite & $8\times4\times8$ & 32 & 150\\
\hline
\end{tabular}
\end{table}

In the PW basis set, we find a plane-wave cutoff of 800 eV can achieve
convergence in the total-energies to below 1.0 meV/atom for all the
five iron oxides and oxyhydroxides considered in our study. For bulk
Fe, a smaller cutoff (600 eV) is able to achieve the same convergence.
In the calculations of isolated $\mathrm{O_{2}}$, $\mathrm{H_{2}}$,
and $\mathrm{H_{2}O}$ molecules, the PW cutoffs are 850, 600, and
850 eV, respectively. With these cutoffs, the difference in total-energies
can be reduced to less than 1 meV/atom, which is the limiting resolution
of the DFT implementation. SIESTA uses a finite real-space grid over
which integrations are performed to calculated energies, forces, stresses,
and dipoles. The fineness of this finite grid is determined by a cutoff
value, which is equivalent to the PW cutoff in the PW basis set. There
are subtle differences between these equivalent settings across the
two basis sets. In the LCAO basis set, wave functions are constructed using
atomic orbitals, and the cutoff should only affect the accuracy of
integration; in the PW basis sets, the plane waves are also used to construct
the valence wave functions, so the cutoff has a larger impact on the
quality of calculations. After the convergence tests, we chose cutoffs
of 5130 eV for bulk Fe, 4080 eV for $\mathrm{O_{2}}$, 2040 eV for
$\mathrm{\mathrm{H}_{2}}$, and 6800 eV for all the 5 iron oxides
and oxyhydroxides, so that the total-energies converge below 2 meV/atom. 

Gaussian (in the PW basis set) or Fermi-Dirac (in the LCAO basis set) distribution
functions are used for electronic occupations for the molecules (H$_{2}$,
O$_{2}$, and H$_{2}$O), iron oxides and oxyhydroxides; Mehfessel-Paxton
functions of order 1 is used for bulk iron. The smearing width or
electronic temperature has been set to 0.03 eV for all the iron oxides
and oxyhydroxides, and 0.05 eV for bulk Fe, in both basis sets; relatively
small value (0.15 eV) is used for the isolated O$_{2}$ molecule,
and large values (0.4 or 0.5 eV) are used for the isolated H$_{2}$
and H$_{2}$O molecules (in both basis sets). These values are chosen
so that the energies diverge by less than 2 meV/atom compared with
smaller smearing widths. 

Energy minimizations to all the structures are conducted using conjugate
gradient (CG) algorithms with the force convergence of 0.005 eV/\AA.
For the iron oxides and oxyhydroxides, geometry optimizations of unit
cells are done with a convergence criterion of 0.005 GPa for the stress
tensor components. For the isolated molecules ($\mathrm{O}_{2}$,
$\mathrm{H}_{2}$ and $\mathrm{H}_{2}\mathrm{O}$), a large super
cell of $10\times10\times10$ \AA$^{3}$ is used (in both basis sets). 

\subsection{Magnetization states}

Iron oxides and oxyhydroxides undergo magnetic phase transitions at
different temperatures. Most of them are antiferromagnetic or ferrimagnetic
at temperatures below their Neel or Curie temperatures. Magnetite
and maghemite are ferrimagnetic; hematite, goethite, and lepidocrocite
are antiferromagnetic.~\cite[p.123]{Cornell2003} In this study,
we consider alternative magnetization states in addition to those
observed experimentally. By comparing the energetic stability of different
magnetization states, we are able to test the validity of our calculations.
In general, a non-spin polarized state, a ferromagnetic state, and
several other initial spin-polarization states are included. However,
we only consider collinear magnetization states, which are most commonly
observed in iron oxides and oxyhydroxides at low temperatures. 

For consistency, we have also included spin polarizations when calculating
the properties of the isolated molecules, even though $\mathrm{H}_{2}$
and $\mathrm{H}_{2}\mathrm{O}$ are non-magnetic (or diamagnetic).
In both PW and LCAO calculations, the net spin moments of $\mathrm{H}_{2}$
and $\mathrm{H}_{2}\mathrm{O}$ are zero, in agreement of experimental
observations. The spin moment of $\mathrm{O}_{2}$ is 2.0 $\mu_{\mathrm{B}}$
using both PW and LCAO basis sets.

\subsection{Elastic properties}

In this study we calculated bulk moduli of each solid material by
fitting to Birch-Murnagham equation of state.~\cite{Birch1947}
In addition to this, we calculate the elasticity tensors of bulk Fe,
iron oxides and oxyhydroxides using a finite-difference method. In
this method, a series of strains are applied to the equilibrium unit
cell, the total-energies of the strained structures are calculated,
and the elasticity tensor components $c_{ij}$ are calculated through:\begin{equation}
E=E_{0}+\frac{1}{2}\sum c_{ij}\epsilon_{i}\epsilon_{j},\end{equation}
where $E$ is the total-energies of strained structures, $E_{0}$
is the total-energy of equilibrium structure with zero stresses, and
$\epsilon$ is the applied strain. The subscripts $i$ and $j$ are
of matrix notations.~\cite[p. 134]{Nye1985} The strains are grouped
into a number of transformations, which are chosen in accordance with
the crystal symmetry of the structures. For each transformation, 6
strains of $\pm0.015$, $\pm0.010$, and $\pm0.005$, in addition
to the equilibrium structure, are used for linear least-square fitting
to calculate the $c_{ij}$ tensor components. 

We developed a computer program to calculate elastic constants of
crystals by using \textit{ab initio }packages as backends. Since this
method only requires total-energies, which can be calculated using many
computation packages, we can make consistent comparisons by 
using the same strains. 
This method and program have been previously tested in calculating elastic 
constants of Co~\cite{Guo2005PRB} and Ni-B alloys~\cite{Kong2010}.

\section{Results and discussions}
\label{sec:results}

In the following sections we will focus
on presenting results of our detailed comparisons between the PW and
LCAO basis sets, as well as the physical comparisons being made in energetic
stability, lattice parameters, elastic properties, and magnetization
states of our collection of iron oxides and oxyhydroxides. 

\subsection{Bulk Fe}

\begin{figure}
\includegraphics[scale=0.7]{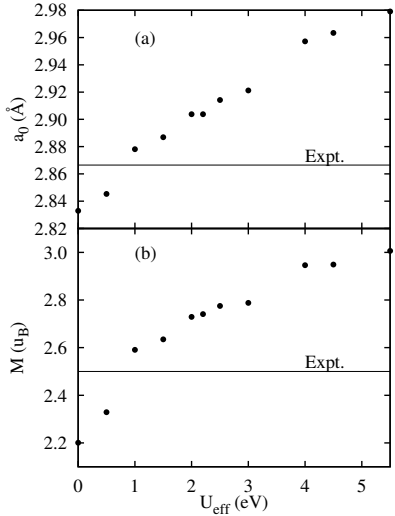} 
\caption{Dependence of (a) lattice parameter and (b) spin moment 
on $U_{\mathrm{eff}}$
in PW-based calculations. Experimental values are marked with horizontal
lines.\label{fig:ueff:fe}}
\end{figure}

The ground state of bulk iron is of body-centered cubic (BCC) structure
(space group $Im\bar{3}m$, No. 229) and ferromagnetic. Fe is a well-behaved
system within the framework of standard DFT-GGA, which predicts correctly
thermodynamic properties such as energetic stability and lattice parameters.
With on-site Coulomb interactions, first-principles calculations can
improve the predictions to electronic band structures. As mentioned
above, the parameter $U_{\mathrm{eff}}$ may vary from below 1 eV
to about 6 eV, depending on the methods and interpretations.~\cite{Pou2002,Postnikov2006,Garcia-Suarez2004,Cococcioni2005,Anisimov1991}
As our focus is on thermodynamic properties, we apply mild on-site
Coulomb interactions with $U_{\mathrm{eff}}=1.0$ eV when calculating
the properties of bulk iron. We choose this value of $U_{\mathrm{eff}}$
because it improves the predictions of the lattice constant and cohesive
energy in PW basis set (see Fig.~\ref{fig:ueff:fe} and Table~\ref{tab:Fe}).
In general, we find that the lattice constant of Fe increases almost
linearly with $U_{\mathrm{eff}}$. This is because on-site interactions
alter charge density around Fe atoms, weakening the metallic bonding
strength, similar to that observed in NiO.~\cite{Dudarev1998} The
spin moment, which is sensitive to changes of atomic volume, also
increases with $U_{\mathrm{eff}}$. 

The calculation results are summarized in Table~\ref{tab:Fe}. Both
calculations using PW and LCAO basis sets reproduce experimental lattice
constants within 1.5\%. All the calculations overestimate cohesive
energy with respect to experimental measurement. LCAO overestimates
by about 1.5 eV. The difference between calculation and experiment
is much smaller in PW basis set. GGA using PW overestimates
by about 0.5 eV, while GGA$+U$ reduce the overestimation to about
0.1 eV. The calculations of the spin polarization moments (not including
orbital moments) compare favorably with experiments, at around 2.5
$\mu_{B}$. 

For the calculations of the bulk moduli, GGA$+U$ using PW best matches
the results from experiments, while we find other methods overestimate
the values by between 8\% (GGA with LCAO) to 18\% (GGA with PW). Both fitting
errors and temperature effect may contribute to the difference between 
calculations and experiments, because bulk
moduli are calculated at ground state from Birch-Murnagham equation
of state, and experiments are conducted at the thermodynamic standard
state. For the calculations of the elasticity tensors, we find that
GGA using LCAO provides the best overall results, and other
methods either underestimate $c_{44}$ or overestimate $c_{11}$ and
$c_{12}$ significantly. In particular, the components of the elasticity
tensor calculated using GGA and the PW basis set can be considerably
different to the experimental values, especially in the case of $c_{11}$,
but they are very close to those in recent calculations using exact
muffin-tin orbitals and PBE functionals.~\cite{Zhang2010} However,
GGA$+U$ tend to considerably underestimate $c_{44}$
using both PW and LCAO. The differences between calculations and experiments
may include defects in single crystal Fe being measured, extrapolation
to ground state, anharmonic effects, and numerical error in the calculations. 

\begin{table}
\caption{Calculation results of ferromagnetic Fe in BCC structure. $a_{0}$:
equilibrium lattice constant; $E_{\mathrm{c}}$: cohesive energy;
$B_{0}$: bulk modulus; $M$: spin polarization moment per Fe atom;
$c_{ij}$: elasticity tensor components. Statistic errors of linear
least square fitting are included for $c_{ij}$.\label{tab:Fe}}
\begin{tabular}{lrrrrr}
\hline 
 & \multicolumn{2}{c}{PW} & \multicolumn{2}{c}{LCAO} & Expt. \\
\cline{2-5} 
 & GGA  & GGA$+U$  & GGA  & GGA$+U$  & \\
\hline 
$a_{0}$\,(\AA{})  & 2.833  & 2.878  & 2.868  & 2.909  & 2.87$^{(a)}$ \\
$E_{\mathrm{c}}$\ (eV) & 4.94 & 4.36 & 5.97 & 5.64 & 4.28$^{(b)}$\\
$M$\,($\mu_{\mathrm{B}}$)  & 2.20  & 2.79 & 2.31 & 2.67  & (about 2.5)\\
$B_{0}$\,(GPa)  & 198.4  & 164.6 & 182.1 & 188.0  & 168.3$^{(c)}$\\
$c_{11}$\ (GPa)  & 302.9$\pm$1.4 & 207.4$\pm$0.1 & 262.3$\pm$8.1 & 230.6$\pm$1.5  & 243.1$^{(d)}$, 239.3$^{(e)}$, 297.8$^{(f)}$\\
$c_{12}$\ (GPa)  & 151.6$\pm$1.5 & 151.0$\pm$0.2 & 126.8$\pm$14.7 & 165.0$\pm$1.8  & 138.1$^{(d)}$, 135.8$^{(e)}$, 141.9$^{(f)}$\\
$c_{44}$\ (GPa)  & 97.8$\pm$1.4 & 58.9$\pm$0.2 & 97.0$\pm$1.8 & 73.1$\pm$1.4  & 121.9$^{(d)}$, 120.7$^{(e)}$, 106.7$^{(f)}$\\
\hline
\multicolumn{6}{l}{(a) Ref.~\cite[p.23]{Kittel1996}.}\\
\multicolumn{6}{l}{(b) Ref.~\cite{Philipsen1996}.}\\
\multicolumn{6}{l}{(c) Ref.~\cite[p.59]{Kittel1996}.}\\
\multicolumn{6}{l}{(d) Ref.~\cite{Rayne1961}.}\\
\multicolumn{6}{l}{(e) Ref.~\cite{Adams2006}.}\\
\multicolumn{6}{l}{(f) Ref.~\cite{Zhang2010}.}\\
\end{tabular}
\end{table}

As shown in Table~\ref{tab:Fe}, GGA+$U$ generally offers a small
improvement over GGA in calculating the lattice constant
and cohesive energy of bulk iron, at the expense of apparent underestimation
of $c_{44}$.

\subsection{Gas molecules}

As stated above, we have calculated the binding energies and bond
lengths of H$_{2}$ and O$_{2}$, and bond angles of H$_{2}$O (see
Table~\ref{tab:gas}). Except for the binding energy of O$_{2}$,
the calculation results match experimental values within 2.5\%. The
significant overestimation of binding energy of oxygen dimer (and
all other first row elements with more-than-half-filled $p$-orbitals)
by DFT is due to the insufficient description of exchange energy,
and lack of error canceling because of different electron shapes of
O and O$_{2}$.~\cite{Behler2004,Gunnarsson1985} Since the energy
of oxygen dimer tend to be canceled out when calculating the energy
differences between different phases, this overestimation is unproblematic
in calculations of compounds, bulk iron oxides and oxyhydroxides. 

\begin{table}
\caption{Calculated properties of gas molecules. $L$ stands for bond length,
$E$ for binding energy, $E_{\mathrm{f}}$ for formation enthalpy,
and $\alpha$ for bond angle of H-O-H in H$_{2}$O.\label{tab:gas}}
\begin{tabular}{lccc}
\hline 
 & PW & LCAO & Expt.$^{(a)}$\\
\hline
$L_{\mathrm{H-H}}$(Å) & 0.7500 & 0.7465 & 0.7414\\
$E_{\mathrm{H-H}}$ (eV/bond) & 4.538 & 4.749 & 4.521\\
$L_{\mathrm{O-O}}$(Å) & 1.2323 & 1.2422 & 1.2074\\
$E_{\mathrm{O-O}}$ (eV/bond) & 6.8074 & 6.2181 & 5.1697\\
$L_{\mathrm{O-H}}$(Å) & 0.9575 & 0.9754 & 0.9575\\
$\alpha_{\mathrm{H-O-H}}$(º) & 104.46 & 104.93 & 104.51\\
$E_{\mathrm{f}}$(H$_{2}$O) (kJ\,mol$^{-1}$) & -243.8 & -234.3 & -241.8\\
\hline
\multicolumn{4}{c}{(a) Ref.~\cite{CRChandbook2007}, pages 9-22, 9-24, 9-57, 9-58.}\\
\end{tabular}
\end{table}

\subsection{Hematite (\texorpdfstring{$\alpha$-Fe$_2$O$_3$}{alpha-Fe2O3})
\label{ss:hmt}}

\begin{figure}
\subfloat[AFM]{\includegraphics[width=0.3\linewidth]{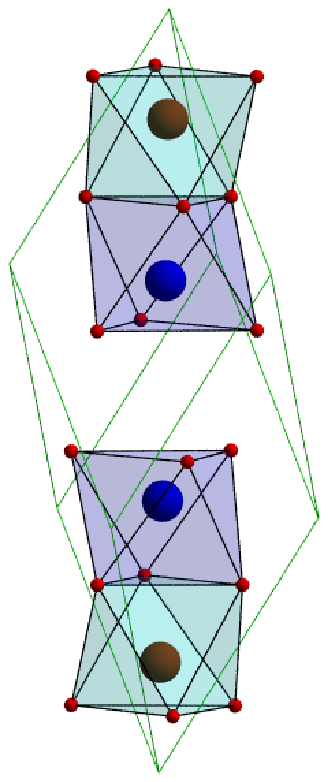}}
\subfloat[AFM']{\includegraphics[width=0.3\linewidth]{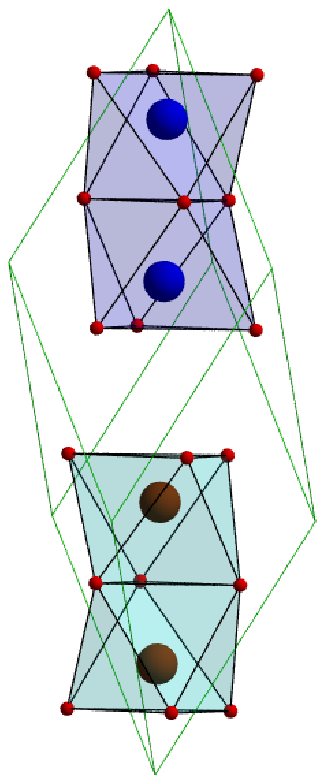}}
\subfloat[AFM'']{\includegraphics[width=0.3\linewidth]{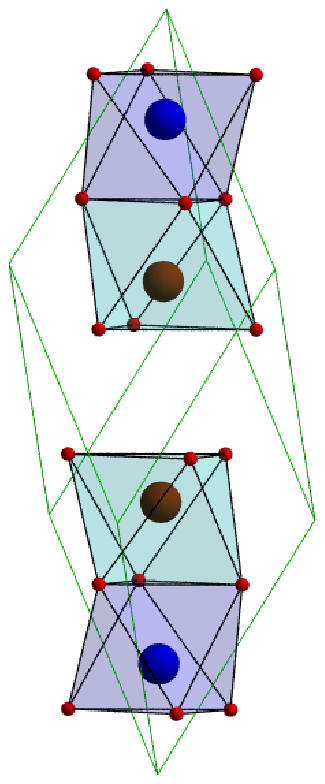}}
\caption{Antiferromagnetic configurations of hematite. Antiparallel spin moments are in blue and cyan colors. The rhombohedral unit cell is marked by green lines (color online). \label{fig:hmtt:mag}}
\end{figure}

Hematite belongs to the trigonal space group of R$\bar{3}$c (No.~167),
and is isostructural with corundum Al$_{2}$O$_{3}$ or Ilmenite (FeTiO$_{3}$).
It is one of the most thermodynamically stable and abundant phases
among all of the iron oxides and oxyhydroxides.~\cite[p.6]{Cornell2003}
Each rhombohedral unit cell contains 4 Fe atoms, distributed over
2 interlayer spaces of cation layers. Hematite is antiferromagnetic
with all Fe ions in the same close-packing layer (perpendicular to
the trigonal axis [0001]) having parallel spin moments, and different layers
having antiparallel spin moments, noted as AFM (see Fig.~\ref{fig:hmtt:mag}). 
At low temperatures below about 250 K,
the spin moments change direction from perpendicular to parallel to
the trigonal axis, keeping the antiferromagnetic configuration;~\cite{Morin1950}
and no reports have found that the crystal structure changes at this
magnetic transition. In order to validate our calculation results
on hematite, we have included another two types of antiferromagnetic
configurations in which Fe ions in the same layer have antiparallel
spin (noted as AFM' and AFM''; see Fig.~\ref{fig:hmtt:mag}), a
ferrimagnetic (FiM), a ferromagnetic (FoM), and a non-magnetic (NM)
configurations.

\begin{figure}
\includegraphics[width=0.9\linewidth]{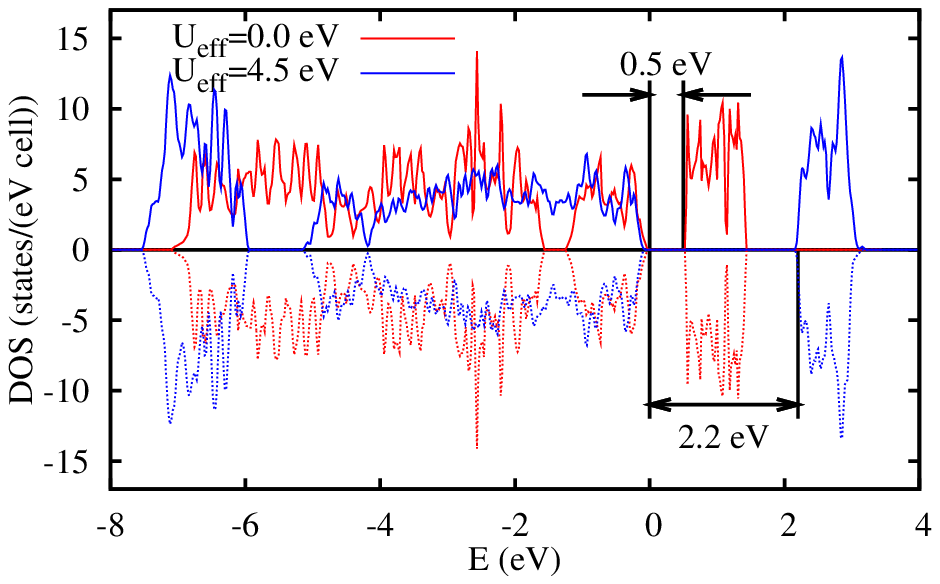} 
\caption{Density of states of antiferromagnetic (AFM) hematite. 
Up- and down-spin states are in solid and dashed lines, respectively. 
The numbers are band gaps for $U_{\mathrm{eff}}=0$\,eV (no correction of 
on-site interaction) and $U_{\mathrm{eff}}=4.5$\,eV. The Fermi energy is
shifted to 0 eV. The density of states is calculated with fully relaxed
structures.\label{fig:hmtt:dos}}
\end{figure}

\begin{figure}
\includegraphics[width=0.9\linewidth]{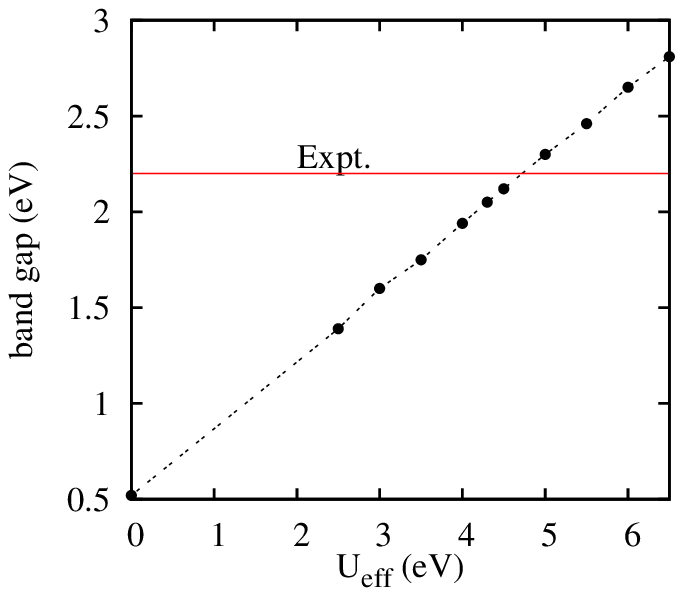} 
\caption{Dependence of band gap of hematite on $U_{\mathrm{eff}}$ in GGA$+U$
calculations.\label{fig:hmtt:bandgap}} 
\end{figure}

Ignoring on-site interactions leads to significant underestimation to the band
gap (0.5 eV in calculation compared with 2.2 eV from experiment; see
Fig.~\ref{fig:hmtt:dos}). We find the calculated band gap linearly increase with
$U_{\mathrm{eff}}$, as shown in Fig.~\ref{fig:hmtt:bandgap}. In choosing the
parameter $U_{\mathrm{eff}}$ in GGA$+U$ calculations, we fit the band gap to
experimental value (about 2.2 eV~\cite{Cornell2003}). To reproduce the
experimental value of 2.2\,eV, $U_{\mathrm{eff}}$ should be between 4.0 eV to 5
eV. We have therefore adopted $U_{\mathrm{eff}}=4.5$\,eV, and used this value
consistently in our calculations to all the iron oxides and oxyhydroxides (in
addition to hematite) using both PW and LCAO basis sets.

The calculated thermodynamic and elastic properties for hematite are
listed in Tables~\ref{tab:hmtt1} and \ref{tab:hmtt2}. We see both
PW and LCAO correctly predict the lowest energy state of the
antiferromagnetic configuration AFM, in agreement with experimental
observations.  The calculated lattice parameters match experimental
values within 3\% for this stable configuration.  In this case we find
PW does a better job of reproducing the lattice parameters
than LCAO. In both basis sets the lattice parameters from GGA+$U$
are slightly larger than those from GGA. The largest
difference between the two basis sets is the spin polarization moment
of the metastable ferromagnetic state (FoM). In PW, the average spin moment is
low (1.00\,$\mu_{B}$), in contrast with that
in LCAO, where the average spin moment is high (3.42\,$\mu_{B}$).
Both basis sets agree on the energetic order of the magnetization states,
predicting that AFM$<$FiM$<$FoM$<$NM (formation energy increasing),
but differ about the two antiferromagnetic states (AFM' and AFM'').

\begin{table}
\caption{Calculated thermodynamic properties of hematite. The numbers in parenthesis
are the errors in per cent compared with experimental values. The
spin polarization is total spin moment of a unit cell (so that
antiferromagnetization has exactly zero spin moment) divided by the number of
Fe atoms. The
energetically most stable magnetic states are in the first row of each basis 
set (PW or LCAO), GGA+$U$ calculations start from the energetically most
stable magnetic states. FoM stands for ferromagnetic, FiM for ferrimagnetic,
AFM for antiferromagnetic, and NM for non-magnetic. GGA$+U$ corresponds to
the magnetization state of the lowest energy in GGA calculations.
\label{tab:hmtt1}}
\begin{tabular}{cccccc}
\hline 
 &  & $a$\,(\AA{}) & $c$\,(\AA{}) & $M$\,($\mu_{\mathrm{B}}$) & $E_{\mathrm{f}}$\,(kJ \,mol$^{-1}$)\\
\hline
 & AFM & 5.005 ($-0.6$) & 13.884 ($+1.0$) & 0.00 & $-628.0$\\
 & AFM' & 4.841 ($-3.8$) & 13.183 ($-4.1$) & 0.00 & $-561.2$\\
 & AFM'' & 5.044 ($+0.2)$ & 13.850 ($+0.7$) & 0.00 & $-588.1$\\
PW & FiM & 4.977 ($-1.1$) & 13.707 ($-0.3$) & 1.50 & $-592.3$\\
 & FoM & 4.783 ($-5.0$) & 13.333 ($-3.0$) & 1.00 & $-569.3$\\
 & NM & 4.733 ($-6.0$) & 13.511 ($-1.8$) & - & $-543.0$\\
 & GGA+$U$ & 5.074 ($+0.8$) & 13.874 ($+0.9$) & 0.00 & -\\
\hline
 & AFM & 5.091 ($+1.1$) & 13.995 ($+1.8$) & 0.00 & $-676.8$\\
 & AFM' & 5.167 ($+2.6$) & 13.781 ($+0.2$) & 0.00 & $-640.1$\\
 & AFM'' & 5.137 ($+2.0$) & 13.955 ($+1.4$) & 0.00 & $-638.5$\\
LCAO & FiM & 5.056 ($+0.4$) & 13.809 ($+0.4$) & 1.50 & $-629.6$\\
 & FoM & 5.026 ($-0.2$) & 13.955 ($+1.5$) & 3.42 & $-585.6$\\
 & NM & 4.751 ($-5.6$) & 13.755 ($-0.0$) & - & $-536.7$\\
 & GGA+$U$ & 5.183 ($+3.0$) & 14.072 ($+2.3$) & 0.00 & -\\
\hline
\multicolumn{2}{c}{Expt.$^{(a)}$} & 5.034 & 13.752 & 0.00 & $-823\sim-828$\\
\hline
\multicolumn{6}{l}{(a) Measured at room temperature and 0.1 MPa. Ref.~\cite[pp. 11 and 187]{Cornell2003}}\\
\end{tabular}
\end{table}

In the case of hematite, the calculated elastic constants using PW
compare favorably with those determined using LCAO.  Both basis sets give
almost zero $c_{14}$ (within numerical errors), while the calculation
results of $c_{44}$ differ by $\sim 30$ GPa between PW and LCAO.
We notice that GGA$+U$ produce larger values of $c_{33}$
than GGA using both PW and LCAO basis sets, indicating on-site interactions
strengthen bonding along the trigonal axis. All the elasticity tensors
satisfy the elastic stability condition, which means hematite is elastically
stable in all the four calculations.

\begin{table}
\caption{Calculated bulk moduli and elastic constants of hematite. Unit: GPa.
\label{tab:hmtt2}}
\begin{tabular}{ccc|cc}
\hline 
 & \multicolumn{2}{c|}{PW} & \multicolumn{2}{c}{LCAO}\\
\cline{2-5} 
 & GGA & GGA+$U$ & GGA & GGA+$U$\\
\hline
$B$ & 174.4 & 190.3 & 173.4 & 176.1\\
$c_{11}$ & $325.0\pm19.0$ & $355.4\pm13.9$ & $310.3\pm12.1$ & $319.6\pm18.6$\\
$c_{12}$  & $131.8\pm5.6$ & $132.1\pm6.4$ & $137.2\pm0.3$ & $125.6\pm4.3$\\
$c_{13}$ & $105.8\pm22.9$ & $116.0\pm4.0$ & $114.6\pm6.9$ & $104.5\pm1.5$\\
$c_{14}$ & $1.2\pm2.9$ & $-5.6\pm6.4$ & $6.1\pm6.9$ & $5.5\pm9.4$\\
$c_{33}$ & $264.2\pm25.1$ & $307.2\pm3.3$ & $255.6\pm3.8$ & $294.4\pm9.8$\\
$c_{44}$ & $103.0\pm6.4$ & $110.6\pm7.0$ & $78.4\pm4.9$ & $80.1\pm5.2$\\
\hline
\end{tabular}
\end{table}

\subsection{Maghemite (\texorpdfstring{$\gamma$-Fe$_{2}$O$_{3}$}{gamma-Fe2O3})}

Maghemite occurs as a weathering product of magnetite, and resembles
magnetite in structure and magnetic properties. The Fe ions are all
in the trivalent state, with balancing vacancies to maintain charge
neutrality. The crystal structure of maghemite has been characterized
to be cubic, the same as magnetite, with partially occupied vacancies
at octahedral sites.~\cite{Greaves1983,Somogyvari2002} Depending
on the ordering of cation vacancies, maghemite may be classified in
either cubic (F$d\bar{3}m$ or P$4_{3}32$) or tetragonal (P$4_{1}2_{1}2$)
space groups. Somogyvári et al reported long-range ordering of vacancies
in powder neutron and XRD of nanocrystalline needle-shaped maghemite,
and classified maghemite to be in P$4_{1}2_{1}2$ space group.~\cite{Somogyvari2002}
Using powder neutron diffraction, Greaves proposed the true symmetry
of maghemite is tetragonal P$4_{3}2_{1}2$ instead of cubic 
P$4_{3}32$.~\cite{Greaves1983}
The lattice parameters of the tetragonal cell are $a=8.3396$\,\AA,
and $c=24.966$\,\AA{} which is slightly smaller $3a$.~\cite{Greaves1983}
Grau-Crespo et al sorted out the energetic order of various possible
vaccancy ordering and found the tetragonal P$4_{1}2_{1}2$ configuration has the
much lower energy (by $\ge 32$\,kJ\,mol$^{-1}$) than non-tetragonal
configurations using classical interatomic potentials.~\cite{Grau-Crespo2010}
The configurations P$4_1 2_1 2$ and P$4_3 2_1 2$ bear much similarity in
structures and thus should have very similiar energetics. In this study we
adopted the configuration proposed by Greaves~\cite{Greaves1983} 
(tetragonal P$4_{3}2_{1}2$ symmetry), each unit cell having 160 (64 Fe and 96 O)
atoms. Maghemite is ferrimagnetic below Curie temperature which is estimated
to be between 820 K and 960 K. The Fe atoms at the tetrahedral sites
(where each Fe forms bonds with 4 nearest O atoms) have antiparallel spin
moments with those at the octahedral sites (where each Fe forms bonds with 6
nearest O atoms). Specifically, the 40 Fe atoms in the $3\times1$ supercell at
positions [1/8,5/8,0], [3/8,1/8,2/24], [1/8,7/8,2/24], [7/8,5/8,2/24],
[3/8,3/8,0], [7/8,7/8,0] consist the majority spin component, and the 24 Fe
atoms at [4/8,6/8,1/24], [0,2/8,1/24], [2/8,4/8,3/24] consist the minority 
spin moment. Measurements of magnetic
moments (spin polarization $+$ orbital moment) showed Fe atoms at the 
octahedral and tetrahedral sites have unequal spin moments: 
3.54 $\mu_{\mathrm{B}}$ versus 4.03 $\mu_{\mathrm{B}}$,~\cite{Somogyvari2002} 
or $4.18\,\mu_{\mathrm{B}}$ versus $4.41\,\mu_{\mathrm{B}}$.~\cite{Greaves1983} 
In addition to this ferrimagnetization, we include ferromagnetic and non-magnetic 
states for validation of the calculation results. 

For the ferrimagnetic state, the calculated lattice constants match
experimental values within 1.7\% (see Table~\ref{tab:mght1}), with
the exception of results from GGA$+U$ using LCAO, in which the errors
are about 3.1\%. In general, the results from PW calculations are
closer to those reported from experiments. If we omit spin polarization,
the lattice parameters simultaneously decrease in both calculations,
and the mismatch in lattice constants between the calculations and
experiments increases to $-4.4$\%. Including on-site interaction
leads to a lattice expansion of about 1\% in both PW and LCAO. Both
calculations reproduce the correct magnetic ordering, predicting that
that ferrimagnetic state has a lower formation energy than non-magnetic
and ferromagnetic states. By comparing Tables~\ref{tab:mght1} and
\ref{tab:hmtt1}, one immediately sees maghemite has a higher formation
energy, and thus less thermodynamically stable than hematite. 

In the case of maghemite, the calculated elastic properties are very
similar, both in trend and numbers, when we compare the PW and LCAO
calculations (see Table~\ref{tab:mght2}). The diagonal components
($c_{11}$ and $c_{33}$) of the elasticity tensor are noticeably
larger in GGA$+U$ than GGA; the shear moduli ($c_{44}$
and $c_{55}$) are also slightly larger when using GGA$+U$. We find
that in all the calculations, $c_{11}\approx c_{33}$, $c_{12}\approx c_{13}$,
and $c_{44}\approx c_{55}$, which are conditions characteristic of
the elasticity tensors of cubic crystals. This in indicative of the
similarity between the tetragonal lattice of maghemite with its cubic
counterpart. Although the long-range ordering of vacancies changes
the symmetry of lattice, the elasticity tensor seems to be insulated
from the change of symmetry. 

\begin{table}
\caption{Calculated thermodynamic properties of maghemite. See Table~I for
annotations.\label{tab:mght1}}
\begin{tabular}{llllll}
\hline 
 &  & $a$\,(\AA{}) & $c$\,(\AA{}) & $M$\,($\mu_{\mathrm{B}}$) & $E_{\mathrm{f}}$\,(kJ \,mol$^{-1}$)\\
\hline
 & FiM & 8.363 ($+0.3$) & 25.034 ($+0.3$) & 1.25 & $-623.9$\\
PW & FoM & 8.192 ($-1.8$) & 24.563 ($-1.6$) & 2.75 & $-568.9$\\
 & NM & 8.026 ($-3.8$) & 24.871 ($-4.4$) & - & $-508.5$\\
 & GGA+$U$ & 8.428 ($+1.1$) & 25.237 ($+1.1$) & 1.25 & -\\
\hline 
 & FiM & 8.480 ($+1.7$) & 25.374 ($+1.6$) & 1.25 & $-661.4$\\
LCAO & FoM & - & - & - & -\\
 & NM & 8.059 ($-3.4$) & 24.153 ($-3.3$) & - & $-496.5$\\
 & GGA+$U$ & 8.598 ($+3.1$) & 25.718 ($+3.1$) & 1.25 & -\\
\hline 
\multicolumn{2}{c}{Expt.$^{(a)}$} & 8.34 & 24.97 &  & -806\textasciitilde{}-813\\
\hline 
\multicolumn{6}{l}{(a) Measured at room temperature and 0.1 MPa. Ref.~\cite[pp. 11 and 187]{Cornell2003}}\\
\end{tabular}
\end{table}

\begin{table}
\caption{Calculated bulk moduli and elastic constants of maghemite. Unit: GPa.
\label{tab:mght2}}
\begin{tabular}{ccc|cc}
\hline 
 & \multicolumn{2}{c|}{PW} & \multicolumn{2}{c}{LCAO}\\
 & FiM & GGA+$U$ & FiM & GGA+$U$\\
\hline
$B_{0}$ & 146.3 & 147.8 & 134.4 & 145.9\\
$c_{11}$ & $264.3\pm27.0$ & $285.0\pm20.8$ & $245.2\pm12.3$ & $266.3\pm31.2$\\
$c_{12}$  & $122.5\pm16.8$ & $120.0\pm12.6$ & $114.2\pm9.0$ & $113.2\pm22.7$\\
$c_{13}$ & $124.4\pm17.9$ & $120.1\pm14.9$ & $113.1\pm1.1$ & $114.2\pm15.4$\\
$c_{33}$ & $265.7\pm10.7$ & $284.1\pm9.4$ & $246.0\pm4.3$ & $266.7\pm12.7$\\
$c_{44}$ & $103.7\pm0.2$ & $106.0\pm0.1$ & $90.9\pm2.7$ & $94.3\pm3.2$\\
$c_{55}$ & $103.0\pm0.2$ & $106.5\pm0.0$ & $92.4\pm1.8$ & $95.8\pm6.5$\\
\hline
\end{tabular}
\end{table}

\subsection{Goethite (\texorpdfstring{$\alpha$-FeOOH}{alpha-FeOOH})}

\begin{figure}
\subfloat[AFM]{\includegraphics[width=0.5\linewidth]{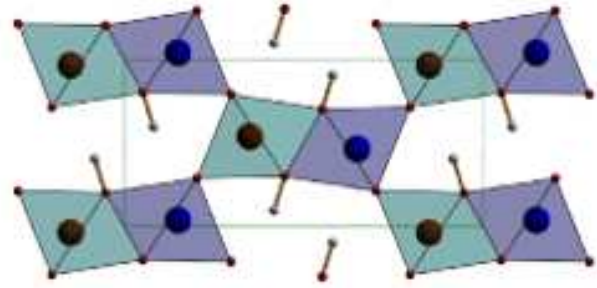}}\\ 
\subfloat[AFM']{\includegraphics[width=0.5\linewidth]{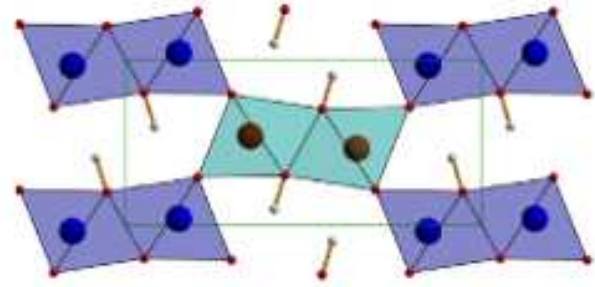}}\\ 
\subfloat[AFM'']{\includegraphics[width=0.5\linewidth]{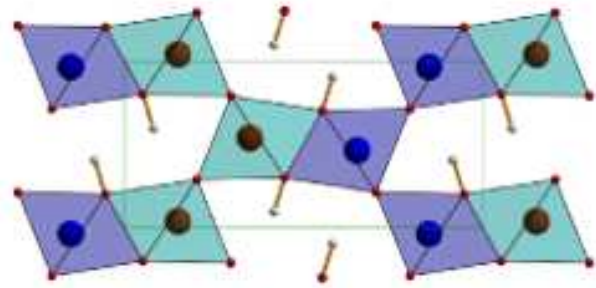}}\\ 
\caption{Antiferromagnetic configurations of goethite. Cyan and blue octahedron
are for Fe atoms with antiparallel spin moments. Red balls are oxygen,
white balls are hydrogen, sticks are hydroxyl bonds. Viewed along
[010] direction.\label{fig:gtht:afm}}
\end{figure}

Goethite is the most thermodynamically stable iron oxyhydroxide, and
has orthorhombic structure (space group Pnma, No. 62).~\cite{Gualtieri1999}
The lattice parameters have been measured by synchrotron powder diffraction
at temperatures between 298 K and 429 K,~\cite{Gualtieri1999}, and
at pressures up to 9 GPa,~\cite{Nagai2003}. Gleason et al.~\cite{Gleason2008}
studied the equation of state of goethite under pressures 0--250 GPa,
and found the equilibrium volume is $138.75\pm0.02$\,\AA{}$^{3}$,
bulk modulus is $140.3\pm3.7$\,GPa, and pressure derivative is $4.6\pm0.4$.
Goethite is antiferromagnetic in its ground state, with edge-sharing
octahedron within a double-chain have antiparallel spin-moments, and
corner-sharing octahedron in two double-chains have antiparallel spin-moments
(see Figure~\ref{fig:gtht:afm}). In addition to this antiferromagnetic
state, we included another two antiferromagnetic states; one has same
spin in a double-chain (noted as AFM'), the other one is similar to
AFM except the corner-sharing octahedron have parallel spin (noted
as AFM''). We have also calculated a ferrimagnetic, a ferromagnetic,
and a non-magnetic state.

The energetic order of antiferromagnetic states are the same in both
PW and LCAO calculations. The energy difference between AFM and AFM'
is only about 3 kJ\,mol$^{-1}$, which is, however, near the limits
of the computation accuracy. The small energy difference between AFM
and AFM' is reproduced in both the GGA$+U$ and GGA calculations,
using both the PW and LCAO basis sets. Despite this reproducibility across
different basis sets, further calculations with high accuracy are required
to distinguish the energetic order of the two antiferromagnetic states.
The energies of AFM and AFM' are lower than AFM'' by about 30 kJ\,mol$^{-1}$
in both PW and LCAO, indicating that corner-sharing octahedron of
antiparallel spins (as in AFM and AFM') are energetically more stable
than that of parallel spins (as in AFM''). In this study, we assume
the AFM state is more energetically stable than AFM', and perform
calculations of elastic properties based on the AFM magnetization
state with and without on-site interaction. 

As we see from the calculations results in previous parts (hematite
and maghemite), the lattice parameters from GGA$+U$ are usually larger
than that of GGA. This trend is violated in the calculation to
lattice parameter $c$. In the PW calculations, the GGA$+U$ result
is smaller than GGA, while in the LCAO results, the calculated
$c$ are almost the same (Table~\ref{tab:gtht1}). This feature
is also seen in the calculation results of $b$-axis of lepidocrocite
(see Section~\ref{sec:lp}). Since the hydrogen bonds are almost
along $c$-axis in goethite (and $b$-axis in lepidocrocite), the
smaller values of lattice parameter $c$ in goethite(and $b$ in lepidocrocite)
indicates strengthening of hydrogen bonds in the GGA$+U$ calculations
compared with the GGA calculations.  The physical origin of this
observation is not clear yet. Possible explanations may be from the
redistribution of the charge density caused by the onsite Coulomb interactions.
Although Fe atoms are not part of the hydrogen bonds (H-O..H), they have an 
influence on the strength of hydrogen bonds by modifying the electron density 
in Fe-O bonds which (in turn) change the electron density around the oxygen 
atoms, which are acceptors of the hydrogen bonds. The onsite
Coulomb repulsion among Fe $3d$ electrons decreases the charge density
in Fe-O bonds, increasing the electron density around the oxygen atoms
and strengthening the hydrogen bonds. The changes in electron density
will be illustrated in more details in a separate paper.

The ferrimagnetic state has the same average spin-polarization moment
in both calculations. Like the results of hematite, the spin moment
of the ferromagnetic state is quite different: low spin in PW, and
high spin in LCAO. With the exception of this difference in spin moments
of ferromagnetic state, the calculation results from the PW and LCAO
basis sets are consistent with each other. 

\begin{table}
\caption{Calculated thermodynamic properties of goethite. See Table~I for
annotations. \label{tab:gtht1}}
\begin{tabular}{ccccccc}
\hline 
 &  & $a$\,(\AA{}) & b\,(\AA{}) & $c$\,(\AA{}) & $M$\,($\mu_{\mathrm{B}}$) & $E_{\mathrm{f}}$\,(kJ \,mol$^{-1}$)\\
\hline
 & AFM & 10.018 ($+0.6$) & 3.017 ($-0.1$) & 4.661 ($+1.2$) & 0.00 & -453.1\\
 & AFM' & 10.025 ($+0.7$) & 3.015 ($-0.2$) & 4.650 ($+0.9$) & 0.00 & -450.4\\
 & AFM'' & 10.045 ($+0.9$) & 3.055 ($+1.1$) & 4.655 ($+1.0$) & 0.00 & -419.6\\
PW & FiM & 10.039 ($+0.8$) & 3.042 ($+0.7$) & 4.664 ($+1.2$) & 2.50 & -432.9\\
 & FoM & 10.103 ($+1.5$) & 2.842 ($-5.9$) & 4.560 ($-1.0$) & 2.63 & -417.6\\
 & NM & 9.529 ($-4.3$) & 2.920 ($-3.3$) & 4.366 ($-5.2$) & - & -415.6\\
 & GGA+$U$ & 10.040 ($+0.8$) & 3.045 ($+0.8$) & 4.628 ($+0.4$) & 0.00 & -\\
\hline
 & AFM & 10.148 ($+1.9$) & 3.060 ($+1.3$) & 4.654 ($+1.0$) & 0.00 & -492.9\\
 & AFM' & 10.149 ($+1.9$) & 3.061 ($+1.3$) & 4.656 ($+1.0$) & 0.00 & -490.0\\
 & AFM'' & 10.171 ($+2.2$) & 3.092 ($+2.3$) & 4.683 ($+1.6$) & 0.00 & -463.7\\
LCAO & FiM & 10.177 ($+2.2$) & 3.079 ($+1.9$) & 4.673 ($+1.4$) & 2.50 & -475.2\\
 & FoM & 10.200 ($+2.4$) & 3.100 ($+2.6$) & 4.695 ($+1.9$) & 4.99 & -456.6\\
 & NM & 9.642 ($-3.2$) & 2.939 ($-2.7$) & 4.353 ($-5.5$) & - & -428.4\\
 & GGA+$U$ & 10.207($+2.5$) & 3.105($+2.8$) & 4.663 ($+1.2$) & 0.00 & -\\
\hline
\multicolumn{2}{c}{Expt.$^{(a)}$} & 9.956 & 3.021 & 4.608 & 0.00 & -559.3, -562.9\\
\hline
\multicolumn{7}{l}{(a) Measured at room temperature and 0.1 MPa. Ref.~\cite[pp. 11 and 187]{Cornell2003}.}\\
\end{tabular}
\end{table}

The calculated bulk moduli and elastic constants of the AFM state
are listed in Table~\ref{tab:gtht2}. We found GGA$+U$ calculations 
produce appreciably larger values of bulk moduli and most tensor components
than the GGA in both PW and LCAO basis sets. The reason for
the strengthening effect of GGA$+U$ is not clear. It may be related
to the hydrogen bonds which are sensitive to the distribution of electron
density, but further work will be needed to understand this definitively.

\begin{table}
\caption{Calculated bulk moduli and elastic constants of goethite. Unit: GPa.
\label{tab:gtht2}}
\begin{tabular}{ccc|cc}
\hline 
 & \multicolumn{2}{c|}{PW} & \multicolumn{2}{c}{LCAO}\\
 & AFM & GGA+$U$ & AFM & GGA+$U$\\
\hline
$B_{0}$ & 93.1 & 114.1 & 98.6 & 109.4\\
$c_{11}$ & $235.4\pm3.8$ & $298.4\pm4.4$ & $231.7\pm8.4$ & $252.2\pm5.3$\\
$c_{12}$  & $89.0\pm1.2$ & $106.2\pm4.8$ & $86.6\pm5.4$ & $89.2\pm0.4$\\
$c_{13}$ & $112.2\pm5.2$ & $117.4\pm3.6$ & $111.4\pm8.5$ & $117.7\pm6.0$\\
$c_{22}$ & $263.8\pm4.3$ & $347.9\pm3.6$ & $234.0\pm10.0$ & $271.7\pm1.4$\\
$c_{23}$ & $96.3\pm6.1$ & $105.7\pm2.0$ & $87.3\pm11.1$ & $99.0\pm1.7$\\
$c_{33}$ & $406.7\pm6.5$ & $414.8\pm3.9$ & $363.3\pm12.0$ & $369.2\pm5.1$\\
$c_{44}$ & $78.9\pm0.1$ & $105.0\pm0.3$ & $58.7\pm0.2$ & $91.0\pm0.4$\\
$c_{55}$ & $122.6\pm0.7$ & $131.4\pm0.0$ & $106.9\pm2.8$ & $122.4\pm2.1$\\
$c_{66}$ & $65.9\pm0.0$ & $97.3\pm0.6$ & $60.1\pm2.1$ & $72.3\pm4.8$\\
\hline
\end{tabular}
\end{table}

\subsection{Lepidocrocite
(\texorpdfstring{$\gamma$-FeOOH}{gamma-FeOOH})\label{sec:lp}}

Lepidocrocite has an orthorhombic structure (space group Cmc$2_{1}$,
No.~36~\cite{Christensen1978}), which consists of double chains
of Fe(O,OH)$_{6}$ octahedron which are aligned perpendicular to $b$-axis.
The double chains form sheets, held together mainly by hydrogen bonds,
which are weaker than covalent or metallic bonds, and may be longer
than normal chemical bonds. Depending on the position of hydrogen
atoms, the crystal structure of lepidocrocite can either be in the
Cmcm space group (No. 63) where the hydrogen atom reside at the middle
of two oxygen atoms in a hydrogen bond,~\cite{Ewing1935,Oles1970}
or in the Cmc$2_{1}$ space group (No.~36) where the hydrogen atom
is closer to one of the two oxygen atoms.~\cite{Christensen1978}
The difference is that Cmc$2_{1}$ is non-centrosymmetric, but is
indistinguishable from the centrosymmetric Cmcm in XRD or neutron
diffraction. The bond distances in the H-bonds in the Cmcm space group
are extraordinarily large, thus the positions of hydrogen atoms may
be averaged positions in neutron diffraction~\cite{Oles1970}, and
the true symmetry may be Cmc$2_{1}$ (which has normal bond distances).
We adopted the proposal in~\cite{Christensen1978} as the starting
structure for our calculations.

\begin{figure}
\subfloat[AFM]{\includegraphics[width=0.3\linewidth]{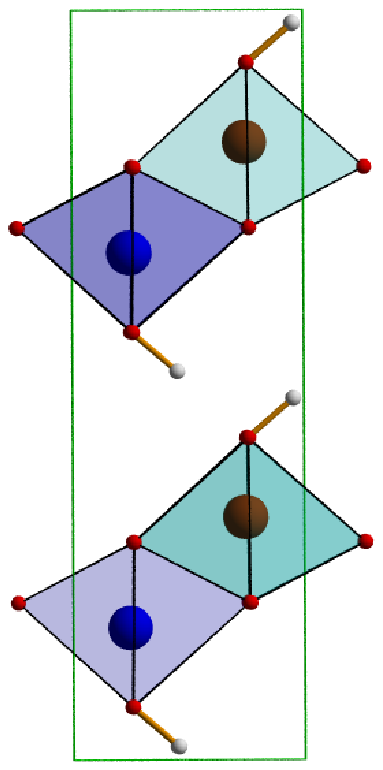}} 
\subfloat[AFM']{\includegraphics[width=0.3\linewidth]{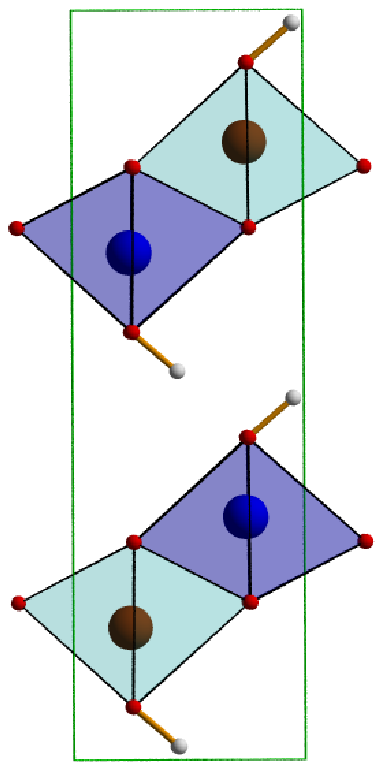}} 
\subfloat[AFM'']{\includegraphics[width=0.3\linewidth]{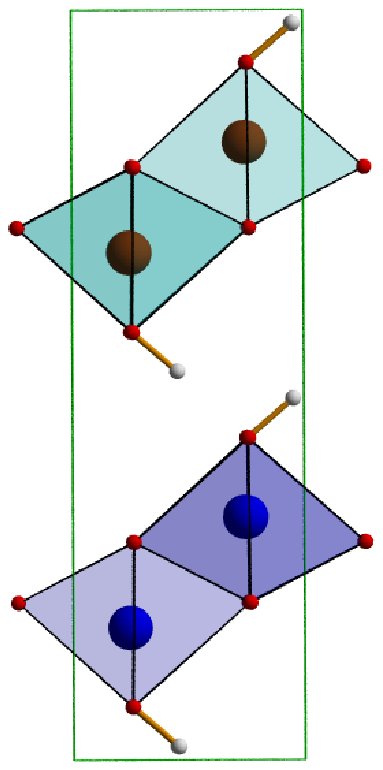}} 
\caption{Three antiferromagnetic configurations of lepidocrocite. Cyan and
blue octahedron are for Fe atoms with antiparallel spin moments. Red
balls are oxygen, white balls are hydrogen, sticks are hydroxyl bonds.
Viewed along [010] direction.\label{fig:lpdc:afm}}
\end{figure}

Each primitive cell contains 2 iron atoms, whose spin moment may align
in parallel (ferromagnetic) or antiparallel (antiferromagnetic) configurations.
More magnetization states may be included if the magnetization state
is stated in a conventional cell which contains 4 iron atoms. Lepidocrocite
is antiferromagnetic with antiparallel spins in the same double layer,
and antiparallel spins linked by hydrogen bonds.~\cite{Christensen1978}
This antiferromagnetic state is noted as AFM in this paper (Figure~\ref{fig:lpdc:afm}).
Another two antiferromagnetic states, noted as AFM' and AFM'' are
also included for comparison in addition to one ferrimagnetic, ferromagnetic,
and non-magnetic states. AFM' is similar to AFM, except the octahedron
linked by hydrogen bonds have parallel spin moments; AFM'' has parallel
spin in a double layer, and antiparallel spin in neighboring double
layers.

The calculated thermodynamic properties of lepidocrocite are listed
in Table~\ref{tab:lpdc1}, where we can see that calculations performed
using the LCAO basis set produce larger error with respect to available
experimental values than those obtained from the PW calculations.
The largest error in our LCAO calculations is the overestimation to
lattice parameter $c$ by about 7.6\%. The magnetization state of
AFM' also deviated from antiferromagnetic, and converged to ferrimagnetic
state during geometry optimization in LCAO calculations. The non-magnetic
state has larger errors in both calculations than other states, which
is consistent with the results obtained for the other iron oxides
and oxyhydroxides, as described in previous sections.
\begin{table}
\caption{Calculated thermodynamic properties of lepidocrocite. See Table~I for
annotations. \label{tab:lpdc1}}
\begin{tabular}{ccccccc}
\hline 
 &  & $a$\,(\AA{}) & b\,(\AA{}) & $c$\,(\AA{}) & $M$\,($\mu_{\mathrm{B}}$) & $E_{\mathrm{f}}$\,(kJ \,mol$^{-1}$)\\
\hline
 & AFM & 3.038 ($-1.4$) & 12.624 ($+1.0$) & 3.896 ($+0.7$) & 0.00 & $-425.8$\\
 & AFM' & 3.046 ($-1.1$) & 12.604 ($+0.8$) & 3.908 ($+1.0$) & 0.00 & $-425.2$\\
 & AFM'' & 3.080 ($+0.0$) & 12.204 ($-2.4$) & 3.866 ($-0.1$) & 0.00 & $-417.4$\\
PW & FiM & 3.066 ($-0.4$) & 12.390 ($-0.9$) & 3.898 ($+0.7$) & 2.03 & $-421.9$\\
 & FoM & 3.086 ($+0.2$) & 12.233 ($-2.1$) & 3.862 ($-0.2$) & 4.17 & $-416.9$\\
 & NM & 2.900 ($-5.9$) & 11.846 ($-5.2$) & 3.795 ($-1.9$) & - & $-398.2$\\
 & GGA+$U$ & 3.074 ($-0.2$) & 12.546 ($+0.4$) & 3.935 ($+1.7$) & 0.00 & -\\
\hline
 & AFM & 3.061 ($-0.6$) & 12.417 ($-0.7$) & 4.165 ($+7.6$) & 0.00 & $-463.5$\\
 & AFM' & 3.107 ($+0.9$) & 12.533 ($+0.3$) & 4.029 ($+4.1$) & 2.47 & $-457.8$\\
 & AFM'' & 3.125 ($+1.5$) & 12.456 ($-0.4$) & 4.021 ($+3.9$) & 0.00 & $-452.7$\\
LCAO & FiM & 3.107 ($+0.9$) & 12.532 ($+0.3$) & 4.029 ($+4.1$) & 2.48 & $-457.8$\\
 & FoM & 3.128 ($+1.5$) & 12.472 ($-0.2$) & 4.021 ($+3.9$) & 4.94 & $-452.0$\\
 & NM & 2.912 ($-5.5$) & 11.745 ($-6.0$) & 3.853 ($-0.4$) & - & $-406.5$\\
 & GGA+$U$ & 3.091 ($+0.4$) & 12.487 ($-0.1$) & 4.011 ($+3.6$) & 0.00 & -\\
\hline
\multicolumn{2}{c}{Expt.$^{(a)}$} & 3.08 & 12.50 & 3.87 & 0.00 & -554.6\\
\hline
\multicolumn{7}{l}{(a) Measured at room temperature and 0.1 MPa. Ref.~\cite[pp. 11 and 187]{Cornell2003}}\\
\end{tabular}
\end{table}

In this case, the energetic order predicted by PW and LCAO calculations
compare very well among all the magnetization states, with the exception
of the AFM' state (which deviates from the initial antiferromagnetic
state) in LCAO calculations. The energy difference between the AFM
and AFM'' states are almost the same, about 10 kJ\,mol$^{-1}$, using
both PW and LCAO. This indicates that the spin moments in a double-layer
is unlikely to be parallel, as in the energetically unstable AFM''.
The AFM'' and FoM states have almost the same formation energies in
both calculations, indicating there is weak correlation between iron
atoms connected by hydrogen bonds. This can also be seen from a comparison
of the AFM and AFM' states, between which the difference is solely
due to the alignment of spin moments of iron atoms linked by hydrogen
bonds. The formations energies of lepidocrocite (Table~\ref{tab:lpdc1})
are higher than those of goethite (Table~\ref{tab:gtht1}), which
is in good agreement with the established thermodynamic stability
of the two iron oxyhydroxide phases (where goethite is known to be
more stable than lepidocrocite).

\begin{figure}
\subfloat[Correct]{\includegraphics[height=5cm]{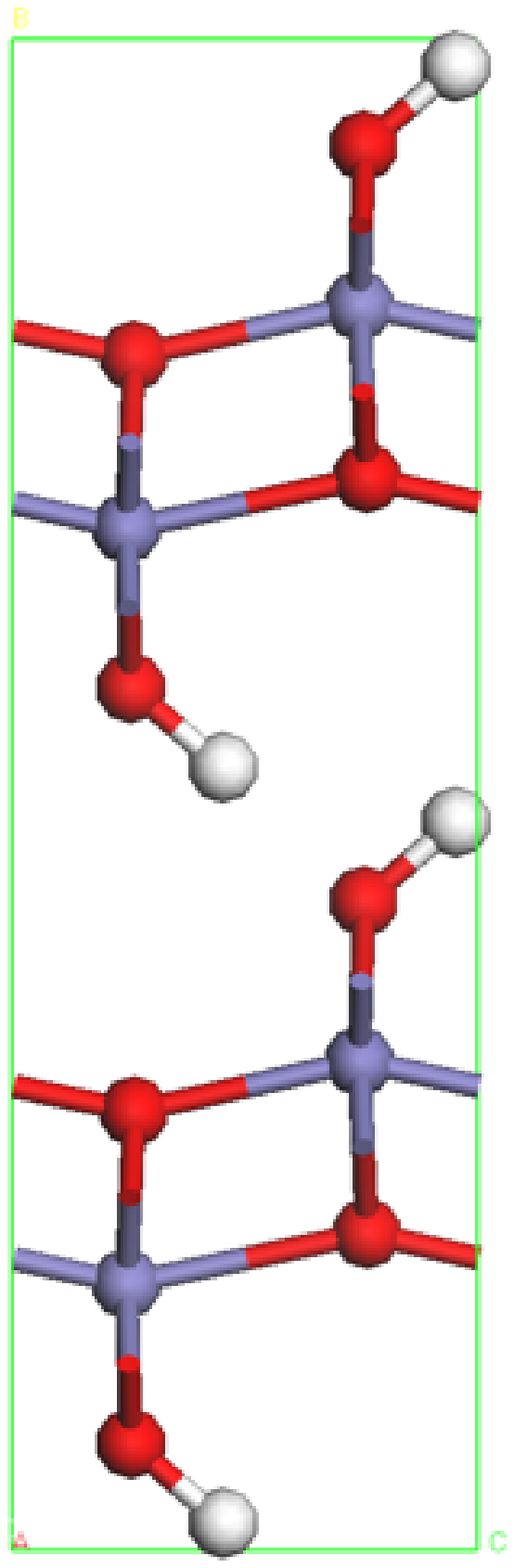}} 
\subfloat[Incorrect]{\includegraphics[height=5cm]{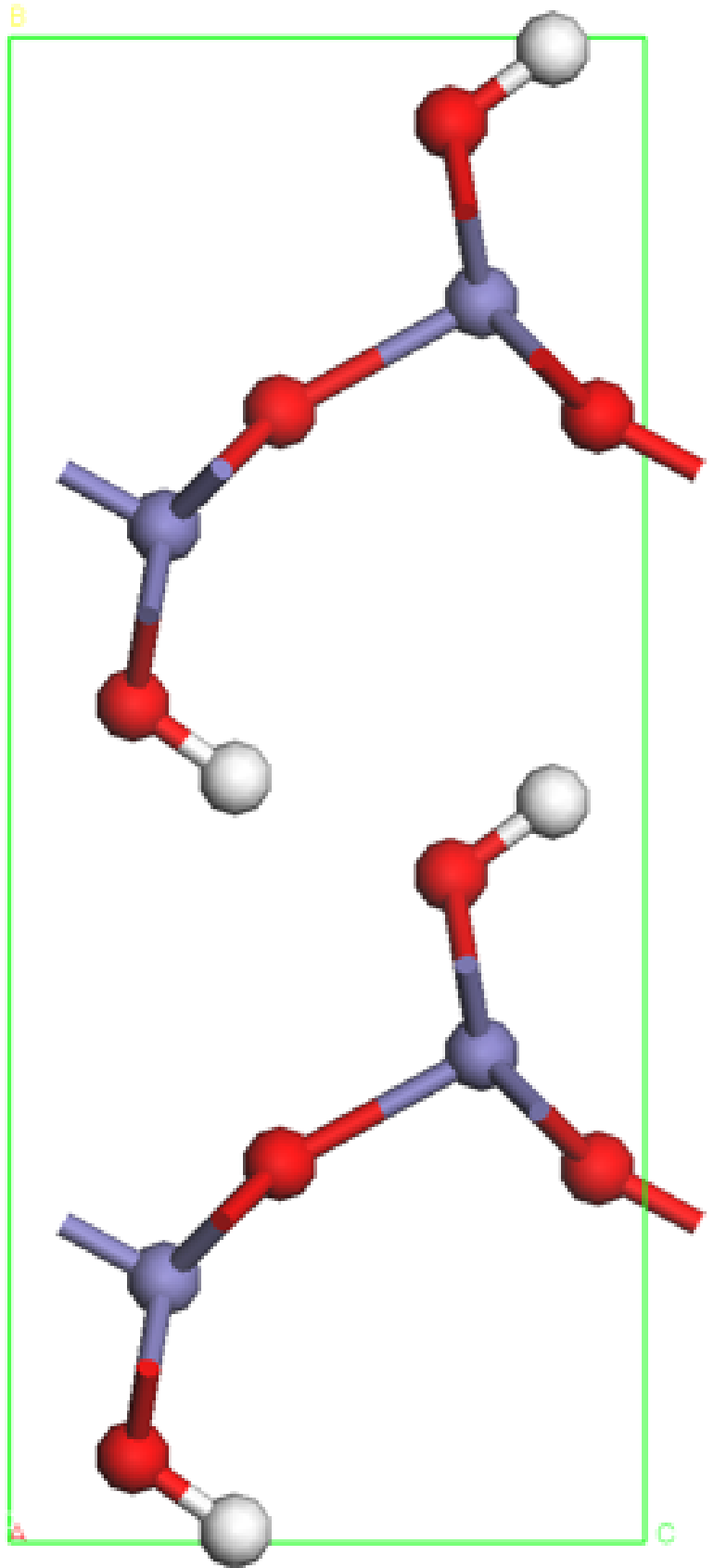}} 
\caption{Structural abnormality in geometry optimization. Gray blue balls are
Fe, red are O, and white are hydrogen. Viewed along [010] direction.
(color online) \label{fig:lpdc:GO} }
\end{figure}

\begin{figure}
\centering
\includegraphics[width=0.9\linewidth]{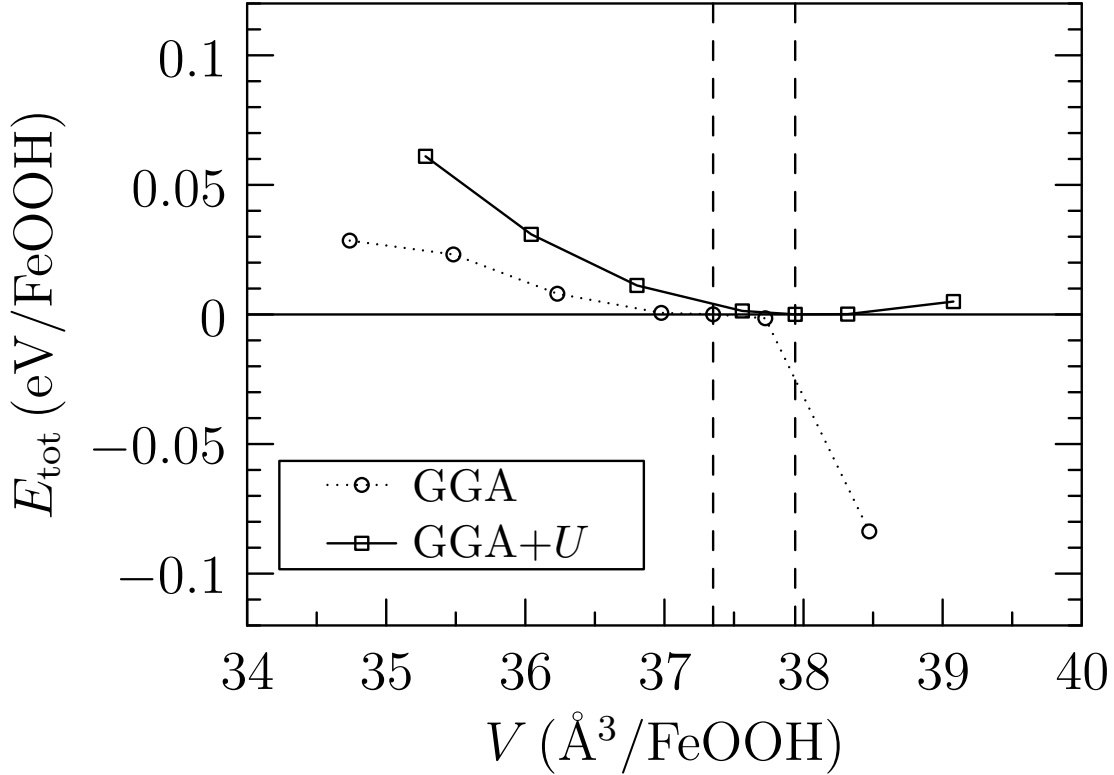} 
\caption{Equation of state of lepidocrocite in GGA and GGA$+U$ calculations, 
both using the PW basis set. The two vertical dashed lines mark the 
equilibrium volumes optimized by GGA and GGA$+U$. The total energies are 
shifted to align the minimum energies of GGA and GGA$+U$.\label{fig:eos}}
\end{figure}

At this point we would like to highlight that a correct description
of the loose, layered structure of lepidocrocite is much more difficult
to obtain in our computations (using both PW and LCAO basis sets) than
other types of oxides and oxyhydroxides. Geometry optimizations often
become trapped in an incorrect structure, as shown in Fig.~\ref{fig:lpdc:GO}.
In the incorrectly optimized structure, the iron atoms are trigonal-bipyramid
coordinated instead of octahedron, while the oxygen atoms that do
not form hydroxyl bonds are bonded to only 3 iron atoms instead of
4. The incorrect structure may have an abnormally small lattice parameter
$a$ (up to about 20\% below experimental value), large $b$ (up to
about 25\% above experimental value), or large $c$ (up to about 40\%
above experimental value). These incorrect structures occurred when
the geometry optimizations began using the structure models that have
hydrogen atoms equidistant between oxygen atoms, and may also occur
with certain computational settings.  We show the equation of state (EOS)
calculated using GGA and GGA$+U$ to demonstrate the sensitivity of geometry
optimization by GGA on the starting structure (Fig.~\ref{fig:eos}). The
incorrect structure is accompanied by the steep energy decrease when 
the cell volume is
slightly larger (3\%; the correct structure can retain up to 1\% volume
increase) than the equilibrium volume. In contrast, 
the GGA$+U$ is robust in geometry optimizations with varying volume in this case. 
We carefully examine the final
structures after geometry optimization, and rigorously tested the
computational settings and initial structures to ensure the double
layer structure of lepidocrocite. 

In the case of lepidocrocite, the calculation results of elastic properties
are more diverged than for other iron oxides and oxyhydroxides in
this paper. As shown in Table~\ref{tab:lpdc2}, the difference between
GGA and GGA$+U$ can be more than 50\% ($c_{13}$, $c_{22}$
and $c_{66}$ in PW, $c_{55}$ and $c_{66}$ in LCAO), and the agreements
between PW and LCAO are acceptable only for components $c_{22}$,
$c_{23}$, and $c_{33}$. This is partly due to the delicacy of lepidocrocite.

\begin{table}
\caption{Calculated bulk moduli and elastic constants of lepidocrocite. Unit:
GPa. \label{tab:lpdc2}}
\begin{tabular}{ccc|cc}
\hline 
 & \multicolumn{2}{c|}{PW} & \multicolumn{2}{c}{LCAO}\\
 & AFM & GGA+$U$ & AFM & GGA+$U$\\
\hline
$B_{0}$ & 75.6 & 74.8 & x & 81.8\\
$c_{11}$ & $219.0\pm22.8$ & $246.9\pm9.8$ & $242.4\pm0.9$ & $264.5\pm12.8$\\
$c_{12}$  & $77.3\pm14.5$ & $84.8\pm6.8$ & $88.0\pm0.2$ & $95.1\pm13.2$\\
$c_{13}$ & $31.9\pm26.8$ & $80.6\pm10.1$ & $79.0\pm1.2$ & $72.2\pm3.8$\\
$c_{22}$ & $221.0\pm11.6$ & $272.2\pm18.7$ & $214.7\pm1.7$ & $246.3\pm10.4$\\
$c_{23}$ & $123.3\pm8.5$ & $137.2\pm22.9$ & $127.3\pm0.9$ & $124.1\pm4.9$\\
$c_{33}$ & $305.4\pm21.9$ & $347.7\pm16.8$ & $303.2\pm1.6$ & $327.4\pm4.5$\\
$c_{44}$ & $121.6\pm0.0$ & $131.4\pm0.0$ & $97.6\pm1.2$ & $119.0\pm0.8$\\
$c_{55}$ & $63.4\pm0.0$ & $64.0\pm0.1$ & $49.2\pm0.8$ & $63.6\pm1.9$\\
$c_{66}$ & $44.8\pm0.0$ & $88.9\pm0.1$ & $73.3\pm0.6$ & $93.6\pm1.5$\\
\hline
\end{tabular}
\end{table}

\subsection{Magnetite (\texorpdfstring{Fe$_3$O$_4$}{Fe3O4})}

Magnetite has a cubic inverse spinel structure (space group F$d\bar{3}m$,
No. 227) at thermodynamic standard state (room temperature, ambient
pressure). Its chemical formula, $\mathrm{Fe}_{3}\mathrm{O}_{4}$,
is often written as $\mathrm{Fe^{3+}[Fe^{3+},Fe^{2+}]}\mathrm{O}_{4}$
to show that tetrahedral sites (A) are occupied by trivalent Fe ions,
and octahedral sites (B) are occupied by equal trivalent and divalent
ions. The spin moments of A- and B-sites align antiparallel, resulting
in a ferrimagnetic state. Magnetite undergoes the Verwey phase transition
at about 125 K, below which the electronic resistivity increases 2
orders of magnitude.~\cite{Verwey1939} This phenomenon was explained
by charge-ordering model in which electron hopping among Fe ions are
frozen below Verwey transition temperature and aligned in an ordered
pattern.~\cite{Verwey1939} However, after 6 decades of study researchers
found the phenomenon is far more complicated than was previously thought.~\cite{Walz2002}
Among various changes (electronic resistivity, band structure, heat
capacity) accompanied by the Verwey transition, the structure slightly
distorted from the room-temperature cubic structure. At low temperatures,
the structure of magnetite was proposed to be orthorhombic from nuclear
magnetic resonance spectroscopy;~\cite{Mizoguchi1978,Mizoguchi1978a,Novak2000}
monoclinic from x-ray diffraction,~\cite{Yoshida1977,Wright2002},
neutron diffraction,~\cite{Iizumi1982}, electron diffraction,~\cite{Zuo1990}
and x-ray resonant scattering;~\cite{Wilkins2009} or even lower
symmetry of triclinic.~\cite{Miyamoto1993} In the present study,
we have restricted our calculation to the room-temperature cubic structure
because the calculated thermodynamic properties at ground state can
be extrapolated to room temperature without discontinuity by the phase
transition.

In the case of magnetite we have tested 3 magnetization states: ferrimagnetic
(FiM), ferromagnetic (FoM), and non-magnetic (NM). Both PW and LCAO
basis sets predict that the FiM state has the lowest formation enthalpy
among all these magnetization states (see Table~\ref{tab:mgnt1}).
The lattice constants of the FiM state also provide better agreement
with the experimental measurements. Using both basis sets, the calculation
results differ from experimental values if we ignore spin polarization.
The error in lattice constants using the LCAO approach is slightly
larger than that using the PW approach. The calculated spin moments
agree well with each other in both PW- and LCAO-based methods.

\begin{table}
\caption{Calculated thermodynamic properties of magnetite. See Table~I for
annotations. \label{tab:mgnt1}}
\begin{tabular}{ccccc}
\hline 
 &  & $a$\,(\AA{}) & $M$\,($\mu_{\mathrm{B}}$) & $E_{\mathrm{f}}$\,(kJ \,mol$^{-1}$)\\
\hline
 & FiM & 8.392 ($-0.0$) & 1.33 & $-871.7$\\
PW & FoM & 8.528 ($+1.6$) & 4.57 & $-747.1$\\
 & NM & 8.049 ($-4.1$) & - & $-666.4$\\
 & GGA+$U$ & 8.481 ($+1.0$) & 1.33 & -\\
\hline 
 & FiM & 8.504 ($+1.3$) & 1.33 & $-930.9$\\
LCAO & FoM & 8.645 ($+3.0$) & 4.67 & $-828.7$\\
 & NM & 8.111 ($-3.3$) & - & $-650.2$\\
 & GGA+$U$ & 8.653 ($+3.1$) & 1.33 & -\\
\hline 
\multicolumn{2}{c}{Expt.$^{(a)}$} & 8.396 &  & \textasciitilde{}$-1120$\\
\hline 
\multicolumn{5}{l}{(a) Measured at room temperature and 0.1 MPa. Ref.~\cite[pp. 11 and 187]{Cornell2003}}\\
\end{tabular}
\end{table}

\begin{figure}
\includegraphics[width=0.7\linewidth]{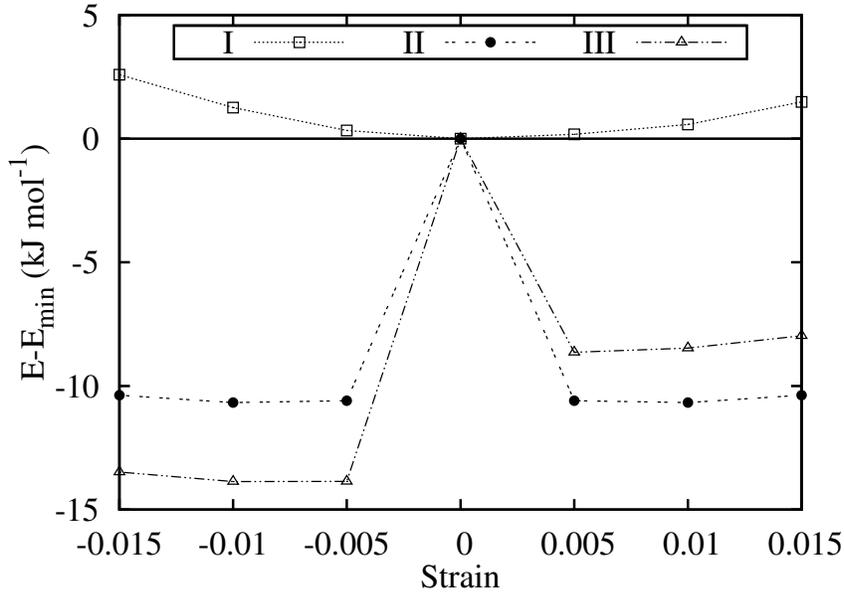} 
\caption[magnetite elastic constants]{\label{fig:mgnt:strainE}
Strain energies in calculations to elastic 
constants of GGA$+U$ using PW. The 3 deformations are 
I (isotropic) $\left( \begin{smallmatrix}
1 & 0 & 0 \\
0 & 1 & 0 \\
0 & 0 & 1 
\end{smallmatrix} \right)$, 
II (orthorhombic)$\left(\begin{smallmatrix}
1 & 0 & 0 \\
0 & -1 & 0 \\
0 & 0 & 0 \end{smallmatrix}\right)$, 
and III (shear) $\left( \begin{smallmatrix} 
0 & \nicefrac{1}{2} & \nicefrac{1}{2} \\
0 & 0 & \nicefrac{1}{2} \\
\nicefrac{1}{2} & \nicefrac{1}{2} & 0 
\end{smallmatrix}\right)$. 
}
\end{figure}

We calculated elastic properties (bulk moduli and elasticity tensor)
of FiM magnetite, as shown in Table~\ref{table:mgnt2}. GGA calculations using PW
and LCAO both predict that cubic magnetite is elastically stable. 
In the case of GGA$+U$ calculations, we were
unable to fit the strain energies with strains to calculate elastic
constants, because the equilibrium cubic structure has higher energy
than strained states. As shown in Figure~\ref{fig:mgnt:strainE},
except the isotropic deformation, the other two deformations have
even lower energy than the zero-strain, {}``equilibrium'' structure
in GGA$+U$ calculations. Increasing $k$-point sampling density does
not solve this problem. This indicates the cubic magnetite is elastically
unstable, which agrees well with experimental observations that low-temperature
(below Verwey transition temperature) structure is monoclinic, but
not cubic. 

\begin{table}
\caption{Calculated bulk moduli and elastic constants of magnetite. Unit: GPa.
The calculation results with GGA$+U$ are discussed in main text.\label{table:mgnt2}}
\begin{tabular}{ccc|cc}
\hline 
 & \multicolumn{2}{c|}{PW} & \multicolumn{2}{c}{LCAO}\\
\hline
 & FiM & GGA+$U$ & FiM & GGA+$U$\\
\hline
$B$ & 187.4 & 173.3 & 165.3 & 168.3\\
$c_{11}$ & $275.4\pm40.9$ & - & $253.6\pm5.7$ & -\\
$c_{12}$  & $155.2\pm60.3$ & - & $128.1\pm10.3$ & -\\
$c_{44}$  & $97.5\pm13.0$ & - & $75.4\pm0.9$ & - \\
\hline
\end{tabular}
\end{table}

\subsection{Energetic order}

As mentioned above, one of the computational challenges in modeling
different iron oxides and oxyhydroxides is the small energy differences
among different solid phases. Since the typical accuracy of DFT calculations
is about several kJ\,mol$^{-1}$, which is comparable to the energy
differences between competing phases, calculations with different
settings may lead to very different energetic order, but may have
very little physical meaning due to numerical inconsistencies. Systematic
and consistent calculations of all the phases are highly desirable
to make comparisons among the different phases, as well as case-by-case
comparisons with experiments. The calculations in this study have
enabled us to assess the PW and LCAO basis sets, but also to make such
comparisons for the first time.

\begin{figure}
\subfloat[Without correction of connection
energy][]{\includegraphics[width=7cm]{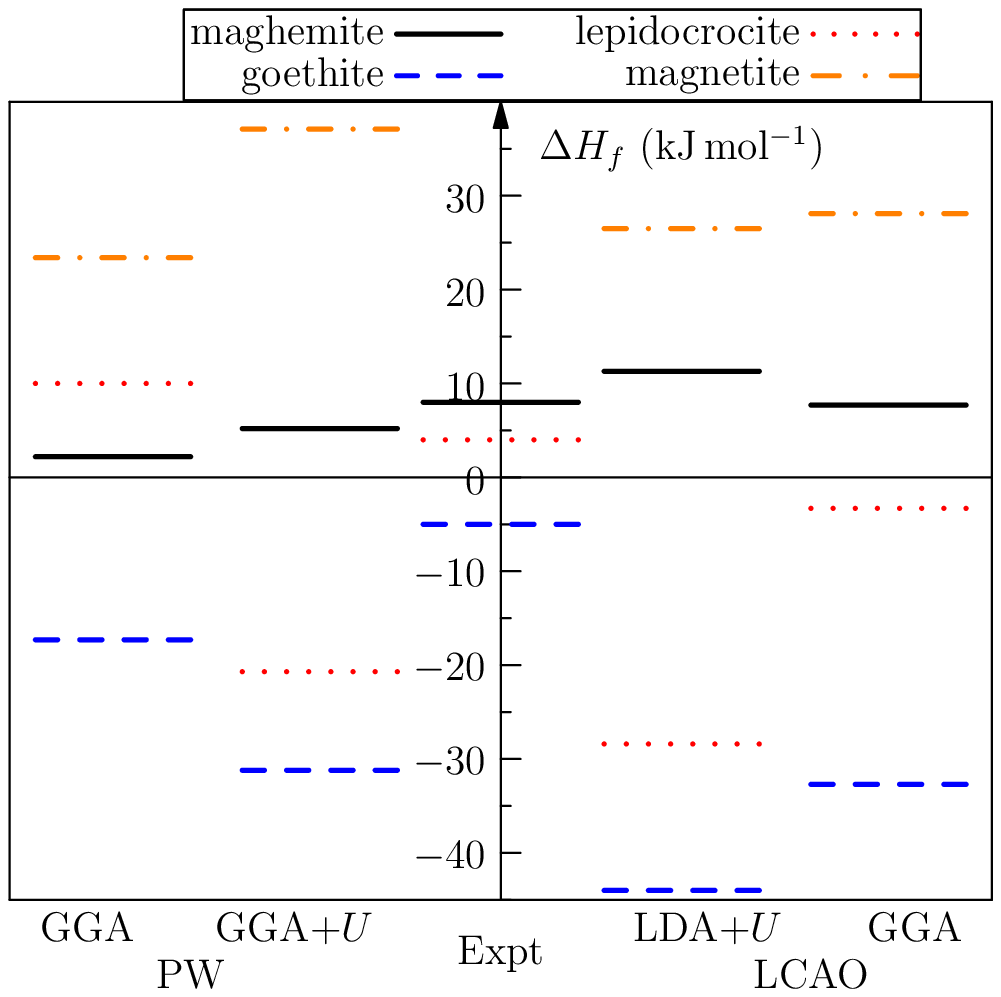}} 
\subfloat[With correction of connection
energy][]{\includegraphics[width=7cm]{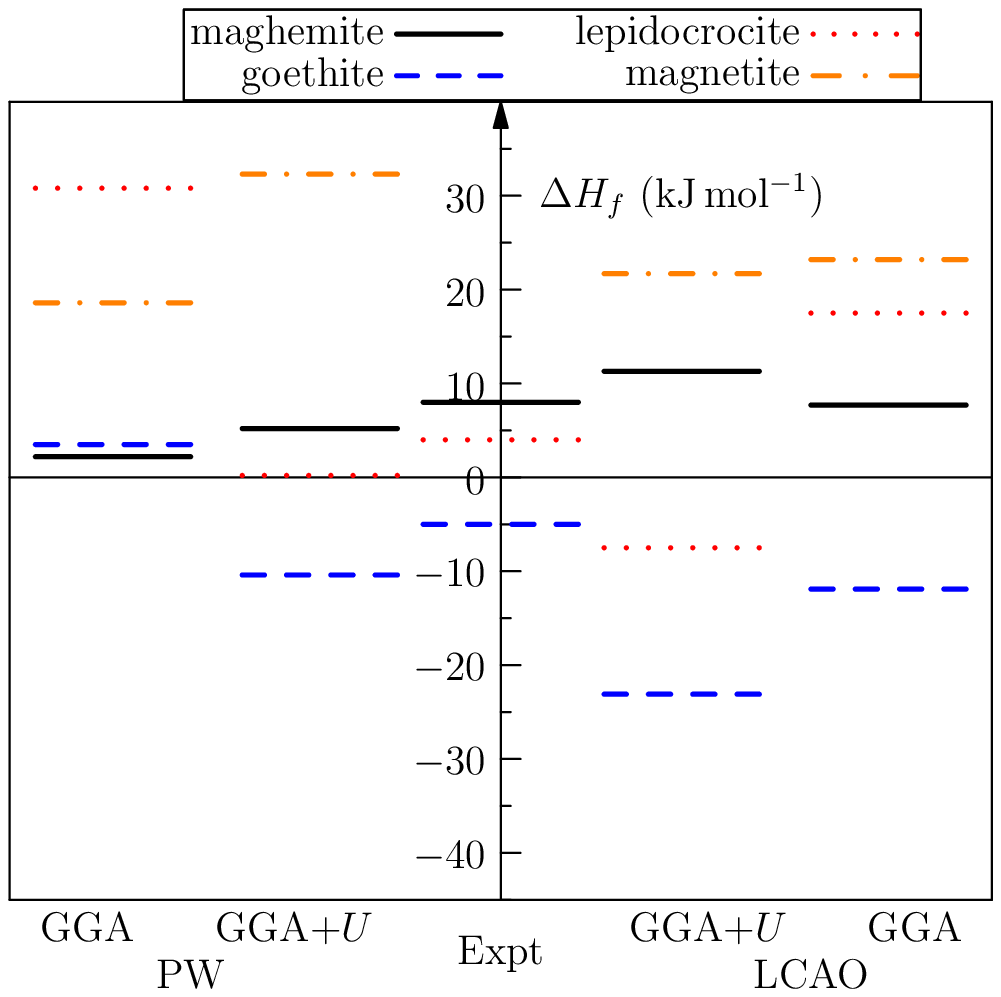}} 
\caption{Calculated formation enthalpies of iron oxides and oxyhydroxides with
respect to hematite and water or oxygen (The formation enthalpies
of goethite and lepidocrocite are with respect to hematite and water;
magnetite is with respect to hematite and oxygen). Expt stands for experimental
values. \label{fig:enthalpy}}
\end{figure}

To begin with, we have calculated the formation enthalpies of the
iron oxides and oxyhydroxides with respect to hematite and water or
oxygen (Figure~\ref{fig:enthalpy}). Available experimental values~\cite{Majzlan2003}
are also shown in figure for comparison. Among the 5 iron oxides and
oxyhydroxides, magnetite is typically excluded from the experiments
because all the other four compounds (goethite, lepidocrocite, maghemite,
and hematite) can be rewritten as hematite$+x\mathrm{H_{2}O}+\Delta H_{f}$
($x=0$ for maghemite, or 0.5 for goethite and lepidocrocite). In
our work, we are able to include magnetite and compare the energetic
order with respect to hematite and oxygen. In this way we can plot
the energetic order of all the 5 iron oxides and oxyhydroxides together,
keeping hematite plus balancing gases (H$_{2}$O and O$_{2}$ in their
standard state) as the reference. 

At this point, it is prudent to point out that our DFT calculations
correspond to the ground state, while experiments were conducted at
thermodynamic standard state. The influence of temperature and pressure
on formation energies of solids are much less than that of gases.
We therefore include the connection energy, which defines the difference
between energies at ground state and standard state, for the gases.~\cite{ZhangW2004}
By this ab-initio thermodynamics scheme, we can extend the calculation
results at ground state to finite temperature and pressures. In this
case, we use the connection energies for the gases which have been
calculated to solve tribochemistry problems.~\cite{Guo2010} 
For solids, the
available thermochemistry data enable one to calculate connection energy by
integrating from 0\,K ground state to the thermodynamic standard state as
\begin{equation}
\Delta \mu^0(T_{\mathrm{r}}) = \int_{0\,K}^{T_{\mathrm{r}}} C_p \mathrm{d}T 
- T \int_{0\,K}^{T_{\mathrm{r}}} \frac{C_p}{T} \mathrm{d}T,  
\end{equation} 
where $\Delta \mu^0(T_{\mathrm{r}})$ is the connection energy at standard
pressure ($p^0=1$\,atm) and room temperature $T_\mathrm{r}=298.15$\,K, $C_p$ is
the molar heat capacity. The heat capacity, enthalpy difference between room
temperature and 0\,K, and entropy can be looked up in thermochemistry tables, eg
NIST-JANAF table.~\cite{Chase1985} Calculation of the connection energies of gases are
usually done through reaction equilibrium with solids. For example, we used the
reaction
\begin{equation}
\mathrm{MgO} + \mathrm{H_2O} \leftrightarrow \mathrm{Mg(OH)_2}
\end{equation}
to calculate the connection energy of gas-phase water because
the thermochemical data of MgO and Mg(OH)$_2$ are available.~\cite{Guo2010}
Once the connection energies for room temperature is known, the chemical
potentials at other temperatures (and pressures for gases) can be readily
calculated using thermodynamics as long as the heat capacity data are available
for the temperature range.

From Fig.~\ref{fig:enthalpy} we see the enthalpy difference between maghemite 
and hematite is apparently underestimated (by about 5 kJ mol$^{-1}$) when 
using GGA with PW basis set, while all other settings
reproduce this energy difference well. For magnetite, the enthalpy
calculated using GGA$+U$ with PW is larger than others, but the energetic
order is consistent in all the calculations. The energy difference
between goethite and lepidocrocite is larger in GGA than
that in GGA$+U$, in both PW and LCAO calculations. It is worth noting
that the corrections using connection energy is ineffective to change
the relative energetic orders of hematite and maghemite, or goethite
and lepidocrocite, because they have same chemical compositions. However,
with the corrections of connection energies, the relative energetic
order between compounds with different composition may change, as
we see the sub-figures in Fig.~\ref{fig:enthalpy}. With the corrections,
the difference between the state for calculations and experiments
are approximately eliminated, enabling us to make fair comparisons.
The calculated energetic order of lepidocrocite, hematite, and maghemite
is very different in the 4 data sets, and we may conclude that
GGA$+U$ with PW implementation best matches experiment. 

The consistent computational settings across different iron oxides
and oxyhydroxides offer us a number of significant advantages, one
of which is that we are in a position to construct phase diagrams.
For this purpose we have chosen to use the calculation results from
GGA$+U$ with the PW implementation, and compute the the free energy
of formation of a compound FeO$_{x}$H$_{y}$ as:

\begin{equation}
\Delta G=\Delta H-\frac{x}{2}\left(\Delta_{\mathrm{O_{2}}}(T)+RT\ln\frac{P_{\mathrm{O_{2}}}}{P^{0}}\right)-\frac{y}{2}\left(\Delta_{\mathrm{H_{2}}}(T)+RT\ln\frac{P_{\mathrm{H_{2}}}}{P^{0}}\right),\label{eq:free}
\end{equation}
where $\Delta H$ is the formation energy at ground state, $\Delta_{\mathrm{O_{2}}}(T)$
and $\Delta_{\mathrm{H_{2}}}(T)$ are connection energies at a certain
temperature for O$_{2}$ and H$_{2}$, respectively, $R$ is the gas
constant. One can write the formation energies with respect to H$_{2}$O
and O$_{2}$ by analogy. The connection energies were calculated using
thermochemistry data in a previous study,~\cite{Guo2010} and $\Delta H$
are from the ground-state calculations in the present study. In a
phase diagram, the phase boundaries determined from equation~\ref{eq:free}
are straight lines in a phase diagram. 

\begin{figure}
\subfloat[H$_{2}$ and
O$_{2}$\label{fig:pd:H}]{\includegraphics[scale=0.7]{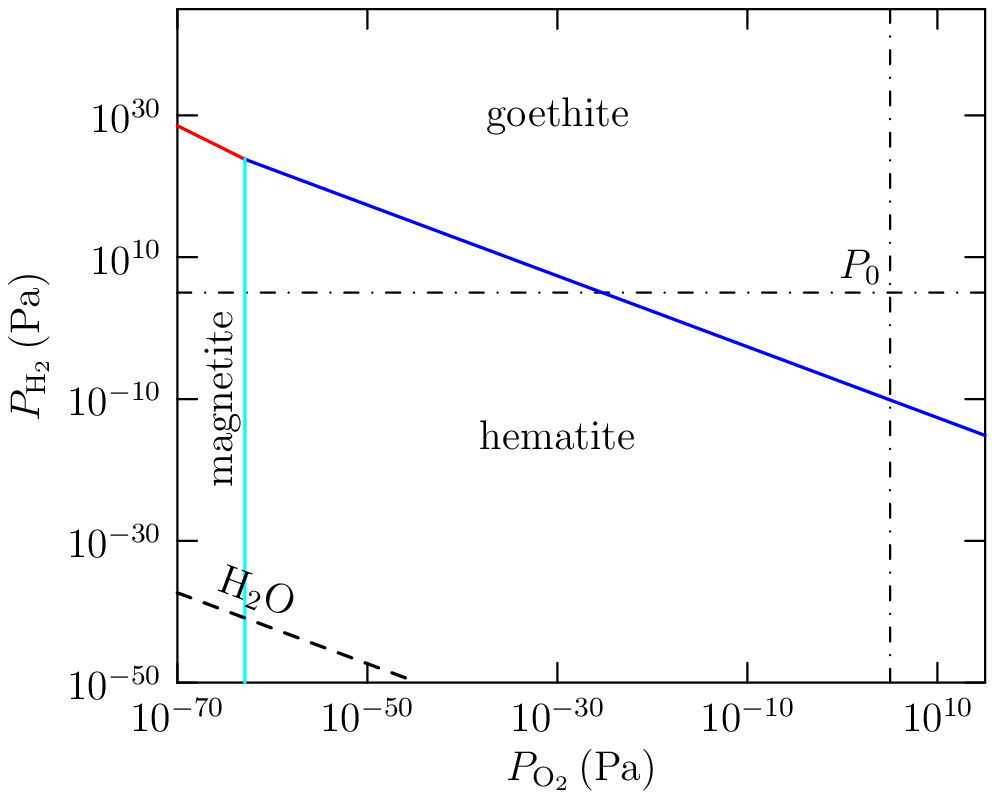}} 
\subfloat[H$_{2}$O and
O$_{2}$\label{fig:pd:H2O}]{\includegraphics[scale=0.7]{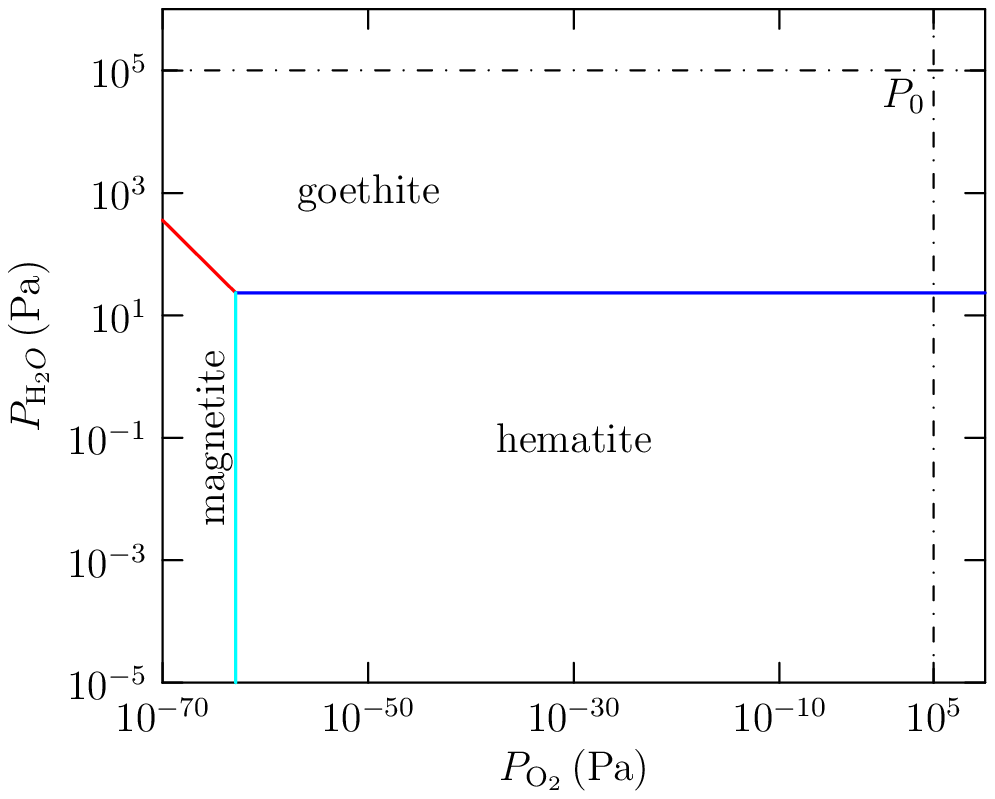}} 
\caption{Phase diagram of iron oxides and oxyhydroxides with partial pressures
of gases at $T=298.15$\,K. Lepidocrocite is metastable with respect
to goethite, and maghemite is metastable to hematite; they are excluded
in this equilibrium phase diagram. Dash-dotted lines indicate standard
pressure (1 atm); the dashed line in (a) indicates formation of water
vapor from H$_{2}$ and O$_{2}$ with no energy gain or loss. Note
the different scales of partial pressures of H$_{2}$in (a) and H$_{2}$O
in (b). \label{fig:pd}}
\end{figure}

Using Equation~\ref{eq:free} we have constructed two phase diagrams,
both corresponding to room temperature (see Fig.~\ref{fig:pd}).
The metastable phases of lepidocrocite and maghemite are not shown,
as these are equilibrium phase diagrams. The two sub-figures refer
to the same systems with respect to the chemical potentials of (a)
H$_{2}$ and O$_{2}$ and (b) H$_{2}$O and O$_{2}$, respectively.
One notices the extremely low partial pressure of oxygen required
for the formation of magnetite instead of hematite. This means that,
at room temperature, magnetite should form under oxygen-poor conditions;
otherwise the more stable hematite phase should prevail in exogenous
environments. This is compatible with the fact that most magnetotactic
bacteria that produce magnetite are either anaerobic or microaerobic.~\cite[p.481-489]{Cornell2003}
Magnetite is able to form from hematite at low temperatures in the
presence of hydrazine,~\cite[p.405-406]{Cornell2003} which removes
dissolved oxygen in the solutions. 

The phase boundary between hematite and goethite has the same slope
of water formation in Fig.~\ref{fig:pd:H}; therefore, in a phase
diagram of Fe-H$_{2}$O-O$_{2}$, it is independent of chemical potential
of H$_{2}$O, as shown in Fig.~\ref{fig:pd:H2O}. The phase diagram
shows that the free energy of goethite is lower than hematite at standard
state, and this agrees with the calorimetry measurements.~\cite{Majzlan2003}
In a wet environment, these phase diagrams predict that the formation
of goethite will be more thermodynamically favorable than hematite;
while dehydration (dry conditions) will cause goethite to transform
into hematite given a suitable driving force. 

\section{Comparison of efficiency}

One of the advantages of LCAO basis sets is efficiency.~\cite{Artacho1999,Junquera2001,Soler2002,Artacho2008}
The atomic orbitals used to expand the wave functions are very economic
(in terms of number of orbitals per atom to achieve an accuracy) compared
with the PW basis set. In the DZP (double-$\zeta$ with one polarization
orbital) scheme which is used in the present study, each Fe atom needs
17 orbitals for the valence electrons, O needs 13, and H needs 3 orbitals.
For a $4\times4\times4$ $k$-point grid for hematite, the number
of atomic orbitals is 8832. In order to achieve similar convergence
in energy calculations, the PW basis set requires about 120000 PW's,
which is about 15 times as that of LCAO basis set. The advantage of less
orbitals will be even more apparent if the computation cell has vacuum
space, such as in surface calculations, since the LCAO's are centered
at ions, and vacuum requires no additional orbitals. In contrast,
the PW's are delocalized, and even vacuum space has similar number-density
of PW's.

In addition to this, the localized nature of LCAO's enables one to
implement the order-N algorithms, which critically rely on localization
of wave functions. In integrating over bands, the Fermi level needs
to reside in the band gap, which should be large enough to cover the
varying chemical potential. This is not true for metals and semiconductors
with narrow band gaps, which include most iron oxides and oxyhydroxides.
Therefore, studies on these systems are not able to benefit from the
order-N algorithms.

\begin{figure}
\includegraphics[scale=0.7]{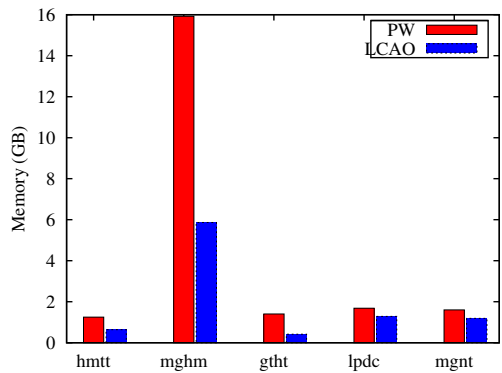} 
\caption{Memory requirement of PW and LCAO in geometry optimizations to the
iron oxides. \label{fig:memory}}
\end{figure}

\begin{figure}
\includegraphics[scale=0.7]{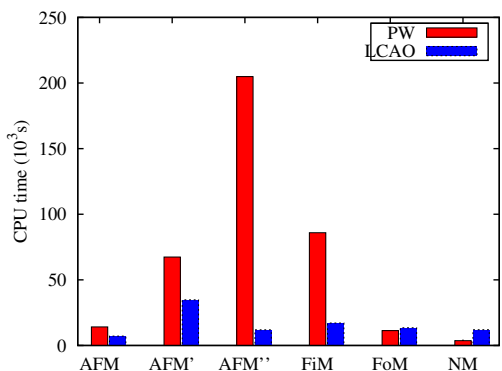} 
\caption{CPU time consumption in geometry optimizations to hematite. \label{fig:CPUt}}
\end{figure}

However, fewer numbers of orbitals should still translate into efficiency
(of computation time and memory usage), even without order-N algorithms.
We find this is true for memory usage, but it is not always true for
computation time. As shown in Fig.~\ref{fig:memory}, the PW basis set
uses more memory than LCAO for calculations of all the iron oxides
and oxyhydroxides included in our study (see main text). It is worth
noting that the memory requirements also depends on parallelization,
and the numbers are extracted from calculations using 8 CPUs for all
the iron oxides except maghemite, which uses 32 CPUs. The CPU time
usage for geometry optimizations for different magnetization states
of hematite (Fig.~\ref{fig:CPUt}) shows the PW basis set may exceed
LCAO in some geometry optimizations, even though the PW basis set uses
much more orbitals. Other factors may affect the computation time,
such as minimization path, so we have taken care to always start from
the same structures, and use the same method (CG) and force convergence
(0.005 eV/\AA) in moving atoms in order to minimise this effect. The
PW and LCAO basis sets also differ in their use of symmetry (as described
above), which leads to differences in the force calculations. In general
we find that the difference in computation time is not as large as
that in number of orbitals. By utilizing symmetrization. As an aside,
we also compared the numbers of self-consistency iterations to reach
the geometry optimization criteria. In most cases, the LCAO basis set
needs more MD steps than the PW basis set to reach the convergence criteria.

At this point it is also worth pointing out that one of the problems
with LCAO's is systematic convergence. Simply increasing radii of
the atomic orbitals does not always lead to better convergence, and
tuning the parameters of the atomic orbitals requires considerably
more effort than is needed for the PW basis set. While increasing the
number of atomic orbitals can increase the accuracy, this comes at
the cost of computation (in the PW basis set as well). Tests of the size
of atomic orbitals have shown high accuracy within the frame work
of DFT can be achieved with multiple-$\zeta$ and multiple polarization
orbitals. DZP, which is used in the present study, is usually a reasonable
compromise between accuracy and efficiency.

\section{Conclusions and discussions}

In summary, by comparing the calculation results of PW and LCAO basis sets,
with and without on-site interactions, as well as among different
magnetization configurations, we presented solutions to the computational
challenge in modeling iron oxides. Consistency is paramount, and this
has been maintained in the comparisons as to energy functionals, convergence
criteria of force, $k$-point mesh, and starting structures for geometry
optimizations. We have shown that both PW and LCAO basis sets can find
the thermodynamically stable magnetization states, and reproduce lattice
parameters well (except lepidocrocite by LCAO which overestimate $c$
by about 7\% in GGA and 4\% in GGA$+U$). However, in most
geometry optimizations, the LCAO basis set is more efficient in CPU time
and memory usage than the PW basis set,
but the accuracy is slightly reduced when comparing with PW basis set.
Several factors contribute to the efficiency difference between the
two implementations, including number of orbitals, molecular dynamics
algorithms in moving ions, electron density mixing, force calculation,
$k$-point density.
Using these basis sets, we evaluated elastic stability of all the materials.
We find that the PW and LCAO basis sets are comparable for most structures
except lepidocrocite, and that the elasticity tensor of maghemite
is close to that of a cubic crystal, though the true symmetry is tetragonal
due to the long-range ordering of vacancies. While GGA predicts
cubic magnetite is elastically stable, GGA+$U$ calculations contradict
the prediction. 

The crystal structure of lepidocrocite consists layers held by H-bonds.
In computational modeling, functionals with general gradient approximation
and hybrid functionals are able to describe the relatively weak interactions
of hydrogen bonds. While van der Waals interactions may also contribute
significantly to inter-layer interactions, as they do in graphite,
they are not included in the present study, as calculations of dispersive
forces are either very expensive or relying on empirical parameters.
In our calculations this delicate structure exhibited some structural
abnormalities, which may be due to the omission of dispersive forces.
This layered structure is not as delicate as that of graphite (in
which the carbon layers are held by even weaker van der Waals interaction),
but still impose an challenge to computational modeling. Accurate
energy functionals that include van der Waals interactions may describe
better the crystal structure of lepidocrocite.

Based on these results (which represent the first consistent set of
ab initio predictions of the elastic, magnetic and thermodynamic properties),
we also present the first phase diagram of 5 iron oxides and oxyhydroxides
designed to predict the relative stability of these materials under
different chemical conditions. Given that chemical conditions are
typically characteristic of specific environments (both during and
post-formation), this phase diagram will be invaluable in understanding
the environmental stability of these important materials, and anticipating
transformations that may be invoked by moving from one environment
to another, or by variations in climatic conditions.

\section*{Acknowledgement}
The authors thank Prof. H.F. Xu and J.D. Gale for fruitful discussions
on iron oxides and DFT modeling using LCAO. The authors acknowledge
NCI National Facility for computational support of project code p00.


\begin{thebibliography}{108}
\expandafter\ifx\csname natexlab\endcsname\relax\def\natexlab#1{#1}\fi
\expandafter\ifx\csname bibnamefont\endcsname\relax
  \def\bibnamefont#1{#1}\fi
\expandafter\ifx\csname bibfnamefont\endcsname\relax
  \def\bibfnamefont#1{#1}\fi
\expandafter\ifx\csname citenamefont\endcsname\relax
  \def\citenamefont#1{#1}\fi
\expandafter\ifx\csname url\endcsname\relax
  \def\url#1{\texttt{#1}}\fi
\expandafter\ifx\csname urlprefix\endcsname\relax\def\urlprefix{URL }\fi
\providecommand{\bibinfo}[2]{#2}
\providecommand{\eprint}[2][]{\url{#2}}

\bibitem[{\citenamefont{Cornell and Schwertmann}(2003)}]{Cornell2003}
\bibinfo{author}{\bibfnamefont{R.~M.} \bibnamefont{Cornell}} \bibnamefont{and}
  \bibinfo{author}{\bibfnamefont{U.}~\bibnamefont{Schwertmann}},
  \emph{\bibinfo{title}{The iron oxides}} (\bibinfo{publisher}{Wiley-VCH},
  \bibinfo{year}{2003}), \bibinfo{edition}{2nd} ed.

\bibitem[{\citenamefont{Navrotsky et~al.}(2008)\citenamefont{Navrotsky,
  Mazeina, and Majzlan}}]{Navrotsky2008}
\bibinfo{author}{\bibfnamefont{A.}~\bibnamefont{Navrotsky}},
  \bibinfo{author}{\bibfnamefont{L.}~\bibnamefont{Mazeina}}, \bibnamefont{and}
  \bibinfo{author}{\bibfnamefont{J.}~\bibnamefont{Majzlan}},
  \bibinfo{journal}{Science} \textbf{\bibinfo{volume}{319}},
  \bibinfo{pages}{1635} (\bibinfo{year}{2008}).

\bibitem[{\citenamefont{McKay et~al.}(1996)\citenamefont{McKay, Gibson,
  ThomasKeprta, Vali, Romanek, Clemett, Chiller, Maechling, and
  Zare}}]{McKay1996}
\bibinfo{author}{\bibfnamefont{D.~S.} \bibnamefont{McKay}},
  \bibinfo{author}{\bibfnamefont{E.~K.} \bibnamefont{Gibson}},
  \bibinfo{author}{\bibfnamefont{K.~L.} \bibnamefont{ThomasKeprta}},
  \bibinfo{author}{\bibfnamefont{H.}~\bibnamefont{Vali}},
  \bibinfo{author}{\bibfnamefont{C.~S.} \bibnamefont{Romanek}},
  \bibinfo{author}{\bibfnamefont{S.~J.} \bibnamefont{Clemett}},
  \bibinfo{author}{\bibfnamefont{X.~D.~F.} \bibnamefont{Chiller}},
  \bibinfo{author}{\bibfnamefont{C.~R.} \bibnamefont{Maechling}},
  \bibnamefont{and} \bibinfo{author}{\bibfnamefont{R.~N.} \bibnamefont{Zare}},
  \bibinfo{journal}{Science} \textbf{\bibinfo{volume}{273}},
  \bibinfo{pages}{924} (\bibinfo{year}{1996}).

\bibitem[{\citenamefont{Thomas-Keprta et~al.}(2000)\citenamefont{Thomas-Keprta,
  Bazylinski, Kirschvink, Clemett, McKay, Wentworth, Vali, Gibson, and
  Romanek}}]{Thomas-Keprta2000}
\bibinfo{author}{\bibfnamefont{K.~L.} \bibnamefont{Thomas-Keprta}},
  \bibinfo{author}{\bibfnamefont{D.~A.} \bibnamefont{Bazylinski}},
  \bibinfo{author}{\bibfnamefont{J.~L.} \bibnamefont{Kirschvink}},
  \bibinfo{author}{\bibfnamefont{S.~J.} \bibnamefont{Clemett}},
  \bibinfo{author}{\bibfnamefont{D.~S.} \bibnamefont{McKay}},
  \bibinfo{author}{\bibfnamefont{S.~J.} \bibnamefont{Wentworth}},
  \bibinfo{author}{\bibfnamefont{H.}~\bibnamefont{Vali}},
  \bibinfo{author}{\bibfnamefont{E.~K.} \bibnamefont{Gibson}},
  \bibnamefont{and} \bibinfo{author}{\bibfnamefont{C.~S.}
  \bibnamefont{Romanek}}, \bibinfo{journal}{Geochimica Et Cosmochimica Acta}
  \textbf{\bibinfo{volume}{64}}, \bibinfo{pages}{4049} (\bibinfo{year}{2000}).

\bibitem[{\citenamefont{Thomas-Keprta et~al.}(2001)\citenamefont{Thomas-Keprta,
  Clemett, Bazylinski, Kirschvink, McKay, Wentworth, Vali, Gibson, McKay, and
  Romanek}}]{Thomas-Keprta2001}
\bibinfo{author}{\bibfnamefont{K.~L.} \bibnamefont{Thomas-Keprta}},
  \bibinfo{author}{\bibfnamefont{S.~J.} \bibnamefont{Clemett}},
  \bibinfo{author}{\bibfnamefont{D.~A.} \bibnamefont{Bazylinski}},
  \bibinfo{author}{\bibfnamefont{J.~L.} \bibnamefont{Kirschvink}},
  \bibinfo{author}{\bibfnamefont{D.~S.} \bibnamefont{McKay}},
  \bibinfo{author}{\bibfnamefont{S.~J.} \bibnamefont{Wentworth}},
  \bibinfo{author}{\bibfnamefont{H.}~\bibnamefont{Vali}},
  \bibinfo{author}{\bibfnamefont{E.~K.} \bibnamefont{Gibson}},
  \bibinfo{author}{\bibfnamefont{M.~F.} \bibnamefont{McKay}}, \bibnamefont{and}
  \bibinfo{author}{\bibfnamefont{C.~S.} \bibnamefont{Romanek}},
  \bibinfo{journal}{PNAS} \textbf{\bibinfo{volume}{98}}, \bibinfo{pages}{2164}
  (\bibinfo{year}{2001}).

\bibitem[{\citenamefont{Gibson et~al.}(2001)\citenamefont{Gibson, McKay,
  Thomas-Keprta, Wentworth, Westall, Steele, Romanek, Bell, and
  Toporski}}]{Gibson2001}
\bibinfo{author}{\bibfnamefont{E.~K.} \bibnamefont{Gibson}},
  \bibinfo{author}{\bibfnamefont{D.~S.} \bibnamefont{McKay}},
  \bibinfo{author}{\bibfnamefont{K.~L.} \bibnamefont{Thomas-Keprta}},
  \bibinfo{author}{\bibfnamefont{S.~J.} \bibnamefont{Wentworth}},
  \bibinfo{author}{\bibfnamefont{F.}~\bibnamefont{Westall}},
  \bibinfo{author}{\bibfnamefont{A.}~\bibnamefont{Steele}},
  \bibinfo{author}{\bibfnamefont{C.~S.} \bibnamefont{Romanek}},
  \bibinfo{author}{\bibfnamefont{M.~S.} \bibnamefont{Bell}}, \bibnamefont{and}
  \bibinfo{author}{\bibfnamefont{J.}~\bibnamefont{Toporski}},
  \bibinfo{journal}{Precambrian Research} \textbf{\bibinfo{volume}{106}},
  \bibinfo{pages}{15} (\bibinfo{year}{2001}).

\bibitem[{\citenamefont{Buseck et~al.}(2001)\citenamefont{Buseck,
  Dunin-Borkowski, Devouard, Frankel, McCartney, Midgley, Rosfai, and
  Weyland}}]{Buseck2001}
\bibinfo{author}{\bibfnamefont{P.~R.} \bibnamefont{Buseck}},
  \bibinfo{author}{\bibfnamefont{R.~E.} \bibnamefont{Dunin-Borkowski}},
  \bibinfo{author}{\bibfnamefont{B.}~\bibnamefont{Devouard}},
  \bibinfo{author}{\bibfnamefont{R.~B.} \bibnamefont{Frankel}},
  \bibinfo{author}{\bibfnamefont{M.~R.} \bibnamefont{McCartney}},
  \bibinfo{author}{\bibfnamefont{P.~A.} \bibnamefont{Midgley}},
  \bibinfo{author}{\bibfnamefont{M.}~\bibnamefont{Rosfai}}, \bibnamefont{and}
  \bibinfo{author}{\bibfnamefont{M.}~\bibnamefont{Weyland}},
  \bibinfo{journal}{Proceedings of the National Academy of Sciences of the
  United States of America} \textbf{\bibinfo{volume}{98}},
  \bibinfo{pages}{13490} (\bibinfo{year}{2001}).

\bibitem[{\citenamefont{Golden et~al.}(2004)\citenamefont{Golden, Ming, Morris,
  Brearley, Lauer, Treiman, Zolensky, Schwandt, Lofgren, and
  McKay}}]{Golden2004}
\bibinfo{author}{\bibfnamefont{D.~C.} \bibnamefont{Golden}},
  \bibinfo{author}{\bibfnamefont{D.~W.} \bibnamefont{Ming}},
  \bibinfo{author}{\bibfnamefont{R.~V.} \bibnamefont{Morris}},
  \bibinfo{author}{\bibfnamefont{A.~J.} \bibnamefont{Brearley}},
  \bibinfo{author}{\bibfnamefont{H.~V.} \bibnamefont{Lauer},
  \bibfnamefont{Jr.}}, \bibinfo{author}{\bibfnamefont{A.~H.}
  \bibnamefont{Treiman}}, \bibinfo{author}{\bibfnamefont{M.~E.}
  \bibnamefont{Zolensky}}, \bibinfo{author}{\bibfnamefont{C.~S.}
  \bibnamefont{Schwandt}}, \bibinfo{author}{\bibfnamefont{G.~E.}
  \bibnamefont{Lofgren}}, \bibnamefont{and}
  \bibinfo{author}{\bibfnamefont{G.~A.} \bibnamefont{McKay}},
  \bibinfo{journal}{Am. Mineral.} \textbf{\bibinfo{volume}{89}},
  \bibinfo{pages}{681} (\bibinfo{year}{2004}).

\bibitem[{\citenamefont{Arat{\'o} et~al.}(2005)\citenamefont{Arat{\'o},
  Sz{\'a}nyi, Flies, Sch{\"u}ler, Frankel, Buseck, and P{\'o}sfai}}]{Arato2005}
\bibinfo{author}{\bibfnamefont{B.}~\bibnamefont{Arat{\'o}}},
  \bibinfo{author}{\bibfnamefont{Z.}~\bibnamefont{Sz{\'a}nyi}},
  \bibinfo{author}{\bibfnamefont{C.}~\bibnamefont{Flies}},
  \bibinfo{author}{\bibfnamefont{D.}~\bibnamefont{Sch{\"u}ler}},
  \bibinfo{author}{\bibfnamefont{R.~B.} \bibnamefont{Frankel}},
  \bibinfo{author}{\bibfnamefont{P.}~\bibnamefont{Buseck}}, \bibnamefont{and}
  \bibinfo{author}{\bibfnamefont{M.}~\bibnamefont{P{\'o}sfai}},
  \bibinfo{journal}{Am. Mineral.} \textbf{\bibinfo{volume}{90}},
  \bibinfo{pages}{1233} (\bibinfo{year}{2005}).

\bibitem[{\citenamefont{Faivre and Zuddas}(2006)}]{Faivre2006}
\bibinfo{author}{\bibfnamefont{D.}~\bibnamefont{Faivre}} \bibnamefont{and}
  \bibinfo{author}{\bibfnamefont{P.}~\bibnamefont{Zuddas}},
  \bibinfo{journal}{Earth and Planetary Science Letters}
  \textbf{\bibinfo{volume}{243}}, \bibinfo{pages}{53} (\bibinfo{year}{2006}).

\bibitem[{\citenamefont{Kohn and Sham}(1965)}]{KS1965}
\bibinfo{author}{\bibfnamefont{W.}~\bibnamefont{Kohn}} \bibnamefont{and}
  \bibinfo{author}{\bibfnamefont{L.~J.} \bibnamefont{Sham}},
  \bibinfo{journal}{Phys. Rev.} \textbf{\bibinfo{volume}{140}},
  \bibinfo{pages}{A1133} (\bibinfo{year}{1965}).

\bibitem[{\citenamefont{Anisimov et~al.}(1991)\citenamefont{Anisimov, Zaanen,
  and Andersen}}]{Anisimov1991a}
\bibinfo{author}{\bibfnamefont{V.~I.} \bibnamefont{Anisimov}},
  \bibinfo{author}{\bibfnamefont{J.}~\bibnamefont{Zaanen}}, \bibnamefont{and}
  \bibinfo{author}{\bibfnamefont{O.~K.} \bibnamefont{Andersen}},
  \bibinfo{journal}{Phys. Rev. B} \textbf{\bibinfo{volume}{44}},
  \bibinfo{pages}{941} (\bibinfo{year}{1991}).

\bibitem[{\citenamefont{Cullen and Callen}(1971)}]{Cullen1971}
\bibinfo{author}{\bibfnamefont{J.~R.} \bibnamefont{Cullen}} \bibnamefont{and}
  \bibinfo{author}{\bibfnamefont{E.}~\bibnamefont{Callen}},
  \bibinfo{journal}{Journal De Physique} \textbf{\bibinfo{volume}{C1}},
  \bibinfo{pages}{C1} (\bibinfo{year}{1971}).

\bibitem[{\citenamefont{Yoshida and Iida}(1977)}]{Yoshida1977}
\bibinfo{author}{\bibfnamefont{J.}~\bibnamefont{Yoshida}} \bibnamefont{and}
  \bibinfo{author}{\bibfnamefont{S.}~\bibnamefont{Iida}}, \bibinfo{journal}{J.
  Phys. Soc. Jpn.} \textbf{\bibinfo{volume}{42}}, \bibinfo{pages}{230}
  (\bibinfo{year}{1977}).

\bibitem[{\citenamefont{Iizumi et~al.}(1982)\citenamefont{Iizumi, Koetzle,
  Shirane, Chikazumi, Matsui, and Todo}}]{Iizumi1982}
\bibinfo{author}{\bibfnamefont{M.}~\bibnamefont{Iizumi}},
  \bibinfo{author}{\bibfnamefont{T.~F.} \bibnamefont{Koetzle}},
  \bibinfo{author}{\bibfnamefont{G.}~\bibnamefont{Shirane}},
  \bibinfo{author}{\bibfnamefont{S.}~\bibnamefont{Chikazumi}},
  \bibinfo{author}{\bibfnamefont{M.}~\bibnamefont{Matsui}}, \bibnamefont{and}
  \bibinfo{author}{\bibfnamefont{S.}~\bibnamefont{Todo}},
  \bibinfo{journal}{Acta Cryst.} \textbf{\bibinfo{volume}{B38}},
  \bibinfo{pages}{2121} (\bibinfo{year}{1982}).

\bibitem[{\citenamefont{Zuo and Spence}(1990)}]{Zuo1990}
\bibinfo{author}{\bibfnamefont{J.~M.} \bibnamefont{Zuo}} \bibnamefont{and}
  \bibinfo{author}{\bibfnamefont{J.~C.~H.} \bibnamefont{Spence}},
  \bibinfo{journal}{Phys. Rev. B} \textbf{\bibinfo{volume}{42}},
  \bibinfo{pages}{8451} (\bibinfo{year}{1990}).

\bibitem[{\citenamefont{Zhang and Satpathy}(1991)}]{Zhang1991}
\bibinfo{author}{\bibfnamefont{Z.}~\bibnamefont{Zhang}} \bibnamefont{and}
  \bibinfo{author}{\bibfnamefont{S.}~\bibnamefont{Satpathy}},
  \bibinfo{journal}{Phys. Rev. B} \textbf{\bibinfo{volume}{44}},
  \bibinfo{pages}{13319} (\bibinfo{year}{1991}).

\bibitem[{\citenamefont{Madsen and Nov{\'a}k}(2005)}]{Madsen2005}
\bibinfo{author}{\bibfnamefont{G.~K.~H.} \bibnamefont{Madsen}}
  \bibnamefont{and}
  \bibinfo{author}{\bibfnamefont{P.}~\bibnamefont{Nov{\'a}k}},
  \bibinfo{journal}{Europhys. Lett.} \textbf{\bibinfo{volume}{69}},
  \bibinfo{pages}{777} (\bibinfo{year}{2005}).

\bibitem[{\citenamefont{Mazo-Zuluaga and Restrepo}(2005)}]{Mazo-Zuluaga2005}
\bibinfo{author}{\bibfnamefont{J.}~\bibnamefont{Mazo-Zuluaga}}
  \bibnamefont{and} \bibinfo{author}{\bibfnamefont{J.}~\bibnamefont{Restrepo}},
  \bibinfo{journal}{Phys. Stat. Sol.} \textbf{\bibinfo{volume}{2}},
  \bibinfo{pages}{3540} (\bibinfo{year}{2005}).

\bibitem[{\citenamefont{Punkkinen et~al.}(1999)\citenamefont{Punkkinen, Kokko,
  Hergert, and V{\"a}yrynen}}]{Punkkinen1999}
\bibinfo{author}{\bibfnamefont{M.~P.~J.} \bibnamefont{Punkkinen}},
  \bibinfo{author}{\bibfnamefont{K.}~\bibnamefont{Kokko}},
  \bibinfo{author}{\bibfnamefont{W.}~\bibnamefont{Hergert}}, \bibnamefont{and}
  \bibinfo{author}{\bibfnamefont{I.~J.} \bibnamefont{V{\"a}yrynen}},
  \bibinfo{journal}{J. Phys.: Condens. Matter} \textbf{\bibinfo{volume}{11}},
  \bibinfo{pages}{2341} (\bibinfo{year}{1999}).

\bibitem[{\citenamefont{Rosso and Rustad}(2001)}]{Rosso2001}
\bibinfo{author}{\bibfnamefont{K.~M.} \bibnamefont{Rosso}} \bibnamefont{and}
  \bibinfo{author}{\bibfnamefont{J.~R.} \bibnamefont{Rustad}},
  \bibinfo{journal}{Am. Mineral.} \textbf{\bibinfo{volume}{86}},
  \bibinfo{pages}{312} (\bibinfo{year}{2001}).

\bibitem[{\citenamefont{Rollmann et~al.}(2004)\citenamefont{Rollmann, Rohrbach,
  Entel, and Hafner}}]{Rollmann2004}
\bibinfo{author}{\bibfnamefont{G.}~\bibnamefont{Rollmann}},
  \bibinfo{author}{\bibfnamefont{A.}~\bibnamefont{Rohrbach}},
  \bibinfo{author}{\bibfnamefont{P.}~\bibnamefont{Entel}}, \bibnamefont{and}
  \bibinfo{author}{\bibfnamefont{J.}~\bibnamefont{Hafner}},
  \bibinfo{journal}{Phys. Rev. B} \textbf{\bibinfo{volume}{69}},
  \bibinfo{pages}{165107} (\bibinfo{year}{2004}).

\bibitem[{\citenamefont{Chamritski and Burns}(2005)}]{Chamritski2005}
\bibinfo{author}{\bibfnamefont{I.}~\bibnamefont{Chamritski}} \bibnamefont{and}
  \bibinfo{author}{\bibfnamefont{G.}~\bibnamefont{Burns}}, \bibinfo{journal}{J.
  Phys. Chem. B} \textbf{\bibinfo{volume}{109}}, \bibinfo{pages}{4965}
  (\bibinfo{year}{2005}).

\bibitem[{\citenamefont{Shiroishi et~al.}(2005)\citenamefont{Shiroishi, Oda,
  Hamada, and Fujima}}]{Shiroishi2005}
\bibinfo{author}{\bibfnamefont{H.}~\bibnamefont{Shiroishi}},
  \bibinfo{author}{\bibfnamefont{T.}~\bibnamefont{Oda}},
  \bibinfo{author}{\bibfnamefont{I.}~\bibnamefont{Hamada}}, \bibnamefont{and}
  \bibinfo{author}{\bibfnamefont{N.}~\bibnamefont{Fujima}},
  \bibinfo{journal}{Polyhedron} \textbf{\bibinfo{volume}{24}},
  \bibinfo{pages}{2472} (\bibinfo{year}{2005}).

\bibitem[{\citenamefont{Kubicki et~al.}(2008)\citenamefont{Kubicki, Pau, and
  Sparks}}]{Kubicki2008}
\bibinfo{author}{\bibfnamefont{J.~D.} \bibnamefont{Kubicki}},
  \bibinfo{author}{\bibfnamefont{K.~W.} \bibnamefont{Pau}}, \bibnamefont{and}
  \bibinfo{author}{\bibfnamefont{D.~L.} \bibnamefont{Sparks}},
  \bibinfo{journal}{Geochem. Trans.} \textbf{\bibinfo{volume}{9}},
  \bibinfo{pages}{4} (\bibinfo{year}{2008}).

\bibitem[{\citenamefont{Pentcheva et~al.}(2008)\citenamefont{Pentcheva, Moritz,
  Rundgren, Frank, Schrupp, and Scheffler}}]{Pentcheva2008}
\bibinfo{author}{\bibfnamefont{R.}~\bibnamefont{Pentcheva}},
  \bibinfo{author}{\bibfnamefont{W.}~\bibnamefont{Moritz}},
  \bibinfo{author}{\bibfnamefont{J.}~\bibnamefont{Rundgren}},
  \bibinfo{author}{\bibfnamefont{S.}~\bibnamefont{Frank}},
  \bibinfo{author}{\bibfnamefont{D.}~\bibnamefont{Schrupp}}, \bibnamefont{and}
  \bibinfo{author}{\bibfnamefont{M.}~\bibnamefont{Scheffler}},
  \bibinfo{journal}{Surf. Sci.} \textbf{\bibinfo{volume}{602}},
  \bibinfo{pages}{1299} (\bibinfo{year}{2008}).

\bibitem[{\citenamefont{Martin et~al.}(2009)\citenamefont{Martin, Cutting,
  Vaughan, and Warren}}]{Martin2009}
\bibinfo{author}{\bibfnamefont{G.~J.} \bibnamefont{Martin}},
  \bibinfo{author}{\bibfnamefont{R.~S.} \bibnamefont{Cutting}},
  \bibinfo{author}{\bibfnamefont{D.~J.} \bibnamefont{Vaughan}},
  \bibnamefont{and} \bibinfo{author}{\bibfnamefont{M.~C.}
  \bibnamefont{Warren}}, \bibinfo{journal}{Am. Mineral.}
  \textbf{\bibinfo{volume}{94}}, \bibinfo{pages}{1341} (\bibinfo{year}{2009}).

\bibitem[{\citenamefont{Pinney et~al.}(2009)\citenamefont{Pinney, Kubicki,
  Middlemiss, Grey, and Morgan}}]{Pinney2009}
\bibinfo{author}{\bibfnamefont{N.}~\bibnamefont{Pinney}},
  \bibinfo{author}{\bibfnamefont{J.~D.} \bibnamefont{Kubicki}},
  \bibinfo{author}{\bibfnamefont{D.~S.} \bibnamefont{Middlemiss}},
  \bibinfo{author}{\bibfnamefont{C.~P.} \bibnamefont{Grey}}, \bibnamefont{and}
  \bibinfo{author}{\bibfnamefont{D.}~\bibnamefont{Morgan}},
  \bibinfo{journal}{Chem. Mater.} \textbf{\bibinfo{volume}{21}},
  \bibinfo{pages}{5727} (\bibinfo{year}{2009}).

\bibitem[{\citenamefont{Russell et~al.}(2009)\citenamefont{Russell, Payne, and
  Ciacchi}}]{Russell2009}
\bibinfo{author}{\bibfnamefont{B.}~\bibnamefont{Russell}},
  \bibinfo{author}{\bibfnamefont{M.}~\bibnamefont{Payne}}, \bibnamefont{and}
  \bibinfo{author}{\bibfnamefont{L.~C.} \bibnamefont{Ciacchi}},
  \bibinfo{journal}{Phys. Rev. B} \textbf{\bibinfo{volume}{79}},
  \bibinfo{pages}{165101} (\bibinfo{year}{2009}).

\bibitem[{\citenamefont{Wilson and Russo}(2009)}]{Wilson2009}
\bibinfo{author}{\bibfnamefont{N.~C.} \bibnamefont{Wilson}} \bibnamefont{and}
  \bibinfo{author}{\bibfnamefont{S.~V.} \bibnamefont{Russo}},
  \bibinfo{journal}{Phys. Rev. B} \textbf{\bibinfo{volume}{79}},
  \bibinfo{pages}{094113} (\bibinfo{year}{2009}).

\bibitem[{\citenamefont{Grau-Crespo et~al.}(2010)\citenamefont{Grau-Crespo,
  Al-Baitai, Saadoune, and de~Leeuw}}]{Grau-Crespo2010}
\bibinfo{author}{\bibfnamefont{R.}~\bibnamefont{Grau-Crespo}},
  \bibinfo{author}{\bibfnamefont{A.~Y.} \bibnamefont{Al-Baitai}},
  \bibinfo{author}{\bibfnamefont{I.}~\bibnamefont{Saadoune}}, \bibnamefont{and}
  \bibinfo{author}{\bibfnamefont{N.~H.} \bibnamefont{de~Leeuw}},
  \bibinfo{journal}{J. Phys.: Condens. Matter} \textbf{\bibinfo{volume}{22}},
  \bibinfo{pages}{255401} (\bibinfo{year}{2010}).

\bibitem[{\citenamefont{Artacho et~al.}(2008)\citenamefont{Artacho, Anglada,
  Di{\'e}guez, Gale, Garc{\'\i}a, Junquera, Martin, Ordej{\'o}n, Pruneda,
  S{\'a}nchez-Portal et~al.}}]{Artacho2008}
\bibinfo{author}{\bibfnamefont{E.}~\bibnamefont{Artacho}},
  \bibinfo{author}{\bibfnamefont{E.}~\bibnamefont{Anglada}},
  \bibinfo{author}{\bibfnamefont{O.}~\bibnamefont{Di{\'e}guez}},
  \bibinfo{author}{\bibfnamefont{J.~D.} \bibnamefont{Gale}},
  \bibinfo{author}{\bibfnamefont{A.}~\bibnamefont{Garc{\'\i}a}},
  \bibinfo{author}{\bibfnamefont{J.}~\bibnamefont{Junquera}},
  \bibinfo{author}{\bibfnamefont{R.~M.} \bibnamefont{Martin}},
  \bibinfo{author}{\bibfnamefont{P.}~\bibnamefont{Ordej{\'o}n}},
  \bibinfo{author}{\bibfnamefont{J.~M.} \bibnamefont{Pruneda}},
  \bibinfo{author}{\bibfnamefont{D.}~\bibnamefont{S{\'a}nchez-Portal}},
  \bibnamefont{et~al.}, \bibinfo{journal}{J. Phys.: Condens. Matter}
  \textbf{\bibinfo{volume}{20}}, \bibinfo{pages}{064208}
  (\bibinfo{year}{2008}).

\bibitem[{\citenamefont{Poulet et~al.}(2003)\citenamefont{Poulet, Sautet, and
  Artacho}}]{Poulet2003}
\bibinfo{author}{\bibfnamefont{G.}~\bibnamefont{Poulet}},
  \bibinfo{author}{\bibfnamefont{P.}~\bibnamefont{Sautet}}, \bibnamefont{and}
  \bibinfo{author}{\bibfnamefont{E.}~\bibnamefont{Artacho}},
  \bibinfo{journal}{Phys. Rev. B} \textbf{\bibinfo{volume}{68}},
  \bibinfo{pages}{075118} (\bibinfo{year}{2003}).

\bibitem[{\citenamefont{Kresse and
  Furthm{\"u}ller}(1996{\natexlab{a}})}]{Kresse1996:VASP1}
\bibinfo{author}{\bibfnamefont{G.}~\bibnamefont{Kresse}} \bibnamefont{and}
  \bibinfo{author}{\bibfnamefont{J.}~\bibnamefont{Furthm{\"u}ller}},
  \bibinfo{journal}{Phys. Rev. B} \textbf{\bibinfo{volume}{54}},
  \bibinfo{pages}{11169} (\bibinfo{year}{1996}{\natexlab{a}}).

\bibitem[{\citenamefont{Kresse and
  Furthm{\"u}ller}(1996{\natexlab{b}})}]{Kresse1996:VASP2}
\bibinfo{author}{\bibfnamefont{G.}~\bibnamefont{Kresse}} \bibnamefont{and}
  \bibinfo{author}{\bibfnamefont{J.}~\bibnamefont{Furthm{\"u}ller}},
  \bibinfo{journal}{Comput. Mater. Sci.} \textbf{\bibinfo{volume}{6}},
  \bibinfo{pages}{15} (\bibinfo{year}{1996}{\natexlab{b}}).

\bibitem[{\citenamefont{Artacho et~al.}(1999)\citenamefont{Artacho,
  S{\'a}nchez-Portal, Ordej{\'o}n, Garc{\'{\i}}a, and Soler}}]{Artacho1999}
\bibinfo{author}{\bibfnamefont{E.}~\bibnamefont{Artacho}},
  \bibinfo{author}{\bibfnamefont{D.}~\bibnamefont{S{\'a}nchez-Portal}},
  \bibinfo{author}{\bibfnamefont{P.}~\bibnamefont{Ordej{\'o}n}},
  \bibinfo{author}{\bibfnamefont{A.}~\bibnamefont{Garc{\'{\i}}a}},
  \bibnamefont{and} \bibinfo{author}{\bibfnamefont{J.~M.} \bibnamefont{Soler}},
  \bibinfo{journal}{Phys. Stat. Sol. (b)} \textbf{\bibinfo{volume}{215}},
  \bibinfo{pages}{809} (\bibinfo{year}{1999}).

\bibitem[{\citenamefont{Soler et~al.}(2002)\citenamefont{Soler, Artacho, Gale,
  Garc{\'\i}a, Junquera, Ordej{\'o}n, and S{\'a}nchez-Portal}}]{Soler2002}
\bibinfo{author}{\bibfnamefont{J.~M.} \bibnamefont{Soler}},
  \bibinfo{author}{\bibfnamefont{E.}~\bibnamefont{Artacho}},
  \bibinfo{author}{\bibfnamefont{J.~D.} \bibnamefont{Gale}},
  \bibinfo{author}{\bibfnamefont{A.}~\bibnamefont{Garc{\'\i}a}},
  \bibinfo{author}{\bibfnamefont{J.}~\bibnamefont{Junquera}},
  \bibinfo{author}{\bibfnamefont{P.}~\bibnamefont{Ordej{\'o}n}},
  \bibnamefont{and}
  \bibinfo{author}{\bibfnamefont{D.}~\bibnamefont{S{\'a}nchez-Portal}},
  \bibinfo{journal}{J. Phys.: Condens. Matter} \textbf{\bibinfo{volume}{14}},
  \bibinfo{pages}{2745} (\bibinfo{year}{2002}).

\bibitem[{\citenamefont{Leung et~al.}(1991)\citenamefont{Leung, Chan, and
  Harmon}}]{Leung1991}
\bibinfo{author}{\bibfnamefont{T.~C.} \bibnamefont{Leung}},
  \bibinfo{author}{\bibfnamefont{C.~T.} \bibnamefont{Chan}}, \bibnamefont{and}
  \bibinfo{author}{\bibfnamefont{B.~N.} \bibnamefont{Harmon}},
  \bibinfo{journal}{Phys. Rev. B} \textbf{\bibinfo{volume}{44}},
  \bibinfo{pages}{2923} (\bibinfo{year}{1991}).

\bibitem[{\citenamefont{Perdew et~al.}(1996)\citenamefont{Perdew, Burke, and
  Ernzerhof}}]{Perdew1996}
\bibinfo{author}{\bibfnamefont{J.~P.} \bibnamefont{Perdew}},
  \bibinfo{author}{\bibfnamefont{K.}~\bibnamefont{Burke}}, \bibnamefont{and}
  \bibinfo{author}{\bibfnamefont{M.}~\bibnamefont{Ernzerhof}},
  \bibinfo{journal}{Phys. Rev. Lett.} \textbf{\bibinfo{volume}{77}},
  \bibinfo{pages}{3865} (\bibinfo{year}{1996}).

\bibitem[{\citenamefont{Latham et~al.}(2006)\citenamefont{Latham, {\"O}berg,
  Briddon, and Louchet}}]{Latham2006}
\bibinfo{author}{\bibfnamefont{C.~D.} \bibnamefont{Latham}},
  \bibinfo{author}{\bibfnamefont{S.}~\bibnamefont{{\"O}berg}},
  \bibinfo{author}{\bibfnamefont{P.~R.} \bibnamefont{Briddon}},
  \bibnamefont{and} \bibinfo{author}{\bibfnamefont{F.}~\bibnamefont{Louchet}},
  \bibinfo{journal}{J. Phys.: Condens. Matter.} \textbf{\bibinfo{volume}{18}},
  \bibinfo{pages}{8859} (\bibinfo{year}{2006}).

\bibitem[{\citenamefont{Trinit{\'e} et~al.}(2008)\citenamefont{Trinit{\'e},
  Vast, and Hayoun}}]{Trinite2008}
\bibinfo{author}{\bibfnamefont{V.}~\bibnamefont{Trinit{\'e}}},
  \bibinfo{author}{\bibfnamefont{N.}~\bibnamefont{Vast}}, \bibnamefont{and}
  \bibinfo{author}{\bibfnamefont{M.}~\bibnamefont{Hayoun}},
  \bibinfo{journal}{J. Phys.: Condens. Matter.} \textbf{\bibinfo{volume}{20}},
  \bibinfo{pages}{235239} (\bibinfo{year}{2008}).

\bibitem[{\citenamefont{Troullier and Martins}(1991)}]{Troullier1991}
\bibinfo{author}{\bibfnamefont{N.}~\bibnamefont{Troullier}} \bibnamefont{and}
  \bibinfo{author}{\bibfnamefont{J.~L.} \bibnamefont{Martins}},
  \bibinfo{journal}{Phys. Rev. B} \textbf{\bibinfo{volume}{43}},
  \bibinfo{pages}{1993} (\bibinfo{year}{1991}).

\bibitem[{\citenamefont{Fagan et~al.}(2003)\citenamefont{Fagan, Mota, da~Silva,
  and Fazzio}}]{Fagan2003}
\bibinfo{author}{\bibfnamefont{S.~B.} \bibnamefont{Fagan}},
  \bibinfo{author}{\bibfnamefont{R.}~\bibnamefont{Mota}},
  \bibinfo{author}{\bibfnamefont{A.~J.~R.} \bibnamefont{da~Silva}},
  \bibnamefont{and} \bibinfo{author}{\bibfnamefont{A.}~\bibnamefont{Fazzio}},
  \bibinfo{journal}{Microelectronics Journal} \textbf{\bibinfo{volume}{34}},
  \bibinfo{pages}{481} (\bibinfo{year}{2003}).

\bibitem[{\citenamefont{Izquierdo et~al.}(2000)\citenamefont{Izquierdo, Vega,
  Balb{\'a}s, S{\'a}nchez, Junquera, Artacho, Soler, and
  Ordej{\'o}n}}]{Izquierdo2000}
\bibinfo{author}{\bibfnamefont{J.}~\bibnamefont{Izquierdo}},
  \bibinfo{author}{\bibfnamefont{A.}~\bibnamefont{Vega}},
  \bibinfo{author}{\bibfnamefont{L.~C.} \bibnamefont{Balb{\'a}s}},
  \bibinfo{author}{\bibfnamefont{D.}~\bibnamefont{S{\'a}nchez}},
  \bibinfo{author}{\bibfnamefont{J.}~\bibnamefont{Junquera}},
  \bibinfo{author}{\bibfnamefont{E.}~\bibnamefont{Artacho}},
  \bibinfo{author}{\bibfnamefont{J.~M.} \bibnamefont{Soler}}, \bibnamefont{and}
  \bibinfo{author}{\bibfnamefont{P.}~\bibnamefont{Ordej{\'o}n}},
  \bibinfo{journal}{Phys. Rev. B} \textbf{\bibinfo{volume}{61}},
  \bibinfo{pages}{13639} (\bibinfo{year}{2000}).

\bibitem[{\citenamefont{Garc{\'\i}a-Su{\'a}rez
  et~al.}(2009)\citenamefont{Garc{\'\i}a-Su{\'a}rez, Newman, Lambert, Pruneda,
  and Ferrer}}]{Garcia-Suarez2008}
\bibinfo{author}{\bibfnamefont{V.~M.} \bibnamefont{Garc{\'\i}a-Su{\'a}rez}},
  \bibinfo{author}{\bibfnamefont{C.~M.} \bibnamefont{Newman}},
  \bibinfo{author}{\bibfnamefont{C.~J.} \bibnamefont{Lambert}},
  \bibinfo{author}{\bibfnamefont{J.~M.} \bibnamefont{Pruneda}},
  \bibnamefont{and} \bibinfo{author}{\bibfnamefont{J.}~\bibnamefont{Ferrer}}
  (\bibinfo{year}{2009}), \bibinfo{note}{2008}.

\bibitem[{\citenamefont{Ortega-Castro et~al.}(2009)\citenamefont{Ortega-Castro,
  Hern{\'a}ndez-Haro, Mu{\~n}oz-Santiburcio, and
  Sainz-D{\'{\i}}az}}]{Ortega-Castro2009}
\bibinfo{author}{\bibfnamefont{J.}~\bibnamefont{Ortega-Castro}},
  \bibinfo{author}{\bibfnamefont{N.}~\bibnamefont{Hern{\'a}ndez-Haro}},
  \bibinfo{author}{\bibfnamefont{A.}~\bibnamefont{Mu{\~n}oz-Santiburcio},
  \bibfnamefont{Hern{\'a}ndez-Laguna}}, \bibnamefont{and}
  \bibinfo{author}{\bibfnamefont{C.~I.} \bibnamefont{Sainz-D{\'{\i}}az}},
  \bibinfo{journal}{Journal of Molecular Structure: THEOCHEM}
  \textbf{\bibinfo{volume}{912}}, \bibinfo{pages}{82} (\bibinfo{year}{2009}).

\bibitem[{\citenamefont{Winkler et~al.}(2008)\citenamefont{Winkler, Gale,
  Refson, Wilson, and Milman}}]{Winkler2008}
\bibinfo{author}{\bibfnamefont{B.}~\bibnamefont{Winkler}},
  \bibinfo{author}{\bibfnamefont{J.~D.} \bibnamefont{Gale}},
  \bibinfo{author}{\bibfnamefont{K.}~\bibnamefont{Refson}},
  \bibinfo{author}{\bibfnamefont{D.~J.} \bibnamefont{Wilson}},
  \bibnamefont{and} \bibinfo{author}{\bibfnamefont{V.}~\bibnamefont{Milman}},
  \bibinfo{journal}{Phys. Chem. Minerals} \textbf{\bibinfo{volume}{35}},
  \bibinfo{pages}{25} (\bibinfo{year}{2008}).

\bibitem[{\citenamefont{Junquera et~al.}(2001)\citenamefont{Junquera, Paz,
  S{\'a}nchez-Portal, and Artacho}}]{Junquera2001}
\bibinfo{author}{\bibfnamefont{J.}~\bibnamefont{Junquera}},
  \bibinfo{author}{\bibfnamefont{{\'O}.}~\bibnamefont{Paz}},
  \bibinfo{author}{\bibfnamefont{D.}~\bibnamefont{S{\'a}nchez-Portal}},
  \bibnamefont{and} \bibinfo{author}{\bibfnamefont{E.}~\bibnamefont{Artacho}},
  \bibinfo{journal}{Phys. Rev. B} \textbf{\bibinfo{volume}{64}},
  \bibinfo{pages}{235111} (\bibinfo{year}{2001}).

\bibitem[{\citenamefont{Anisimov and Gunnarsson}(1991)}]{Anisimov1991}
\bibinfo{author}{\bibfnamefont{V.~I.} \bibnamefont{Anisimov}} \bibnamefont{and}
  \bibinfo{author}{\bibfnamefont{O.}~\bibnamefont{Gunnarsson}},
  \bibinfo{journal}{Phys. Rev. B} \textbf{\bibinfo{volume}{43}},
  \bibinfo{pages}{7570} (\bibinfo{year}{1991}).

\bibitem[{\citenamefont{Dudarev et~al.}(1998)\citenamefont{Dudarev, Botton,
  Savrasov, Hmphreys, and Sutton}}]{Dudarev1998}
\bibinfo{author}{\bibfnamefont{S.~L.} \bibnamefont{Dudarev}},
  \bibinfo{author}{\bibfnamefont{G.~A.} \bibnamefont{Botton}},
  \bibinfo{author}{\bibfnamefont{S.~Y.} \bibnamefont{Savrasov}},
  \bibinfo{author}{\bibfnamefont{C.~J.} \bibnamefont{Hmphreys}},
  \bibnamefont{and} \bibinfo{author}{\bibfnamefont{A.~P.}
  \bibnamefont{Sutton}}, \bibinfo{journal}{Phys. Rev. B}
  \textbf{\bibinfo{volume}{57}}, \bibinfo{pages}{1505} (\bibinfo{year}{1998}).

\bibitem[{\citenamefont{Becke}(1993)}]{Becke1993}
\bibinfo{author}{\bibfnamefont{A.~D.} \bibnamefont{Becke}},
  \bibinfo{journal}{J. Chem. Phys.} \textbf{\bibinfo{volume}{98}},
  \bibinfo{pages}{1372} (\bibinfo{year}{1993}).

\bibitem[{\citenamefont{Perdew and Ernzerhof}(1996)}]{Perdew1996a}
\bibinfo{author}{\bibfnamefont{J.~P.} \bibnamefont{Perdew}} \bibnamefont{and}
  \bibinfo{author}{\bibfnamefont{M.}~\bibnamefont{Ernzerhof}},
  \bibinfo{journal}{J. Chem. Phys.} \textbf{\bibinfo{volume}{105}},
  \bibinfo{pages}{9982} (\bibinfo{year}{1996}).

\bibitem[{\citenamefont{Perdew and Zunger}(1981)}]{Perdew1981}
\bibinfo{author}{\bibfnamefont{J.~P.} \bibnamefont{Perdew}} \bibnamefont{and}
  \bibinfo{author}{\bibfnamefont{A.}~\bibnamefont{Zunger}},
  \bibinfo{journal}{Phys. Rev. B} \textbf{\bibinfo{volume}{23}},
  \bibinfo{pages}{5048} (\bibinfo{year}{1981}).

\bibitem[{\citenamefont{Svane and Gunnarsson}(1990)}]{Svane1990}
\bibinfo{author}{\bibfnamefont{A.}~\bibnamefont{Svane}} \bibnamefont{and}
  \bibinfo{author}{\bibfnamefont{O.}~\bibnamefont{Gunnarsson}},
  \bibinfo{journal}{Phys. Rev. Lett.} \textbf{\bibinfo{volume}{65}},
  \bibinfo{pages}{1148} (\bibinfo{year}{1990}).

\bibitem[{\citenamefont{Moreira et~al.}(2002)\citenamefont{Moreira, Illas, and
  Martin}}]{Moreira2002}
\bibinfo{author}{\bibfnamefont{I.}~\bibnamefont{Moreira},
  \bibfnamefont{de~P.~R.}},
  \bibinfo{author}{\bibfnamefont{F.}~\bibnamefont{Illas}}, \bibnamefont{and}
  \bibinfo{author}{\bibfnamefont{R.~L.} \bibnamefont{Martin}},
  \bibinfo{journal}{Phys. Rev. B} \textbf{\bibinfo{volume}{65}},
  \bibinfo{pages}{155102} (\bibinfo{year}{2002}).

\bibitem[{\citenamefont{Kudin et~al.}(2002)\citenamefont{Kudin, Schimka, and
  Martin}}]{Kudin2002}
\bibinfo{author}{\bibfnamefont{K.~N.} \bibnamefont{Kudin}},
  \bibinfo{author}{\bibfnamefont{L.}~\bibnamefont{Schimka}}, \bibnamefont{and}
  \bibinfo{author}{\bibfnamefont{R.~L.} \bibnamefont{Martin}},
  \bibinfo{journal}{Phys. Rev. Lett.} \textbf{\bibinfo{volume}{89}},
  \bibinfo{pages}{266402} (\bibinfo{year}{2002}).

\bibitem[{\citenamefont{Hay et~al.}(2006)\citenamefont{Hay, Martin, Uddin, and
  Scuseria}}]{Hay2006}
\bibinfo{author}{\bibfnamefont{P.~J.} \bibnamefont{Hay}},
  \bibinfo{author}{\bibfnamefont{R.~L.} \bibnamefont{Martin}},
  \bibinfo{author}{\bibfnamefont{J.}~\bibnamefont{Uddin}}, \bibnamefont{and}
  \bibinfo{author}{\bibfnamefont{G.~E.} \bibnamefont{Scuseria}},
  \bibinfo{journal}{J. Chem. Phys.} \textbf{\bibinfo{volume}{125}},
  \bibinfo{pages}{034712} (\bibinfo{year}{2006}).

\bibitem[{\citenamefont{Prodan et~al.}(2005)\citenamefont{Prodan, Scuseria,
  Sordo, Kudin, and Martin}}]{Prodan2005}
\bibinfo{author}{\bibfnamefont{I.~D.} \bibnamefont{Prodan}},
  \bibinfo{author}{\bibfnamefont{G.~E.} \bibnamefont{Scuseria}},
  \bibinfo{author}{\bibfnamefont{J.~A.} \bibnamefont{Sordo}},
  \bibinfo{author}{\bibfnamefont{K.~N.} \bibnamefont{Kudin}}, \bibnamefont{and}
  \bibinfo{author}{\bibfnamefont{R.~L.} \bibnamefont{Martin}},
  \bibinfo{journal}{J. Chem. Phys.} \textbf{\bibinfo{volume}{123}},
  \bibinfo{pages}{014703} (\bibinfo{year}{2005}).

\bibitem[{\citenamefont{Rivero et~al.}(2009)\citenamefont{Rivero, Moreira,
  Scuseria, and Illas}}]{Rivero2009}
\bibinfo{author}{\bibfnamefont{P.}~\bibnamefont{Rivero}},
  \bibinfo{author}{\bibfnamefont{I.}~\bibnamefont{Moreira},
  \bibfnamefont{de~P.~R.}}, \bibinfo{author}{\bibfnamefont{G.~E.}
  \bibnamefont{Scuseria}}, \bibnamefont{and}
  \bibinfo{author}{\bibfnamefont{F.}~\bibnamefont{Illas}},
  \bibinfo{journal}{Phys. Rev. B} \textbf{\bibinfo{volume}{79}},
  \bibinfo{pages}{245129} (\bibinfo{year}{2009}).

\bibitem[{\citenamefont{Csonka et~al.}(2009)\citenamefont{Csonka, Perdew,
  Ruzsinszky, Philipsen, Lebegue, Paier, Vydrov, and Angyan}}]{Csonka2009}
\bibinfo{author}{\bibfnamefont{G.~I.} \bibnamefont{Csonka}},
  \bibinfo{author}{\bibfnamefont{J.~P.} \bibnamefont{Perdew}},
  \bibinfo{author}{\bibfnamefont{A.}~\bibnamefont{Ruzsinszky}},
  \bibinfo{author}{\bibfnamefont{P.~H.~T.} \bibnamefont{Philipsen}},
  \bibinfo{author}{\bibfnamefont{S.}~\bibnamefont{Lebegue}},
  \bibinfo{author}{\bibfnamefont{J.}~\bibnamefont{Paier}},
  \bibinfo{author}{\bibfnamefont{O.~A.} \bibnamefont{Vydrov}},
  \bibnamefont{and} \bibinfo{author}{\bibfnamefont{J.~G.}
  \bibnamefont{Angyan}}, \bibinfo{journal}{Phys. Rev. B}
  \textbf{\bibinfo{volume}{79}}, \bibinfo{pages}{155107}
  (\bibinfo{year}{2009}).

\bibitem[{\citenamefont{Yang et~al.}(2010)\citenamefont{Yang, Zheng, Zhao, and
  Truhlar}}]{Yang2010}
\bibinfo{author}{\bibfnamefont{K.}~\bibnamefont{Yang}},
  \bibinfo{author}{\bibfnamefont{J.}~\bibnamefont{Zheng}},
  \bibinfo{author}{\bibfnamefont{Y.}~\bibnamefont{Zhao}}, \bibnamefont{and}
  \bibinfo{author}{\bibfnamefont{D.~G.} \bibnamefont{Truhlar}},
  \bibinfo{journal}{J. Chem. Phys.} \textbf{\bibinfo{volume}{132}},
  \bibinfo{pages}{164117} (\bibinfo{year}{2010}).

\bibitem[{\citenamefont{Vydrov and Scuseria}(2006)}]{Vydrov2006}
\bibinfo{author}{\bibfnamefont{O.~A.} \bibnamefont{Vydrov}} \bibnamefont{and}
  \bibinfo{author}{\bibfnamefont{G.~E.} \bibnamefont{Scuseria}},
  \bibinfo{journal}{J. Chem. Phys.} \textbf{\bibinfo{volume}{125}},
  \bibinfo{pages}{234109} (\bibinfo{year}{2006}).

\bibitem[{\citenamefont{Krukau et~al.}(2008)\citenamefont{Krukau, Scuseria,
  Perdew, and Savin}}]{Krukau2008}
\bibinfo{author}{\bibfnamefont{A.~V.} \bibnamefont{Krukau}},
  \bibinfo{author}{\bibfnamefont{G.~E.} \bibnamefont{Scuseria}},
  \bibinfo{author}{\bibfnamefont{J.~P.} \bibnamefont{Perdew}},
  \bibnamefont{and} \bibinfo{author}{\bibfnamefont{A.}~\bibnamefont{Savin}},
  \bibinfo{journal}{J. Chem. Phys.} \textbf{\bibinfo{volume}{129}},
  \bibinfo{pages}{124103} (\bibinfo{year}{2008}).

\bibitem[{\citenamefont{Henderson et~al.}(2008)\citenamefont{Henderson,
  Janesko, and Scuseria}}]{Henderson2008}
\bibinfo{author}{\bibfnamefont{T.~M.} \bibnamefont{Henderson}},
  \bibinfo{author}{\bibfnamefont{B.~G.} \bibnamefont{Janesko}},
  \bibnamefont{and} \bibinfo{author}{\bibfnamefont{G.}~\bibnamefont{Scuseria}},
  \bibinfo{journal}{J. Phys. Chem.} \textbf{\bibinfo{volume}{112}},
  \bibinfo{pages}{12530} (\bibinfo{year}{2008}).

\bibitem[{\citenamefont{Heyd et~al.}(2003)\citenamefont{Heyd, Scuseria, and
  Ernzerhof}}]{Heyd2003}
\bibinfo{author}{\bibfnamefont{J.}~\bibnamefont{Heyd}},
  \bibinfo{author}{\bibfnamefont{G.~E.} \bibnamefont{Scuseria}},
  \bibnamefont{and}
  \bibinfo{author}{\bibfnamefont{M.}~\bibnamefont{Ernzerhof}},
  \bibinfo{journal}{J. Chem. Phys.} \textbf{\bibinfo{volume}{118}},
  \bibinfo{pages}{8207} (\bibinfo{year}{2003}).

\bibitem[{\citenamefont{Heyd and Scuseria}(2004)}]{Heyd2004}
\bibinfo{author}{\bibfnamefont{J.}~\bibnamefont{Heyd}} \bibnamefont{and}
  \bibinfo{author}{\bibfnamefont{G.~E.} \bibnamefont{Scuseria}},
  \bibinfo{journal}{J. Chem. Phys.} \textbf{\bibinfo{volume}{121}},
  \bibinfo{pages}{1187} (\bibinfo{year}{2004}).

\bibitem[{\citenamefont{Peralta et~al.}(2006)\citenamefont{Peralta, Heyd,
  Scuseria, and Martin}}]{Peralta2006}
\bibinfo{author}{\bibfnamefont{J.~E.} \bibnamefont{Peralta}},
  \bibinfo{author}{\bibfnamefont{J.}~\bibnamefont{Heyd}},
  \bibinfo{author}{\bibfnamefont{G.}~\bibnamefont{Scuseria}}, \bibnamefont{and}
  \bibinfo{author}{\bibfnamefont{R.~L.} \bibnamefont{Martin}},
  \bibinfo{journal}{Phys. Rev. B} \textbf{\bibinfo{volume}{74}},
  \bibinfo{pages}{073101} (\bibinfo{year}{2006}).

\bibitem[{\citenamefont{Brothers et~al.}(2008)\citenamefont{Brothers, Izmaylov,
  Normand, Barone, and Scuseria}}]{Brothers2008}
\bibinfo{author}{\bibfnamefont{E.~N.} \bibnamefont{Brothers}},
  \bibinfo{author}{\bibfnamefont{A.~F.} \bibnamefont{Izmaylov}},
  \bibinfo{author}{\bibfnamefont{J.~O.} \bibnamefont{Normand}},
  \bibinfo{author}{\bibfnamefont{V.}~\bibnamefont{Barone}}, \bibnamefont{and}
  \bibinfo{author}{\bibfnamefont{G.}~\bibnamefont{Scuseria}},
  \bibinfo{journal}{J. Chem. Phys.} \textbf{\bibinfo{volume}{129}},
  \bibinfo{pages}{011102} (\bibinfo{year}{2008}).

\bibitem[{\citenamefont{Perdew et~al.}(2005)\citenamefont{Perdew, Ruzsinszky,
  Tao, Staroverov, Scuseria, and Csonka}}]{Perdew2005}
\bibinfo{author}{\bibfnamefont{J.~P.} \bibnamefont{Perdew}},
  \bibinfo{author}{\bibfnamefont{A.}~\bibnamefont{Ruzsinszky}},
  \bibinfo{author}{\bibfnamefont{J.~M.} \bibnamefont{Tao}},
  \bibinfo{author}{\bibfnamefont{V.~N.} \bibnamefont{Staroverov}},
  \bibinfo{author}{\bibfnamefont{G.~E.} \bibnamefont{Scuseria}},
  \bibnamefont{and} \bibinfo{author}{\bibfnamefont{G.~I.}
  \bibnamefont{Csonka}}, \bibinfo{journal}{J. Chem. Phys.}
  \textbf{\bibinfo{volume}{123}}, \bibinfo{pages}{062201}
  (\bibinfo{year}{2005}).

\bibitem[{\citenamefont{Rohrbach et~al.}(2004)\citenamefont{Rohrbach, Hafner,
  and Kresse}}]{Rohrbach2004}
\bibinfo{author}{\bibfnamefont{A.}~\bibnamefont{Rohrbach}},
  \bibinfo{author}{\bibfnamefont{J.}~\bibnamefont{Hafner}}, \bibnamefont{and}
  \bibinfo{author}{\bibfnamefont{G.}~\bibnamefont{Kresse}},
  \bibinfo{journal}{Phys. Rev. B} \textbf{\bibinfo{volume}{70}},
  \bibinfo{pages}{125426} (\bibinfo{year}{2004}).

\bibitem[{\citenamefont{Cococcioni and de~Gironcoli}(2005)}]{Cococcioni2005}
\bibinfo{author}{\bibfnamefont{M.}~\bibnamefont{Cococcioni}} \bibnamefont{and}
  \bibinfo{author}{\bibfnamefont{S.}~\bibnamefont{de~Gironcoli}},
  \bibinfo{journal}{Phys. Rev. B} \textbf{\bibinfo{volume}{71}},
  \bibinfo{pages}{035105} (\bibinfo{year}{2005}).

\bibitem[{\citenamefont{Monkhorst and Pack}(1976)}]{Monkhorst1976}
\bibinfo{author}{\bibfnamefont{H.~J.} \bibnamefont{Monkhorst}}
  \bibnamefont{and} \bibinfo{author}{\bibfnamefont{J.~D.} \bibnamefont{Pack}},
  \bibinfo{journal}{Phys. Rev. B} \textbf{\bibinfo{volume}{13}},
  \bibinfo{pages}{5188} (\bibinfo{year}{1976}).

\bibitem[{\citenamefont{Birch}(1947)}]{Birch1947}
\bibinfo{author}{\bibfnamefont{F.}~\bibnamefont{Birch}},
  \bibinfo{journal}{Phys. Rev.} \textbf{\bibinfo{volume}{71}},
  \bibinfo{pages}{809} (\bibinfo{year}{1947}).

\bibitem[{\citenamefont{Nye}(1985)}]{Nye1985}
\bibinfo{author}{\bibfnamefont{J.~F.} \bibnamefont{Nye}},
  \emph{\bibinfo{title}{Physical properties of crystals}}
  (\bibinfo{publisher}{Oxford University Press}, \bibinfo{year}{1985}),
  \bibinfo{note}{sub title: their representation by tensors and matrices}.

\bibitem[{\citenamefont{Guo et~al.}(2005)\citenamefont{Guo, Li, Kong, and
  Liu}}]{Guo2005PRB}
\bibinfo{author}{\bibfnamefont{H.~B.} \bibnamefont{Guo}},
  \bibinfo{author}{\bibfnamefont{J.~H.} \bibnamefont{Li}},
  \bibinfo{author}{\bibfnamefont{L.~T.} \bibnamefont{Kong}}, \bibnamefont{and}
  \bibinfo{author}{\bibfnamefont{B.~X.} \bibnamefont{Liu}},
  \bibinfo{journal}{Phys. Rev. B} \textbf{\bibinfo{volume}{72}},
  \bibinfo{pages}{132102} (\bibinfo{year}{2005}).

\bibitem[{\citenamefont{Kong et~al.}(2010)\citenamefont{Kong, Xiong, Guo, Sun,
  Du, and Zhou}}]{Kong2010}
\bibinfo{author}{\bibfnamefont{Y.}~\bibnamefont{Kong}},
  \bibinfo{author}{\bibfnamefont{W.}~\bibnamefont{Xiong}},
  \bibinfo{author}{\bibfnamefont{H.}~\bibnamefont{Guo}},
  \bibinfo{author}{\bibfnamefont{W.}~\bibnamefont{Sun}},
  \bibinfo{author}{\bibfnamefont{Y.}~\bibnamefont{Du}}, \bibnamefont{and}
  \bibinfo{author}{\bibfnamefont{Y.}~\bibnamefont{Zhou}},
  \bibinfo{journal}{CALPHAD-Computer Coupling of Phase Diagrams And
  Thermochemistry} \textbf{\bibinfo{volume}{34}}, \bibinfo{pages}{245}
  (\bibinfo{year}{2010}).

\bibitem[{\citenamefont{Pou et~al.}(2002)\citenamefont{Pou, Flores, Ortega,
  P{\'e}rez, and Yeyati}}]{Pou2002}
\bibinfo{author}{\bibfnamefont{P.}~\bibnamefont{Pou}},
  \bibinfo{author}{\bibfnamefont{F.}~\bibnamefont{Flores}},
  \bibinfo{author}{\bibfnamefont{J.}~\bibnamefont{Ortega}},
  \bibinfo{author}{\bibfnamefont{R.}~\bibnamefont{P{\'e}rez}},
  \bibnamefont{and} \bibinfo{author}{\bibfnamefont{A.~L.}
  \bibnamefont{Yeyati}}, \bibinfo{journal}{J. Phys.: Condens. Matter}
  \textbf{\bibinfo{volume}{14}}, \bibinfo{pages}{L421} (\bibinfo{year}{2002}).

\bibitem[{\citenamefont{Postnikov et~al.}(2006)\citenamefont{Postnikov,
  Bihlmayer, and Bl{\"u}gel}}]{Postnikov2006}
\bibinfo{author}{\bibfnamefont{A.~V.} \bibnamefont{Postnikov}},
  \bibinfo{author}{\bibfnamefont{G.}~\bibnamefont{Bihlmayer}},
  \bibnamefont{and}
  \bibinfo{author}{\bibfnamefont{S.}~\bibnamefont{Bl{\"u}gel}},
  \bibinfo{journal}{Comput. Mater. Sci.} \textbf{\bibinfo{volume}{36}},
  \bibinfo{pages}{91} (\bibinfo{year}{2006}).

\bibitem[{\citenamefont{Garc{\'{\i}}a-Su{\'a}rez
  et~al.}(2004)\citenamefont{Garc{\'{\i}}a-Su{\'a}rez, Newman, Lambert,
  Pruneda, and Ferrer}}]{Garcia-Suarez2004}
\bibinfo{author}{\bibfnamefont{V.~M.} \bibnamefont{Garc{\'{\i}}a-Su{\'a}rez}},
  \bibinfo{author}{\bibfnamefont{C.~M.} \bibnamefont{Newman}},
  \bibinfo{author}{\bibfnamefont{C.~J.} \bibnamefont{Lambert}},
  \bibinfo{author}{\bibfnamefont{J.~M.} \bibnamefont{Pruneda}},
  \bibnamefont{and} \bibinfo{author}{\bibfnamefont{J.}~\bibnamefont{Ferrer}},
  \bibinfo{journal}{Eur. Phys. J. B} \textbf{\bibinfo{volume}{40}},
  \bibinfo{pages}{371} (\bibinfo{year}{2004}).

\bibitem[{\citenamefont{Zhang et~al.}(2010)\citenamefont{Zhang, Punkkinen,
  Johansson, Hertzman, and Vitos}}]{Zhang2010}
\bibinfo{author}{\bibfnamefont{H.}~\bibnamefont{Zhang}},
  \bibinfo{author}{\bibfnamefont{M.~P.~J.} \bibnamefont{Punkkinen}},
  \bibinfo{author}{\bibfnamefont{B.}~\bibnamefont{Johansson}},
  \bibinfo{author}{\bibfnamefont{S.}~\bibnamefont{Hertzman}}, \bibnamefont{and}
  \bibinfo{author}{\bibfnamefont{L.}~\bibnamefont{Vitos}},
  \bibinfo{journal}{Phys. Rev. B} \textbf{\bibinfo{volume}{81}},
  \bibinfo{pages}{184105} (\bibinfo{year}{2010}).

\bibitem[{\citenamefont{Kittel}(1996)}]{Kittel1996}
\bibinfo{author}{\bibfnamefont{C.}~\bibnamefont{Kittel}},
  \emph{\bibinfo{title}{Introduction to Solid State Physics}}
  (\bibinfo{publisher}{John Wiley \& Sons}, \bibinfo{address}{New York},
  \bibinfo{year}{1996}), \bibinfo{edition}{7th} ed.

\bibitem[{\citenamefont{Philipsen and Baerends}(1996)}]{Philipsen1996}
\bibinfo{author}{\bibfnamefont{P.~H.~T.} \bibnamefont{Philipsen}}
  \bibnamefont{and} \bibinfo{author}{\bibfnamefont{E.~J.}
  \bibnamefont{Baerends}}, \bibinfo{journal}{Phys. Rev. B}
  \textbf{\bibinfo{volume}{54}}, \bibinfo{pages}{5326} (\bibinfo{year}{1996}).

\bibitem[{\citenamefont{Rayne and Chandrasekhar}(1961)}]{Rayne1961}
\bibinfo{author}{\bibfnamefont{J.~A.} \bibnamefont{Rayne}} \bibnamefont{and}
  \bibinfo{author}{\bibfnamefont{B.~S.} \bibnamefont{Chandrasekhar}},
  \bibinfo{journal}{Phys. Rev.} \textbf{\bibinfo{volume}{122}},
  \bibinfo{pages}{1714} (\bibinfo{year}{1961}).

\bibitem[{\citenamefont{Adams et~al.}(2006)\citenamefont{Adams, Agosta,
  Leisure, and Ledbetter}}]{Adams2006}
\bibinfo{author}{\bibfnamefont{J.~J.} \bibnamefont{Adams}},
  \bibinfo{author}{\bibfnamefont{D.~S.} \bibnamefont{Agosta}},
  \bibinfo{author}{\bibfnamefont{R.~G.} \bibnamefont{Leisure}},
  \bibnamefont{and}
  \bibinfo{author}{\bibfnamefont{H.}~\bibnamefont{Ledbetter}},
  \bibinfo{journal}{J. Appl. Phys.} \textbf{\bibinfo{volume}{100}},
  \bibinfo{pages}{113530} (\bibinfo{year}{2006}).

\bibitem[{\citenamefont{Behler}(2004)}]{Behler2004}
\bibinfo{author}{\bibfnamefont{J.}~\bibnamefont{Behler}}, Ph.D. thesis,
  \bibinfo{school}{Von der Fakult\"at II-Mathematik und Naturwissenschaften der
  Technischen Universit\"at Berlin} (\bibinfo{year}{2004}).

\bibitem[{\citenamefont{Gunnarsson and Jones}(1985)}]{Gunnarsson1985}
\bibinfo{author}{\bibfnamefont{O.}~\bibnamefont{Gunnarsson}} \bibnamefont{and}
  \bibinfo{author}{\bibfnamefont{R.~O.} \bibnamefont{Jones}},
  \bibinfo{journal}{Phys. Rev. B} \textbf{\bibinfo{volume}{31}},
  \bibinfo{pages}{7588} (\bibinfo{year}{1985}).

\bibitem[{\citenamefont{Lide}(2007)}]{CRChandbook2007}
\bibinfo{editor}{\bibfnamefont{D.~R.} \bibnamefont{Lide}}, ed.,
  \emph{\bibinfo{title}{CRC handbook of chemistry and physics}}
  (\bibinfo{publisher}{Taylor and Francis}, \bibinfo{address}{Boca Raton, FL},
  \bibinfo{year}{2007}), \bibinfo{edition}{internet version 2007, (87th
  edition)} ed.

\bibitem[{\citenamefont{Morin}(1950)}]{Morin1950}
\bibinfo{author}{\bibfnamefont{F.~J.} \bibnamefont{Morin}},
  \bibinfo{journal}{Phys. Rev.} \textbf{\bibinfo{volume}{78}},
  \bibinfo{pages}{819} (\bibinfo{year}{1950}).

\bibitem[{\citenamefont{Greaves}(1983)}]{Greaves1983}
\bibinfo{author}{\bibfnamefont{C.}~\bibnamefont{Greaves}},
  \bibinfo{journal}{Journal of Solid State Chemistry}
  \textbf{\bibinfo{volume}{49}}, \bibinfo{pages}{325} (\bibinfo{year}{1983}).

\bibitem[{\citenamefont{Somogyv{\'a}ri
  et~al.}(2002)\citenamefont{Somogyv{\'a}ri, Sv{\'a}b, M{\'e}sz{\'a}ros,
  Kreznov, Nedkov, Saj{\'o}, and Bour{\'e}e}}]{Somogyvari2002}
\bibinfo{author}{\bibfnamefont{Z.}~\bibnamefont{Somogyv{\'a}ri}},
  \bibinfo{author}{\bibfnamefont{E.}~\bibnamefont{Sv{\'a}b}},
  \bibinfo{author}{\bibfnamefont{G.}~\bibnamefont{M{\'e}sz{\'a}ros}},
  \bibinfo{author}{\bibfnamefont{K.}~\bibnamefont{Kreznov}},
  \bibinfo{author}{\bibfnamefont{I.}~\bibnamefont{Nedkov}},
  \bibinfo{author}{\bibfnamefont{I.}~\bibnamefont{Saj{\'o}}}, \bibnamefont{and}
  \bibinfo{author}{\bibfnamefont{F.}~\bibnamefont{Bour{\'e}e}},
  \bibinfo{journal}{Appl. Phys. A} \textbf{\bibinfo{volume}{74}},
  \bibinfo{pages}{S1077} (\bibinfo{year}{2002}).

\bibitem[{\citenamefont{Gualtieri and Venturelli}(1999)}]{Gualtieri1999}
\bibinfo{author}{\bibfnamefont{A.~F.} \bibnamefont{Gualtieri}}
  \bibnamefont{and}
  \bibinfo{author}{\bibfnamefont{P.}~\bibnamefont{Venturelli}},
  \bibinfo{journal}{Am. Mineral.} \textbf{\bibinfo{volume}{84}},
  \bibinfo{pages}{895} (\bibinfo{year}{1999}).

\bibitem[{\citenamefont{Nagai et~al.}(2003)\citenamefont{Nagai, Kagi, and
  Yamanaka}}]{Nagai2003}
\bibinfo{author}{\bibfnamefont{T.}~\bibnamefont{Nagai}},
  \bibinfo{author}{\bibfnamefont{H.}~\bibnamefont{Kagi}}, \bibnamefont{and}
  \bibinfo{author}{\bibfnamefont{T.}~\bibnamefont{Yamanaka}},
  \bibinfo{journal}{Am. Mineral.} \textbf{\bibinfo{volume}{88}},
  \bibinfo{pages}{1423} (\bibinfo{year}{2003}).

\bibitem[{\citenamefont{Gleason et~al.}(2008)\citenamefont{Gleason, Jeanloz,
  and Kunz}}]{Gleason2008}
\bibinfo{author}{\bibfnamefont{A.~E.} \bibnamefont{Gleason}},
  \bibinfo{author}{\bibfnamefont{R.}~\bibnamefont{Jeanloz}}, \bibnamefont{and}
  \bibinfo{author}{\bibfnamefont{M.}~\bibnamefont{Kunz}}, \bibinfo{journal}{Am.
  Mineral.} \textbf{\bibinfo{volume}{93}}, \bibinfo{pages}{1882}
  (\bibinfo{year}{2008}).

\bibitem[{\citenamefont{Christensen and Christensen}(1978)}]{Christensen1978}
\bibinfo{author}{\bibfnamefont{H.}~\bibnamefont{Christensen}} \bibnamefont{and}
  \bibinfo{author}{\bibfnamefont{A.~N.} \bibnamefont{Christensen}},
  \bibinfo{journal}{Acta Chem. Scand. A} \textbf{\bibinfo{volume}{32}},
  \bibinfo{pages}{87} (\bibinfo{year}{1978}).

\bibitem[{\citenamefont{Ewing}(1935)}]{Ewing1935}
\bibinfo{author}{\bibfnamefont{E.~J.} \bibnamefont{Ewing}},
  \bibinfo{journal}{J. Chem. Phys.} \textbf{\bibinfo{volume}{3}},
  \bibinfo{pages}{420} (\bibinfo{year}{1935}).

\bibitem[{\citenamefont{Ole{\'s} et~al.}(1970)\citenamefont{Ole{\'s},
  Szytu{\l}a, and Wanic}}]{Oles1970}
\bibinfo{author}{\bibfnamefont{A.}~\bibnamefont{Ole{\'s}}},
  \bibinfo{author}{\bibfnamefont{A.}~\bibnamefont{Szytu{\l}a}},
  \bibnamefont{and} \bibinfo{author}{\bibfnamefont{A.}~\bibnamefont{Wanic}},
  \bibinfo{journal}{Phys. Stat. Sol.} \textbf{\bibinfo{volume}{41}},
  \bibinfo{pages}{173} (\bibinfo{year}{1970}), \bibinfo{note}{the journal
  should be physica status solidi (b).}

\bibitem[{\citenamefont{Verwey}(1939)}]{Verwey1939}
\bibinfo{author}{\bibfnamefont{E.~J.~W.} \bibnamefont{Verwey}},
  \bibinfo{journal}{Nature} \textbf{\bibinfo{volume}{144}},
  \bibinfo{pages}{327} (\bibinfo{year}{1939}).

\bibitem[{\citenamefont{Walz}(2002)}]{Walz2002}
\bibinfo{author}{\bibfnamefont{F.}~\bibnamefont{Walz}}, \bibinfo{journal}{J.
  Phys.: Condens. Matter} \textbf{\bibinfo{volume}{14}}, \bibinfo{pages}{R285}
  (\bibinfo{year}{2002}).

\bibitem[{\citenamefont{Mizoguchi}(1978{\natexlab{a}})}]{Mizoguchi1978}
\bibinfo{author}{\bibfnamefont{M.}~\bibnamefont{Mizoguchi}},
  \bibinfo{journal}{J. Phys. Soc. Jpn.} \textbf{\bibinfo{volume}{44}},
  \bibinfo{pages}{1501} (\bibinfo{year}{1978}{\natexlab{a}}).

\bibitem[{\citenamefont{Mizoguchi}(1978{\natexlab{b}})}]{Mizoguchi1978a}
\bibinfo{author}{\bibfnamefont{M.}~\bibnamefont{Mizoguchi}},
  \bibinfo{journal}{J. Phys. Soc. Jpn.} \textbf{\bibinfo{volume}{44}},
  \bibinfo{pages}{1512} (\bibinfo{year}{1978}{\natexlab{b}}).

\bibitem[{\citenamefont{Nov{\'a}k et~al.}(2000)\citenamefont{Nov{\'a}k,
  {\u{S}t\u{e}p\'{a}nkov\'{a}}, Englich, and Kohout}}]{Novak2000}
\bibinfo{author}{\bibfnamefont{P.}~\bibnamefont{Nov{\'a}k}},
  \bibinfo{author}{\bibfnamefont{H.}~\bibnamefont{{\u{S}t\u{e}p\'{a}nkov\'{a}}}},
  \bibinfo{author}{\bibfnamefont{J.}~\bibnamefont{Englich}}, \bibnamefont{and}
  \bibinfo{author}{\bibfnamefont{J.}~\bibnamefont{Kohout}},
  \bibinfo{journal}{Phys. Rev. B} \textbf{\bibinfo{volume}{61}},
  \bibinfo{pages}{1256} (\bibinfo{year}{2000}).

\bibitem[{\citenamefont{Wright et~al.}(2002)\citenamefont{Wright, Attfield, and
  Radaelli}}]{Wright2002}
\bibinfo{author}{\bibfnamefont{J.~P.} \bibnamefont{Wright}},
  \bibinfo{author}{\bibfnamefont{J.~P.} \bibnamefont{Attfield}},
  \bibnamefont{and} \bibinfo{author}{\bibfnamefont{P.~G.}
  \bibnamefont{Radaelli}}, \bibinfo{journal}{Phys. Rev. B}
  \textbf{\bibinfo{volume}{66}}, \bibinfo{pages}{214422}
  (\bibinfo{year}{2002}).

\bibitem[{\citenamefont{Wilkins et~al.}(2009)\citenamefont{Wilkins, Matteo,
  Beale, Joly, Mazzoli, Hatton, Bencok, Yakhou, and Barbers}}]{Wilkins2009}
\bibinfo{author}{\bibfnamefont{S.~B.} \bibnamefont{Wilkins}},
  \bibinfo{author}{\bibfnamefont{S.~D.} \bibnamefont{Matteo}},
  \bibinfo{author}{\bibfnamefont{T.~A.~W.} \bibnamefont{Beale}},
  \bibinfo{author}{\bibfnamefont{Y.}~\bibnamefont{Joly}},
  \bibinfo{author}{\bibfnamefont{C.}~\bibnamefont{Mazzoli}},
  \bibinfo{author}{\bibfnamefont{P.~D.} \bibnamefont{Hatton}},
  \bibinfo{author}{\bibfnamefont{P.}~\bibnamefont{Bencok}},
  \bibinfo{author}{\bibfnamefont{F.}~\bibnamefont{Yakhou}}, \bibnamefont{and}
  \bibinfo{author}{\bibfnamefont{V.~A.~M.} \bibnamefont{Barbers}},
  \bibinfo{journal}{Phys. Rev. B} \textbf{\bibinfo{volume}{79}},
  \bibinfo{pages}{201102R} (\bibinfo{year}{2009}).

\bibitem[{\citenamefont{Miyamoto and Shindo}(1993)}]{Miyamoto1993}
\bibinfo{author}{\bibfnamefont{Y.}~\bibnamefont{Miyamoto}} \bibnamefont{and}
  \bibinfo{author}{\bibfnamefont{M.}~\bibnamefont{Shindo}},
  \bibinfo{journal}{J. Phys. Soc. Jpn.} \textbf{\bibinfo{volume}{62}},
  \bibinfo{pages}{1423} (\bibinfo{year}{1993}).

\bibitem[{\citenamefont{Majzlan et~al.}(2003)\citenamefont{Majzlan, Greavel,
  and Navrotsky}}]{Majzlan2003}
\bibinfo{author}{\bibfnamefont{J.}~\bibnamefont{Majzlan}},
  \bibinfo{author}{\bibfnamefont{K.-D.} \bibnamefont{Greavel}},
  \bibnamefont{and}
  \bibinfo{author}{\bibfnamefont{A.}~\bibnamefont{Navrotsky}},
  \bibinfo{journal}{Am. Mineral.} \textbf{\bibinfo{volume}{88}},
  \bibinfo{pages}{855} (\bibinfo{year}{2003}).

\bibitem[{\citenamefont{Zhang et~al.}(2004)\citenamefont{Zhang, Smith, and
  Wang}}]{ZhangW2004}
\bibinfo{author}{\bibfnamefont{W.}~\bibnamefont{Zhang}},
  \bibinfo{author}{\bibfnamefont{J.~R.} \bibnamefont{Smith}}, \bibnamefont{and}
  \bibinfo{author}{\bibfnamefont{X.-G.} \bibnamefont{Wang}},
  \bibinfo{journal}{Phys. Rev. B} \textbf{\bibinfo{volume}{70}},
  \bibinfo{pages}{024103} (\bibinfo{year}{2004}).

\bibitem[{\citenamefont{Guo and Qi}(2010)}]{Guo2010}
\bibinfo{author}{\bibfnamefont{H.}~\bibnamefont{Guo}} \bibnamefont{and}
  \bibinfo{author}{\bibfnamefont{Y.}~\bibnamefont{Qi}},
  \bibinfo{journal}{Modelling and Simulation in Materials Science and
  Engineering} \textbf{\bibinfo{volume}{18}}, \bibinfo{pages}{034008}
  (\bibinfo{year}{2010}).

\bibitem[{\citenamefont{Chase}(1985)}]{Chase1985}
\bibinfo{editor}{\bibfnamefont{M.~W.} \bibnamefont{Chase}}, ed.,
  \emph{\bibinfo{title}{JANAF thermochemical tables}} (\bibinfo{publisher}{ACS
  Publishing}, \bibinfo{address}{Washington, D. C.}, \bibinfo{year}{1985}),
  \bibinfo{edition}{3rd} ed.

\end{thebibliography}

\end{document}